\documentclass[amsmath, amssymb, aps, showpacs, prd, twocolumn, superscriptaddress, nofootinbib]{revtex4-1}
\usepackage[T1]{fontenc}
\usepackage{ae,aecompl}
\usepackage{enumerate}
\usepackage[colorlinks=true, linkcolor=blue, citecolor=red]{hyperref}
\usepackage{graphicx}
\usepackage{color}
\usepackage[normalem]{ulem}
\usepackage{empheq}
\usepackage[justification=centerlast, singlelinecheck=false, caption=false]{subfig}

\newcommand{\rd}{\mathrm{d}}
\newcommand{\re}{\mathrm{e}}
\newcommand{\eV}{\,\mathrm{eV}}
\newcommand*\widefbox[1]{\fbox{#1}}

\renewcommand{\vec}{\boldsymbol}

\begin{document}

\title{The Universe as a Cosmic String}

\author{Florian Niedermann}
\email[]{florian.niedermann@physik.uni-muenchen.de}
\affiliation{Arnold Sommerfeld Center for Theoretical Physics, Ludwig-Maximilians-Universit\"at, Theresienstra{\ss}e 37, 80333 Munich, Germany}
\author{Robert Schneider}
\email[]{robert.bob.schneider@physik.uni-muenchen.de}
\affiliation{Arnold Sommerfeld Center for Theoretical Physics, Ludwig-Maximilians-Universit\"at, Theresienstra{\ss}e 37, 80333 Munich, Germany}
\author{Stefan Hofmann}
\email[]{stefan.hofmann@physik.uni-muenchen.de}
\affiliation{Arnold Sommerfeld Center for Theoretical Physics, Ludwig-Maximilians-Universit\"at, Theresienstra{\ss}e 37, 80333 Munich, Germany}
\author{Justin Khoury}
\email[]{jkhoury@sas.upenn.edu}
\affiliation{Center for Particle Cosmology, Department of Physics and Astronomy, University of Pennsylvania, Philadelphia, PA 19104, USA}
\date{\today}

\begin{abstract}

The cosmology of brane induced gravity in six infinite dimensions is investigated. It is shown that a brane with Friedmann-Robertson-Walker symmetries necessarily acts as a source of cylindrically symmetric gravitational waves, so called Einstein-Rosen waves. Their existence essentially distinguishes this model from its codimension-one counterpart and necessitates solving the nonlinear system of bulk and brane-matching equations. A numerical analysis is performed and two qualitatively different and dynamically separated classes of cosmologies are derived: degravitating solutions for which the Hubble parameter settles to zero despite the presence of a non-vanishing energy density on the brane and super-accelerating solutions for which Hubble grows unbounded. The parameter space of both the stable and unstable regime is derived and observational consequences are discussed: It is argued that the degravitating regime does not allow for a phenomenologically viable cosmology. On the other hand, the super-accelerating solutions are potentially viable, however, their unstable behavior questions their physical relevance.  

\end{abstract}


\maketitle

\section{Introduction}
\label{sec:introduction}

We are in the golden age of observational cosmology, in which General Relativity (GR) is being put to the test at the largest observable distances~\cite{Jain:2010ka,Joyce:2014kja}. Consequently, it has become an important task to develop consistent competitor theories which modify $\Lambda$CDM predictions on cosmological scales. Moreover, there is still no fundamental understanding of the dark sector, which constitutes the main part of the energy budget in the $\Lambda$CDM model. The most pressing issue from a theory standpoint is the cosmological constant problem (see \cite{Weinberg:1988cp} for a seminal work and \cite{Burgess:2011va} for a more recent discussion). This provides a strong motivation to look for consistent infrared modifications of gravity.

A prominent candidate is the model of \textit{brane induced gravity} (BIG) \cite{Dvali:2000hr, Dvali:2000xg} according to which our four dimensional universe (the brane) and all its matter content is localized in a $d$-dimensional infinite space-time (the bulk). Despite the fact that the extra dimensions are infinite in extent, 4D gravity is nevertheless recovered at short enough distances on the brane, thanks to an intrinsic Einstein-Hilbert term (or brane induced gravity term) on the brane. This results in a modification of gravity characterized by a single length scale $r_c$ which discriminates between two gravitational regimes: a conventional 4D regime on scales $\ell\ll r_c$, for which the Newtonian potential is proportional to $1/r$ up to small corrections; and a $d$-dimensional regime on scales $\ell\gg r_c$, for which gravity on the brane is effectively weakened and the scaling becomes $1/r^{d-3}$. In order to be in accordance with gravitational measurements on solar system scales, the \textit{cross-over} scale $r_c$ has to be large enough, {\it e.g.}, for $d=5$ lunar laser ranging experiments demand  $r_c^{(5)} \gtrsim 0.04 H_0^{-1}$~\cite{Afshordi:2008rd}.  Thus, cosmology represents the ideal playground for testing these theories. 

Brane induced gravity models are interesting also for other reasons. At the linear level, the effective 4D graviton is a resonance, {\it i.e.}, an infinite superposition of massive graviton states. Historically it turned out to be notoriously difficult to give a mass to the 4D graviton on a nonlinear level without introducing Boulware-Deser ghost instabilities (for recent reviews, see~\cite{Hinterbichler:2011tt,deRham:2014zqa}). This has been achieved recently with dRGT gravity~\cite{deRham:2010kj}. Extra dimensional constructions, such as BIG, offer promising arenas to devise ghost-free examples. Another motivation comes from the {\it degravitation} approach to the cosmological constant problem~\cite{Dvali:2002pe, Dvali:2002fz, ArkaniHamed:2002fu, Dvali:2007kt, deRham:2007rw}. The massive/resonant graviton leads to a weakening of the gravitational force law at large distances, which makes gravity effectively insensitive to a large cosmological constant. There are linear~\cite{deRham:2007rw} and nonlinear~\cite{Charmousis:2001hg} indications for that claim.

The best-known and most extensively studied example is the Dvali-Gabadadze-Porrati (DGP) model~\cite{Dvali:2000hr}, corresponding to $d=5$. The cross-over scale in this case is given by $r_c^{(5)} = \frac{M_{\rm Pl}^2}{2M_5^3}$, where
$M_5$ is the bulk Planck scale. For cosmology, the DGP setup gives rise to a modified Friedmann equation~\cite{Deffayet:2000uy}, $H^2 \pm \frac{H}{r_c^{(5)}} = \frac{\rho}{3M_{\rm Pl}^2}$, featuring an additional term controlled by $r_c$. Accordingly, the modification can be neglected for early times and large curvature ($H \gg 1/r_c^{(5)}$), whereas it becomes significant at late times and small curvature ($H \lesssim 1/r_c^{(5)}$). The plus and minus sign correspond to two different branches of solutions, the ``normal'' and  the ``self-accelerating'' branch, respectively. The former is characterized by a weakening of gravity since the energy density gets effectively reduced, while the latter describes a gravitational enhancement. The self-accelerated branch is widely believed to suffer from perturbative ghost instabilities~\cite{Luty:2003vm,Nicolis:2004qq,Koyama:2005tx,Charmousis:2006pn,Gregory:2007xy,Gorbunov:2005zk}. The normal branch is perturbatively stable. Confronting DGP with cosmological observations yields a rather stringent bound on the cross-over scale: $r_c^{(5)} \gtrsim  3 H_0^{-1}$~\cite{Lombriser:2009}.

A natural generalization of the DGP model are higher-codimension scenarios ($d>5$) \cite{Dvali:2000xg}. Several difficulties have impeded their development: 

\begin{itemize}

\item According to claims in the literature, the model propagates a linear ghost on a Minkowski background~\cite{Dubovsky:2002jm,Hassan:2010ys}, which questions the quantum consistency of the whole theory.

\item Bulk fields are generically divergent at the position of a higher-codimension brane and require a regularization prescription.

\item  A non-trivial cosmology on the brane implies the existence of gravitational waves which are emitted into the bulk. (In $d=5$, the symmetries of the geometry imply a static bulk, because there is a generalization of Birkhoff's theorem to planar symmetry~\cite{Taub:1951}. However, no such theorem exists for cylindrical symmetry, and Einstein-Rosen waves \cite{EinsteinRosen1937} are in fact a counter-example.) Including these waves in the dynamical description makes it much more difficult to solve the full system.
 
\end{itemize}
 
The first point, which clearly would be the most severe, was recently proven to be wrong~\cite{Berkhahn:2012wg}. Through a detailed constraint analysis, it was shown rigorously in~\cite{Berkhahn:2012wg} that the would-be ghost mode is {\it not} dynamical and is instead subject to a constraint. This is analogous to the conformal mode of standard 4D GR. For $d=6$ the positive definiteness of the Hamiltonian was explicitly shown in~\cite{Berkhahn:2012wg}. Consequently, in a weakly coupling regime on a Minkowski background the model is healthy. This result offered a new window of opportunity for investigating consistently modified cosmologies at the largest observable scales.  

In the present paper we explore cosmological solutions in the simplest case: brane induced gravity in $ d=6 $ dimensions. Those solutions are obviously interesting for observational purposes, but they also test the non-perturbative stability of the model. 

To overcome the second issue listed above, we introduce in Sec.~\ref{sec:model} a regularization which replaces the infinitely thin brane by a hollow cylinder of finite size $R$. We stabilize this size by introducing an appropriate azimuthal pressure. The microscopical origin of this pressure component is not specified, but we check {\it a posteriori} whether the required source is physically reasonable ({\it i.e.}, whether it satisfies the standard energy conditions). 

We first check the consistency of our framework by deriving known solutions for a static cosmic string in 6D in Sec.~\ref{sec:static_sol}. Based on these solutions the geometry of the setup is illustrated and a distinction between sub- and super-critical branes is motivated. 

According to the third issue listed above, which is discussed in more detail in Sec.~\ref{sec:Brane_Bulk}, a key feature of the higher codimensional models is the existence of bulk gravitational waves which are emitted by the brane and affect its dynamics. For $d=6$ they correspond to a higher-dimensional generalization of Einstein-Rosen waves. Consequently, we must resort to numerics, introduced in Sec.~\ref{sec:num_impl}, to find the most general solutions. 

We then solve Einstein's field equations in the bulk in the presence of FRW matter (and brane induced gravity terms) on the brane and present the results in Sec.~\ref{sec:num_sol}. We stress that these solutions have been derived from the full system of nonlinear Einstein equations without making any approximations or additional assumptions other than having FRW symmetries on the brane and a source-free bulk. This result makes it possible for the first time to discuss the phenomenological viability of the six dimensional BIG model with respect to cosmological observations.

Depending on the model parameters, we find two qualitatively different classes of solutions: 

\begin{itemize}

\item { \it Degravitating} solutions for which the system approaches the static cosmic string solution, {\it i.e.}, the 4D Hubble parameter becomes zero despite the presence of a non-vanishing on-brane source. 

\item { \it Super-accelerating} solutions for which Hubble grows unbounded for late times.

\end{itemize}

The solution of the first type constitutes the first example of a dynamically realized degravitation mechanism. Accordingly, the brane tension is shielded from a 4D observer by exclusively contributing to extrinsic curvature.
We dismiss the second type due to its pathological run-away behavior.  In addition, the effective energy density that sources 6D gravity turns negative for these solutions. This bears strong resemblance with the self-accelerating branch in the DGP model and thus questions their perturbative quantum stability.

It is shown that the degravitating and super-accelerating solutions are separated by a physical singularity. Thus, it is not possible to dynamically evolve from one regime to the other. We derive an {\it analytic} expression for the separating surface in parameter space. This in turn allows us to derive a necessary condition to be in the degravitating regime:
\begin{equation}\label{eq:stab_bound}
	\left( H r_c\right)^2 < \frac{3}{2} \left|H\right| R\,,
\end{equation}
with $2 \pi R$ the circumference of the cylinder and $r_c$ the crossover scale\footnote{Here and henceforth, $ r_c $ refers to the 6D crossover scale, defined below in \eqref{eq:defCrossover}.}. However, a phenomenologically viable solution has to fulfill two requirements: First, $H r_c \gg 1$ for early times which ensures that the deviation from standard Friedmann cosmology is small. Second,  $H R \ll 1$ in order to be insensitive to unknown UV physics that led to the formation of the brane. Obviously, these two conditions are  incompatible with the bound~\eqref{eq:stab_bound}. {\it As a consequence of these considerations, the degravitating solutions are ruled out phenomenologically.} 

We conclude in Sec.~\ref{sec:conclusion} with some remarks on super-critical energy densities. 
A number of technical results have been relegated to a series of appendices. In particular, we repeat the analysis with a different regularization scheme in Appendix~\ref{ap:dynReg} to check the insensitivity of our results to the regularization details.

We adopt the following notational conventions: capital Latin indices $ A, B, \dots $ denote six-dimensional, small Latin indices $ a, b, \ldots $ five-dimensional, and Greek indices $ \alpha, \beta, \ldots $ four-dimensional  space-time indices. Small Latin indices $ i, j, \ldots $ run over the three large spatial on-brane dimensions and corresponding vectors are written in boldface.
The space-time dimensionality $ d $ of some quantity $ Q $ is sometimes made explicit by writing $ Q^{(d)} $.
Our sign conventions are ``$ +++ $'' as defined (and adopted) in~\cite{Misner}.
We work in units in which $ c = \hbar = 1 $.
\section{The model}
\label{sec:model}

The action of the BIG model in $D=4+n$ dimensions is the sum of three terms:
\begin{align} 
	\label{eq:ActionBIG}
	\mathcal{S}=
	\mathcal{S}_{\rm EH}+
	\mathcal{S}_{\rm BIG}+
	\mathcal{S}_{\rm m}[h]\;.
\end{align}
The first term,
\begin{align}
	\mathcal{S}_{\rm EH}=
	M_{D}^{D-2}\int {\rm d}^D X\;\sqrt{-g}\; \mathcal{R}^{(D)}\,,
\end{align}
describes Einstein-Hilbert gravity in $D$ infinite space-time dimensions. The bulk Planck scale is denoted by $M_D$.
The bulk is assumed to be source-free; in particular, the bulk cosmological constant is set to zero for simplicity. 
The second term is the induced gravity term on a codimension-$n$ brane:
\begin{align}
	\label{eq:S_BIG_delta_0}
	\mathcal{S}_{\rm BIG}=M_{\rm Pl}^2\int {\rm d}^4 x\,\sqrt{- h}\; \mathcal{R}^{(4)} \,.
\end{align}
This describes intrinsic gravity on the brane, with $h_{\mu\nu}$ denoting the induced metric. To match standard GR in the 4D regime, $M_{\rm Pl}$ is identified as the usual 4D Planck scale. 
From the effective field theory point of view, the BIG term can be thought to arise from integrating out heavy matter fields on the brane. The last term in~\eqref{eq:ActionBIG}, $\mathcal{S}_{\rm m}[h]$,
is the action for matter fields localized on the brane, which by definition couple to $h_{\mu\nu}$. 

Henceforth we will focus on $D=6$, corresponding to the codimension $n=2$ case.

\subsection{Regularization schemes}
\label{sec:regul_schemes}

In general, a localized codimension-two source leads to a singular geometry, {\it i.e.}, the bulk metric diverges logarithmically at the position of the brane. This is well known for static solutions, reviewed in Sec.~\ref{sec:static_sol}. For the pure tension case, the space-time develops a conical singularity---the bulk geometry stays flat arbitrarily close to the brane but diverges exactly at the brane. For more general static and non-static solutions we have to deal with curvature singularities other than the purely conical one. These singularities can be properly dealt with by introducing a certain brane width.

In this work, we adopt a regularization which consists of blowing up the brane to a circle of circumference $2 \pi R$~\cite{Kaloper:2007ap, Burgess:2008yx}. In other words, the brane is now a codimension-one object, with topology ${\cal M}_4\times \mathcal{S}_1$. The matter fields are smeared out on the $\mathcal{S}_1$. This amounts to the substitution 
\begin{equation}
\label{reg_sub}
\mathcal{S}_{\rm BIG} ~\longrightarrow~ M_5^3\int_{\mathcal{M}_4 \times \mathcal{S}_1}\!\!\!\!\! {\rm d}^5 x\,\sqrt{- h^{(5)}}\; \mathcal{R}^{(5)}	\;,
\end{equation}  
where $M_5^3=\frac{M_{\rm Pl}^2}{2 \pi R}$, and $h^{(5)}_{ab}$ is the five dimensional induced metric. 

Furthermore, in the main body of the paper, we follow a \textit{static regularization} scheme, which makes the evolution completely insensitive to the geometry inside the regularized brane.
This scheme can be viewed from two, equivalent perspectives:

\begin{itemize}

\item The brane is a boundary of space-time, and there is no interior geometry to speak of. This is the {\it hollow cylinder} perspective. In this case, the equations of motion consist of Einstein's field equations in the exterior, supplemented by Israel's junction conditions~\cite{Israel:1966, Israel:1967} at the brane,
\begin{eqnarray}
\nonumber
	T^{(5) a}_{\hphantom{(5)a} b}-M_5^3 G^{(5)a}_{\hphantom{[h]a} b} &=& M_6^4 \big(K^{~c}_{{\rm out}\,c} \delta^{a}_{\hphantom{a}b} - K_{{\rm out}\,b}^{~a}\big) \\
	&-& \frac{1}{R} \left( \delta^{a}_{\hphantom{a}b} - \delta^a_{~\phi}\delta^{\phi}_{\hphantom{\phi}b}\right)\,,
\label{israel}
\end{eqnarray}
where $K_{{\rm out}\,ab}$ is the extrinsic curvature tensor. In the second line, we have extracted from $T^{(5) a}_{\hphantom{(5)a} b}$ a cosmological constant
along $\mathcal{M}_4$. This is necessary to ensure that the deficit angle vanishes when $T^{(5) a}_{\hphantom{(5)a} b} \rightarrow 0$.

\item The brane has an interior geometry, such that the junction condition now becomes
\begin{equation}
T^{(5) a}_{\hphantom{(5)a} b}-M_5^3 G^{(5)a}_{\hphantom{[h]a} b} = M_6^4 \big([K^c_{~c}] \delta^{a}_{\hphantom{a}b} - [K_{~b}^{a}]\big)\,,
\label{israel2}
\end{equation}
where $[K_{ab}] \equiv K_{{\rm out}\, ab} - K_{{\rm in}\, ab}$. However, to ensure that the interior region does not introduce any dynamics on the brane,
we demand that $K_{ab}^{\rm in}$ is equal to a constant value corresponding to a static cylinder:
\begin{align}
K^{~\phi}_{{\rm in}\,\phi}=\frac{1}{R}\,; \qquad K^{~0}_{{\rm in}\,0} = K^{~i}_{{\rm in}\,j}= 0 \,.
\label{eq:RegII}
\end{align}  
With this choice, the junction condition~\eqref{israel2} agrees with~\eqref{israel}, and the two descriptions give identical brane geometry and
exterior space-time. We will not be concerned with the brane interior.

\end{itemize} 

{\it A priori} one naturally expects that the solutions thus obtained should not depend sensitively on the details of the regularization,
as long as the characteristic time scale ($H^{-1}$, in the case of interest) is much longer than the radius of the circle, {\it i.e.},
\begin{equation}\label{UV_IR_Limit}
H^{-1}\gg R\;. 
\end{equation}
We explicitly check this expectation in Appendix~\ref{ap:dynReg}, by studying a different regularization scheme called \textit{dynamical regularization}.
In this scheme, the gravitational dynamics are fully resolved inside the cylinder. We find that the time-averaged Hubble evolution
on the brane agrees with the static regularization result in the limit~\eqref{UV_IR_Limit}. 

Let us stress that only by performing this fully self-consistent GR analysis, which in particular implements regularity at the symmetry axis, was it possible to quantify the effect of having some interior dynamics and thus to show that our results are regularization independent. Moreover, this analysis revealed that the static regularization corresponds to the favorable case where the effects of the interior dynamics are minimized and perfectly smoothed out. The presentation in the main part of the paper therefore uses the simpler static regularization.
The interested reader is referred to the Appendix~\ref{ap:dynReg} for more details.

\subsection{Bulk geometry}

The assumed symmetries are homogeneity, isotropy and (for simplicity) spatial flatness along the three spatial brane dimensions, as well as axial symmetry about the brane. 
As shown in Appendix~\ref{ERcoords}, given these symmetries and the fact that the space-time is empty away from the brane, the bulk metric can be brought to the form:
\begin{align}
\label{eq:met_cyl_symm_2}
	\rd s^2_6 &= \re^{2(\eta - 3\alpha)} \left( -\rd t^2 \!+ \rd r^2 \right) + \re^{2\alpha} \rd  \vec{x}^2 + \re^{-6\alpha} r^2 \rd\phi^2 \; .
\end{align}
Note that by formally replacing $3\alpha \rightarrow \alpha$ in the first and last term and $\vec{x} \rightarrow z$, we recover the ansatz that was used by Einstein and Rosen to derive the existence of cylindrically symmetric waves in GR \cite{EinsteinRosen1937} (see also, \textit{e.g.},~\cite{Marder1958}). The additional factor 3 in the generalized case simply counts the dimensionality of the symmetry axis. In the remainder of the paper we will refer to \eqref{eq:met_cyl_symm_2} as the \textit{Einstein-Rosen coordinates}.

The Einstein field equations in the exterior (vacuum) region become
\begin{subequations}
\begin{empheq}[box=\widefbox]{align}
		\partial_t^2 \alpha &= \partial_r^2 \alpha + \frac{1}{r}\partial_r\alpha \label{eq:2D_wave}\\
		\partial_r\eta &= 6r\Big((\partial_r\alpha)^2 + (\partial_t\alpha)^2\Big) \label{eq:etaPrime_vac}\\
		\partial_t\eta  &= 12 r \, \partial_r \alpha \, \partial_t\alpha \;. \label{eq:etaDot_vac}
\end{empheq}
\label{eq:einstein_vacuum}
\end{subequations}
The fact that $\alpha$ obeys the linear\footnote{Despite the linearity of this equation, the complete brane-bulk system is still highly nonlinear due to the junction conditions, discussed below.} 2D wave equation \eqref{eq:2D_wave} makes the coordinate choice \eqref{eq:met_cyl_symm_2} unique and especially convenient for numerical implementation.

\subsection{Brane geometry}

The induced cosmological metric on the brane is
\begin{equation}\label{eq:induced_g_3}
	\rd s^2_5	=-\rd \tau^2 +  \re^{2\alpha_0} \rd  \vec{x}^2 + R^2 \rd\phi^2\;,
\end{equation}
where the subscript ``0'' denotes evaluation at the brane position. The scale factor is recognized
as $a(\tau) \equiv \re^{\alpha_0}$, with Hubble parameter $H \equiv \rd \alpha_0/\rd \tau$. 
The proper time $\tau$ is related to the ``bulk'' time via
\begin{equation}\label{eq:dTauDT}
\rd \tau = \frac{\re^{-3\alpha_0}}{\gamma}\rd t \,,
\end{equation}
where 
\begin{equation}\label{eq:defGamma}
	\gamma \equiv \frac{\re^{-\eta_0}}{\sqrt{1 - \left(\frac{{\rm d}r_0}{{\rm d}t}\right)^2}} = \sqrt{\re^{-2\eta_0} +\dot{r}_0^2\re^{-6\alpha_0} }  \;,
\end{equation}
with $r_0(t)$ describing the position of the brane in the extra-dimensional space, and $\dot{r}_0 \equiv \frac{{\rm d}r_0}{{\rm d}\tau}$.
Here and henceforth, dots refer to ${\rm d}/{\rm d}\tau$. 

To recover 4D gravity in the appropriate regime, we assume that the proper circumference (divided by $ 2\pi $) is stabilized:
\begin{equation}
\label{eq:defR}
	R \equiv r_0 \re^{-3\alpha_0} = \text{const.}
\end{equation}
The justification is clear: A realistic defect would have some underlying bulk forces to keep its core stable. Technically, this is imposed
by introducing a suitable azimuthal pressure component $P_\phi $. We must of course check {\it a posteriori} whether the pressure thus inferred
satisfies physically reasonable energy conditions, such as the Null Energy Condition. 

As an immediate consequence of the stabilization condition, the 4D Planck mass, 
\begin{equation}
	M_{\rm Pl}^2 = 2\pi R M_5^3\,,
	\label{M4}
\end{equation}
is constant. Moreover,~\eqref{eq:defR} implies $\dot{r}_0 = 3Hr_0$, which allows us to rewrite~\eqref{eq:defGamma} as
\begin{equation}
	\gamma = \sqrt{\re^{-2\eta_0} +  9H^2R^2} \, .
\label{gammaredef}
\end{equation}

The symmetries of our system allow for a fluid ansatz of the localized 5D energy-momentum tensor
\begin{align}
	\label{eq:EMT}
	T^{(5) a}_{\hphantom{(5)a} b}=\frac{1}{2\pi R} {\rm diag}(-\rho, P,P,P,P_{\phi})\;,
\end{align}
where the overall factor is such that $T_{ab} = 2\pi R T^{(5)}_{ab}$ defines a 4D energy-momentum tensor. Fixing $R$ also implies that
the energy density and pressure satisfy the standard 4D conservation equation
\begin{empheq}[box=\widefbox]{equation}
\dot \rho  + 3H \left( \rho + P \right) = 0\,.
\label{eq:enCons4D}
\end{empheq}

\subsection{Junction conditions} \label{sec:junction_cond}

In the next step, we explicitly evaluate the junction conditions~\eqref{israel}. The outward-pointing unit normal vector is given by $n^A  =  \re^{3\alpha_0} \left(3HR, \gamma, 0,0,0,0 \right)$.
It is straightforward to show that $K_{{\rm out}\,ab}$ has components
\begin{subequations}
	\label{eq:Kcompsout}
	\begin{align}
		K^{~0}_{{\rm out}\,0} &=  \frac{3R}{\gamma}\left(\dot H +H \dot \eta_0 \right) +n^A\partial_A\left(\eta - 3\alpha\right)\vert_0 \,, \\
		K^{~i}_{{\rm out}\,j} &=  \delta^i_{~j} n^A\partial_A \alpha\vert_0 \,  \,,\\
		K^{~\phi}_{{\rm out}\,\phi} &= \frac{\gamma}{R} - 3 n^A\partial_A \alpha\vert_0 \,.
	\end{align}
\end{subequations}

Using~\eqref{eq:defR},~\eqref{M4} and~\eqref{eq:EMT}, the $(0,0)$ component of the junction conditions gives a modified Friedmann equation
\begin{empheq}[box=\widefbox]{equation}
H^2  =\frac{\rho}{3 M_{\rm Pl}^2} +  \frac{1}{r_c^2} \left( \gamma - 1 \right)\,,
\label{eq:rhoJunctCond}
\end{empheq}
where $\gamma$ is given by~\eqref{gammaredef}, and $r_c$ denotes the cross-over scale
\begin{equation}
\label{eq:defCrossover}
	r_c^2 \equiv \frac{3 M_{\rm Pl}^2}{2\pi M_6^4}\,.
\end{equation}
The modification to the standard Friedmann equation is controlled by this cross-over scale. Assuming $|\gamma-1| \sim 1$, one can already tell that in the regime where $ H \gg r_c^{-1} $ the modification is negligible and the model reproduces the standard 4D evolution. When $ H $ becomes of order $ r_c^{-1} $, however, the modification becomes important and we expect a transition to a 6D regime. This is of course the way the model was engineered to work in the first place. It is also very similar to the 5D (DGP) case, where the modification term is simply $ \pm H / r_c^{(5)} $, with the appropriate 5D crossover scale $ r_c^{\rm DGP} = \frac{M_{\rm Pl}^2}{2M_5^3}$. But the crucial difference is that in the 6D case, the modification term cannot be directly expressed in terms of on-brane quantities like $ H $. It knows something about the bulk geometry through its dependence on $ \eta_0 $, and in order to make quantitative predictions one has to solve the bulk Einstein equations \eqref{eq:einstein_vacuum} as well.

The $(i,j)$ component of the junction conditions, combined with the vacuum Einstein equations \eqref{eq:etaPrime_vac} and \eqref{eq:etaDot_vac} in the limit $r \rightarrow r_0^+$, can be expressed as
\begin{equation}
	\label{eq:pJunctCondRConst}
\boxed{\dot H  =  -\frac{3}{2f(\tau)}\left[  \frac{P}{3 M_{\rm Pl}^2} + H^2  - \frac{1}{r_c^2} \Big( \gamma\, g(\xi, \chi)-1\Big )\right]\,,}
\end{equation}
where 
\begin{equation}
f(\tau) \equiv 1- \frac{9R^2}{2r_c^2\gamma} \, , 
\label{eq:def_f}
\end{equation}
and
\begin{subequations}
	\begin{align}
		g(\xi, \chi) &\equiv 1 + 2 \left (9\chi - 1 \right ) \bigl [ 3\chi + \xi \left ( 3\xi - 2 \right ) \left ( 9\chi - 1 \right ) \bigr ] \, , \label{eq:def_g}\\
		\xi & \equiv r \partial_r \alpha\vert_0 \, , \qquad
		\chi \equiv \frac{H^2R^2}{\gamma^2} \,.
	\end{align}
\end{subequations}
In our analysis, we will see that the sign of $f(\tau)$
allows to discriminate between a stable and an unstable class of solutions. 

The closed set of equations describing the bulk-brane system comprises the bulk equations of motion~\eqref{eq:einstein_vacuum}, the energy conservation equation~\eqref{eq:enCons4D}
and the Friedmann equation~\eqref{eq:rhoJunctCond}. The $\dot H$ equation~\eqref{eq:pJunctCondRConst} follows from these, as usual. For the purpose of numerical implementation, 
however, we will integrate the $\dot H$ equation. The Friedmann equation will only be implemented at the initial time and later on will serve as a numerical consistency check. 

Finally, the $(\phi,\phi)$ component of the junction conditions can be used to determine the azimuthal pressure:
\begin{eqnarray}
\nonumber
	\frac{P_\phi}{3M_{\rm Pl}^2}  &=& -\dot{H} \left( 1 - \frac{3R^2}{r_c^2\gamma} \right) - 2H^2 \\
	& & + \frac{6\gamma}{r_c^2} \left\{\chi + \bigl[ 3\chi - \xi ( 9\chi - 1 ) \bigr]^2 \right\}
\label{eq:pPhiJunctCond}
\end{eqnarray}
In our analysis, we will compute $P_\phi$ explicitly to check, for instance, whether the equation of state along the azimuthal direction satisfies the Null Energy Condition.

Before investigating the dynamical solutions, let us pause to recover the well-known static solutions from our setup.
\section{Static Solutions}
\label{sec:static_sol}
 
The static case constitutes an important check of the above equations and will provide a first physical insight into the geometry of the system\footnote{Note that in this case the static regularization (used in the main text) and the dynamical one (discussed in Appendix~\ref{ap:dynReg}) coincide by construction. Indeed, the only non-singular static geometry inside the cylinder is Minkowski space, hence the extrinsic curvature at the inner boundary is exactly the one given by~\eqref{eq:RegII}.}. 

For a purely static solution $\dot r_0=0$ and all metric functions solely depend on $r$.
The exterior field equations~\eqref{eq:einstein_vacuum} yield the solution
\begin{equation}
	\alpha=c \log{\frac{r}{r_0}}+\alpha_0		
	\quad\text{and}\quad		
	\eta=6\, c^2 \log{\frac{r}{r_0}}+\eta_0\;.
\label{staticprofile}
\end{equation}
By rescaling coordinates tangential to the brane, we can set $\alpha_0=0$ without loss of generality.
The remaining constants $c$ and $\eta_0$ are determined by the junction conditions~\eqref{eq:rhoJunctCond} and~\eqref{eq:pJunctCondRConst}: 
\begin{subequations}
\begin{align}
\eta_0&=-\log{\left(1-\frac{\rho}{\rho_{\rm crit}} \right)}\;,\label{eq:eta0}\\
	c& = \frac{1}{3} \left( 1 - \sqrt{\frac{2 \rho_{\rm crit} + (1 + 3w) \rho }{2(\rho_{\rm crit} - \rho)}} \right )\;, \label{eq:c}
\end{align}
\end{subequations}
where $w = P/\rho$ is the equation of state. Here we have introduced the critical density $\rho_{\rm crit} \equiv 2\pi M_6^4$. The third junction condition~\eqref{eq:pPhiJunctCond} then becomes
\begin{equation}
P_{\phi}=6 c^2 \left(\rho_{\rm crit}-\rho \right)\;.
\end{equation}

Note that~\eqref{eq:eta0} is ill-defined for $\rho>\rho_{\rm crit}$; we will come back to this point shortly. The line element for the exterior reads
\begin{eqnarray}
\nonumber
\rd s^2 &=&  \re^{2\eta_0} \left(\frac{r}{r_0}\right)^{12c^2-6c}\!\!\!\!\left( -\rd t^2 \!+ \rd r^2 \right)\\
&+& \left(\frac{r}{r_0}\right)^{2c} \rd  \vec{x}^2 + \left(\frac{r}{r_0}\right)^{-6c} r^2 \rd\phi^2 \; .
\end{eqnarray}
Since the brane induced terms vanish identically for static configurations, this solution is the direct generalization of the exterior metric of a static cylinder in 4D, first derived by Levi-Civita~\cite{Levi:1919} and later reviewed for example in \cite{Thorne:1965}. 

Consider the case of pure 4D tension on the brane:
\begin{subequations}
\begin{gather}
	\rho = - P \equiv \lambda \\
	\Rightarrow\, c = 0 = P_\phi \,.
\label{puretension}
\end{gather}
\end{subequations}
The coordinate rescaling $( \bar t,\bar r)=( \re^{\eta_0} t, \re^{\eta_0} (r-r_0)+r_0)$ yields the famous wedge geometry in Gaussian normal coordinates, characterized by the deficit angle $\delta \equiv \lambda / M_6^4$:
\begin{equation}\label{eq:ds_wedge}
\rd s^2 = -\rd \bar t^2 \!+ \rd \bar r^2 + \rd  \vec{x}^2 + W(\bar r)^2  \rd\phi^2 \; ,
\end{equation}
where
\begin{equation}
	W(\bar r)=
		\begin{cases}
		\bar r 										& \mbox{for } \bar r\leq r_0\\
		\frac{\delta}{2\pi}r_0+\left(1-\frac{\delta}{2\pi}\right)\bar r 	& \mbox{for } \bar r>r_0\;.
		\end{cases}
\end{equation}
Note that this solution corresponds to the generalization of the cosmic string geometry \cite{Vilenkin:1981zs, Hiscock:1985uc} to 6D. The coordinates cover again the whole space-time including the interior. A well-known fact about this solution is that the intrinsic brane geometry is flat and the energy on the brane only affects the extrinsic curvature, thereby creating a deficit angle. This property makes the higher codimensional models in particular interesting with respect to the cosmological constant problem because $\lambda$ is effectively ``filtered out''  from the perspective of a brane observer; see~\cite{Burgess:2011va} and~\cite{Dvali:2002pe} in the case of large or infinite extra dimensions, respectively. 

\begin{figure*}
	\subfloat[$ \delta<\delta_{\rm crit}$]{
		\includegraphics[width=0.3\textwidth]{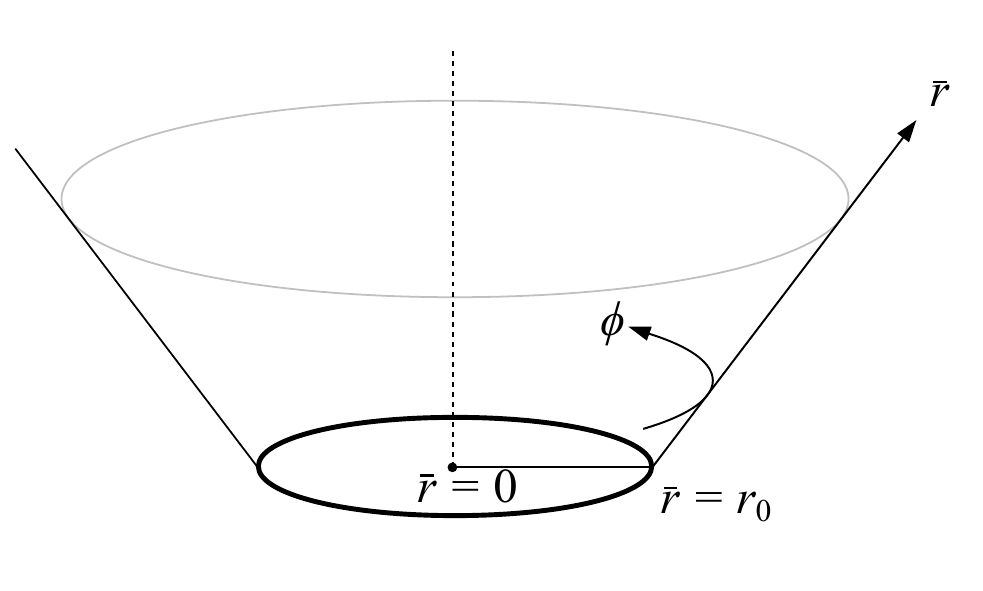}
	}
	\hfil
	\subfloat[$ \delta=\delta_{\rm crit} $]{
		\includegraphics[width=0.3\textwidth]{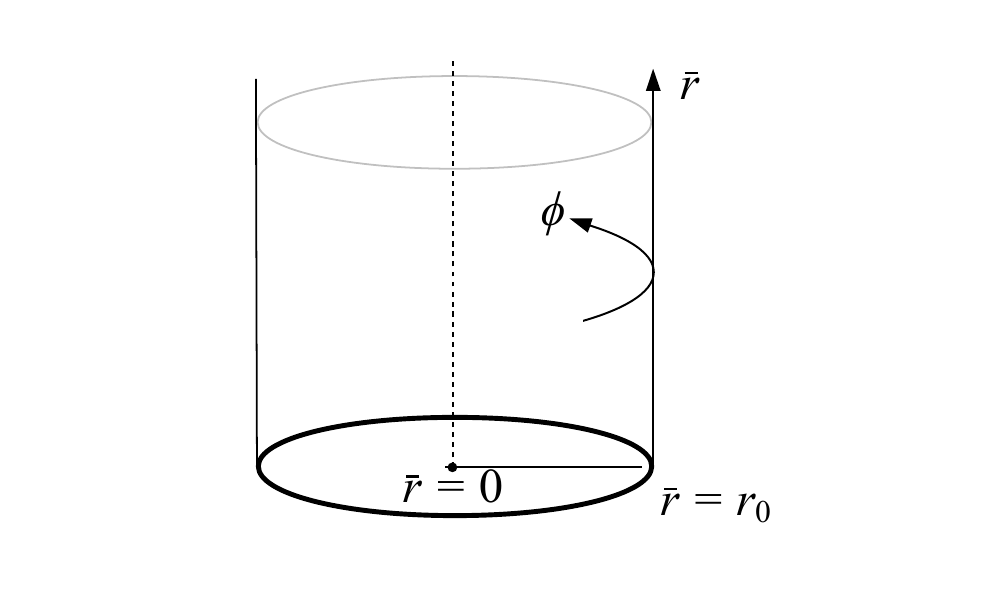}
	}
	\hfil
	\subfloat[$ \delta>\delta_{\rm crit} $]{
		\includegraphics[width=0.3\textwidth]{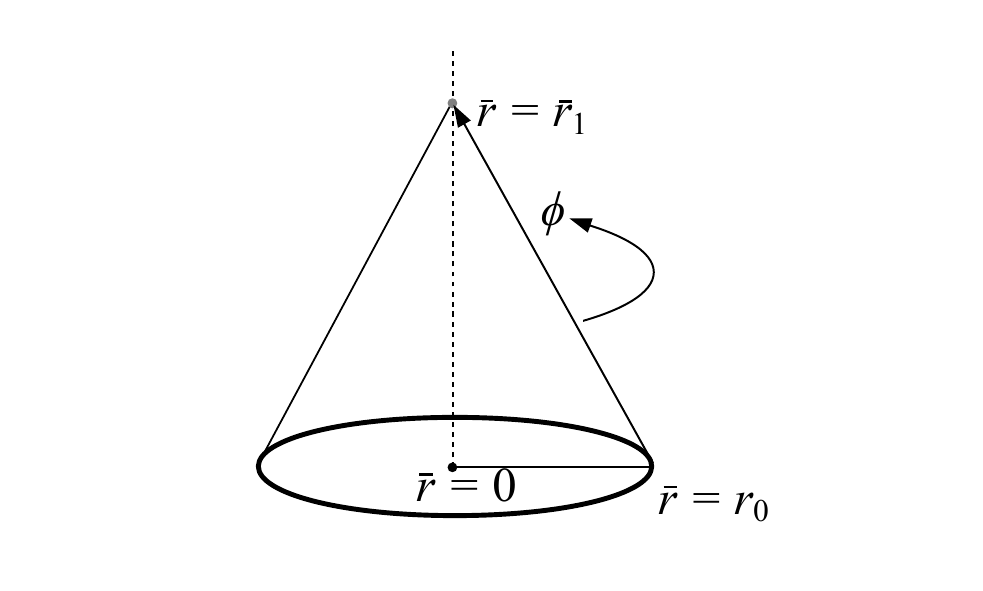}
	}
	\caption{Embedding diagrams of the regularized static geometry in the case of a pure tension brane. The circle at $\bar r=r_0$ describes the brane. As the tension approaches the critical value, the deficit angle approaches $2\pi$, and the bulk geometry becomes cylindrical (b). For super-critical tensions, a naked singularity develops in the bulk a finite distance away from the brane.} \label{fig:embedding_geometry}
\end{figure*}

For sub-critical tensions $\delta<2\pi$ we find for the ratio of physical radius and circumference: $ \bar r/W(\bar r)=1$ for $\bar r\leq r_0$ and $\bar r/W(\bar r)>1$ for $\bar r>r_0$. In an embedding picture this corresponds to a capped cone, as shown in Fig.~\ref{fig:embedding_geometry}. In the critical limit $\delta\rightarrow2\pi$, the embedding geometry becomes  ``cylindrical''.

In the super-critical case, $\delta>2\pi$, the circumference $2\pi W(\bar r)$ decreases for $\bar r>r_0$ and vanishes for a certain radius $\bar r_1$, implying the existence of a second axis. However, in general the geometry is not elementary flat at that position, {\it i.e.}, $W^{\prime}(\bar r_1) \neq 1$, which indicates the existence of a naked singularity. It has been argued that this (conical) singularity is an artifact of the static approximation and is resolved once the full dynamics are taken into account \cite{Cho:1998xy}.

The derivation of the junction conditions in the Einstein-Rosen language is not compatible with the super-critical scenario. This is clear in the static case, as already mentioned, since~\eqref{eq:eta0} does not allow a real solution for $\eta_0$ in the super-critical regime. See Appendix~\ref{ERcoords} for a more detailed discussion of this point in the context of dynamical solutions, and~\cite{Niedermann:2014yka} for a detailed investigation of super-critical cosmic strings. We henceforth exclude the super-critical regime from our analysis.
\section{Interlude: Bulk-brane dynamics}
\label{sec:Brane_Bulk}

The analysis of cosmological solutions on the brane is greatly complicated by the fact that the assumed symmetries allow
for axially symmetric gravitational waves propagating in the bulk. This is unlike the much-studied codimension-one case,
where the assumption of planar symmetry enforces a version of Birkhoff's theorem~\cite{Taub:1951}: The only vacuum 5D solutions
are Minkowski or Schwarzschild. The Schwarzschild mass parameter enters the brane Friedmann equation as the coefficient
of a ``dark radiation'' term. In particular, the brane Friedmann equation is completely local.

The codimension-two case of interest is qualitatively very different. The bulk field equations \eqref{eq:einstein_vacuum} explicitly show that in this case gravitational waves are in fact compatible with all the symmetries. As a consequence, it would be possible to prepare a wave packet in the bulk that reaches the brane at some arbitrary time. Since the amplitude of the wave is given by the metric function $ \alpha(t, r) $, while the 4D scale factor is determined by $ \alpha_0(t) \equiv \alpha(t, r_0) $, the 4D cosmological evolution will inevitably be influenced by such a wave packet. As a result, it cannot be possible to derive a closed local on-brane evolution equation for $ \alpha_0 $, without imposing additional restrictions on the bulk geometry.

What could these restrictions be? As a first guess, one could try to assume a flat bulk geometry, just as could be done in the DGP case. After all, this is also what happens in the static pure tension solution. However, it turns out that this is no longer possible after one demands $ \alpha_0  $ to have non-trivial dynamics. To show this, let us try to set the $(t,x^1,t,x^1)$- and $(t,r,t,r)$-components of the Riemann tensor to zero, which is a necessary condition for flatness. This in turn demands
\begin{equation}
\left(\partial_t \alpha\right)^2- \left(\partial_r \alpha\right)^2=0 \qquad \text{and} \qquad r\,\partial_r \alpha=0\,.
\end{equation}
The only solution to these equations is indeed the trivial configuration $\alpha = \mathrm{constant}$.

So a dynamical codimension-two brane inevitably curves the extra-dimensional space-time, and since the on brane geometry will be time-dependent, so will be the bulk geometry. In other words, gravitational waves are not only possible for a non-trivial cosmology in this setup, but in fact necessary.

One could still try to arrive at a closed on-brane system by implementing an ``outgoing wave condition'' at the outer boundary of the brane to exclude incoming bulk waves. Physically, this is clearly a necessary condition because we assume a source-free, infinite bulk. However, it is well known that such a condition is necessarily non-local (in time) in the case of cylindrically symmetric waves (see \cite{Givoli:1991} for a review, and \cite{Hofmann:2013zea} for a discussion in the context of GR). 
Moreover, because the coordinate position of the brane $ r_0(t) $ will in general be time dependent, the resulting on-brane system would be non-local both in space \emph{and} time. It is clear that solving such a non-local system would not be any easier than solving the full bulk system from the start. In other words, if one tried to accommodate for all allowed bulk configurations in the on-brane system, one would end up with not only one, but infinitely many ``constants of integration''. This is what makes the codimension-two problem much harder to solve.

Therefore, there seems to be no way around solving the full bulk geometry in order to see what 4D cosmology emerges in the codimension-two BIG model. This can in general only be done numerically, and we will do so in the next sections. 
\section{Numerical implementation and Initial Data}
\label{sec:num_impl}

We now turn to the numerical implementation of the full brane-bulk system~\eqref{eq:einstein_vacuum},~\eqref{eq:enCons4D} and~\eqref{eq:rhoJunctCond}.
Solutions were obtained by specifying initial data, as explained below, and numerically integrating this initial value problem forward in time. Since the dynamical bulk equation~\eqref{eq:2D_wave} is nothing but the standard (flat space) cylindrically symmetric scalar wave equation, it is straightforward to find a stable integration scheme for the PDE part of the problem. There is only a slight complication stemming from the matching procedure. Even though the physical brane size $ R $ is fixed, its coordinate position $ r_0 $ is generally time-dependent. Therefore, if one chooses a fixed spatial grid size in the bulk (as we do), one has to allow $ r_0 $ to lie in between those grid points. We deal with this problem by using some suitable interpolation scheme. The details of the numerical implementation can be found in Appendix~\ref{ap:numImpl}.

The numerical integration starts at some initial time $ t = t_i $, $ \tau = \tau_i $. Let us denote all functions evaluated at this time with a subscript $ i $. 
Through a global rescaling of coordinates, we can always set $\alpha = 0$ on the brane initially, {\it i.e.},
\begin{equation}
\label{eq:initCondAlpha0}
	\left(\alpha_{0}\right)_{i} = 0\,.
\end{equation}
Consequently, the initial brane position is 
\begin{equation}
\left(r_{0}\right)_{i} = R \,.
\end{equation}
In the bulk we must specify the initial radial profile $\alpha_i( r)$ and its time derivative $\partial_t \alpha_i( r)$. 
To be definite, as initial profile we choose the static profile given by~\eqref{staticprofile}, namely
\begin{equation}
\label{eq:initProfileAlpha}
	\alpha_{i}(r) = c \ln\left( \frac{r}{R} \right) \, ,
\end{equation}
where the constant $ c $ is given by~\eqref{eq:c} with $\rho \rightarrow \rho_i$. In particular, for a cosmological constant ($ w=-1 $), we get $ c = 0 $, and hence $ \alpha_i(r ) = 0 $. Note that by choosing the static profile we are not putting any potential energy into the bulk gravitational field initially.

At the brane position, the velocity profile is related to the initial Hubble parameter $ H_i $ via
\begin{eqnarray}
\nonumber
	\partial_{ t} \alpha_{0i} &=& \frac{\rd \alpha_{0i} }{\rd t}- \frac{\rd r_{0i} }{\rd t}\,\partial_{ r} \alpha_{0i}  \\
\nonumber
	&=& \frac{\rd \alpha_{0i} }{\rd t}\left( 1 - 3 \partial_{ r} \alpha_{0i} \right) \\
	&=& \frac{H_i}{\gamma_i}(1-3c)\;.
\label{eq:alpha0TildeDot_init}
\end{eqnarray}
Extending this to the bulk, we write
\begin{equation}
	\partial_t \alpha_i(r) = \frac{H_i}{\gamma_i} (1 - 3c) F(r) \,,
\end{equation}
where $ F( r) $ is some profile function satisfying the boundary condition $F(R) = 1$.
To minimize the amount of kinetic energy put into the gravitational field initially, which could impact the brane cosmology for long times, 
we will choose profile functions which are sharply localized around the brane. For definiteness, we will focus on a Gaussian profile of width $\sigma$,
\begin{equation}
	F(r) = \exp\left[ - \frac{(r-R)^2}{\sigma^2} \right]\,,
\label{GaussianF}
\end{equation}
With these choices, we expect the on-brane evolution to rapidly become insensitive to the initial conditions.

This completes the specification of initial data. Indeed, the remaining variable, $ \eta_{0i} $, is fixed by the constraint~\eqref{eq:rhoJunctCond}, together with the relation~\eqref{gammaredef}\footnote{The full radial profile $ \eta(r) $ can be calculated from \eqref{eq:etaPrime_vac}, but is actually not needed for the evolution of $ \alpha $. Only $ \eta_0 $ enters through the junction conditions, and it can be calculated at later times from its initial value using \eqref{eq:etaPrime_vac} and \eqref{eq:etaDot_vac} only locally at the brane position.}. Specifically,
\begin{equation}
	\frac{\rho_i}{\rho_{\rm crit}} = r_c^2 H_i^2 + 1 - \sqrt{ \re^{-2\eta_{0i}} + 9 H_i^2 R^2 } \, .
\end{equation}
Note that this equation does not always have a (real) solution for $ \eta_{0i} $. The existence of a real solution places an upper bound on the energy density:
\begin{equation}
\label{eq:criticalityBound}
	\frac{\rho}{\rho_{\rm crit}} < r_c^2 H^2 + 1 - 3 \left|H\right| R \, .
\end{equation}
Since the constraint has to hold for all times, we were able to drop the subscript $i$. We will refer to this as the \textit{criticality bound}, separating the sub- and super-critical regimes. As soon as~\eqref{eq:criticalityBound} is violated, the initial constraint cannot be fulfilled. 
The reason is that in this parameter regime the Einstein-Rosen coordinates as used in our derivation are no longer valid. The interested reader is referred to Appendix~\ref{ERcoords}
for more details. Since this super-critical regime is not compatible with our coordinate choice, it will not be considered in this paper.

As a check on~ \eqref{eq:criticalityBound}, note that it correctly reproduces the static criticality bound $ \rho < \rho_{\text{crit}} = 2\pi M_6^4 $ in the static limit $ H \to 0 $. In the dynamical case, however, the bound is more general. In particular, for $ r_c = 0 $, {\it i.e.}, without the induced gravity terms, the bound becomes stronger---the critical point is reached for a smaller value of $ \rho $ than in the static case. 
Physically, the reason is that for $ H \neq 0 $, there is additional kinetic energy in the system. For $ r_c \neq 0 $, on the other hand, the induced gravity terms can absorb (or ``shield'') part of the energy density from the bulk, thereby allowing much larger values for $ \rho $ than in the static case.

The final ingredient is the choice of grid spacing for the numerical calculation. We use a scheme in which the temporal and radial grid spacing is the same and constant:
\begin{align}
	\Delta t & = \Delta r \equiv \epsilon \,.
\label{stepsize}
\end{align}
The system can then be evolved forward in time using \eqref{eq:einstein_vacuum} and \eqref{eq:pJunctCondRConst} for any given $ H_i $, $ \sigma $, $ R $, $ r_c $, $ \rho_i $ and equation of state parameter $ w $. (In fact, the quantities $ H_i $, $ R $ and $ r_c $ enter the equations only in the combinations $ H_i r_c $ and $ H_i R $, so only two of them need to be specified while the third one is degenerate.)
The constraint equation \eqref{eq:rhoJunctCond} can be used as an important consistency check for the numerical solver.
Further details of the numerical implementation are given in Appendix~\ref{ap:numImpl}. In what follows we will present the results.
\section{Numerical Solutions}
\label{sec:num_sol}

\begin{figure*}[htb]
	\subfloat[The Hubble parameter on the brane exhibits degravitation. It starts out positive and asymptotically tends to zero.]{
		\includegraphics[width=0.45\textwidth]{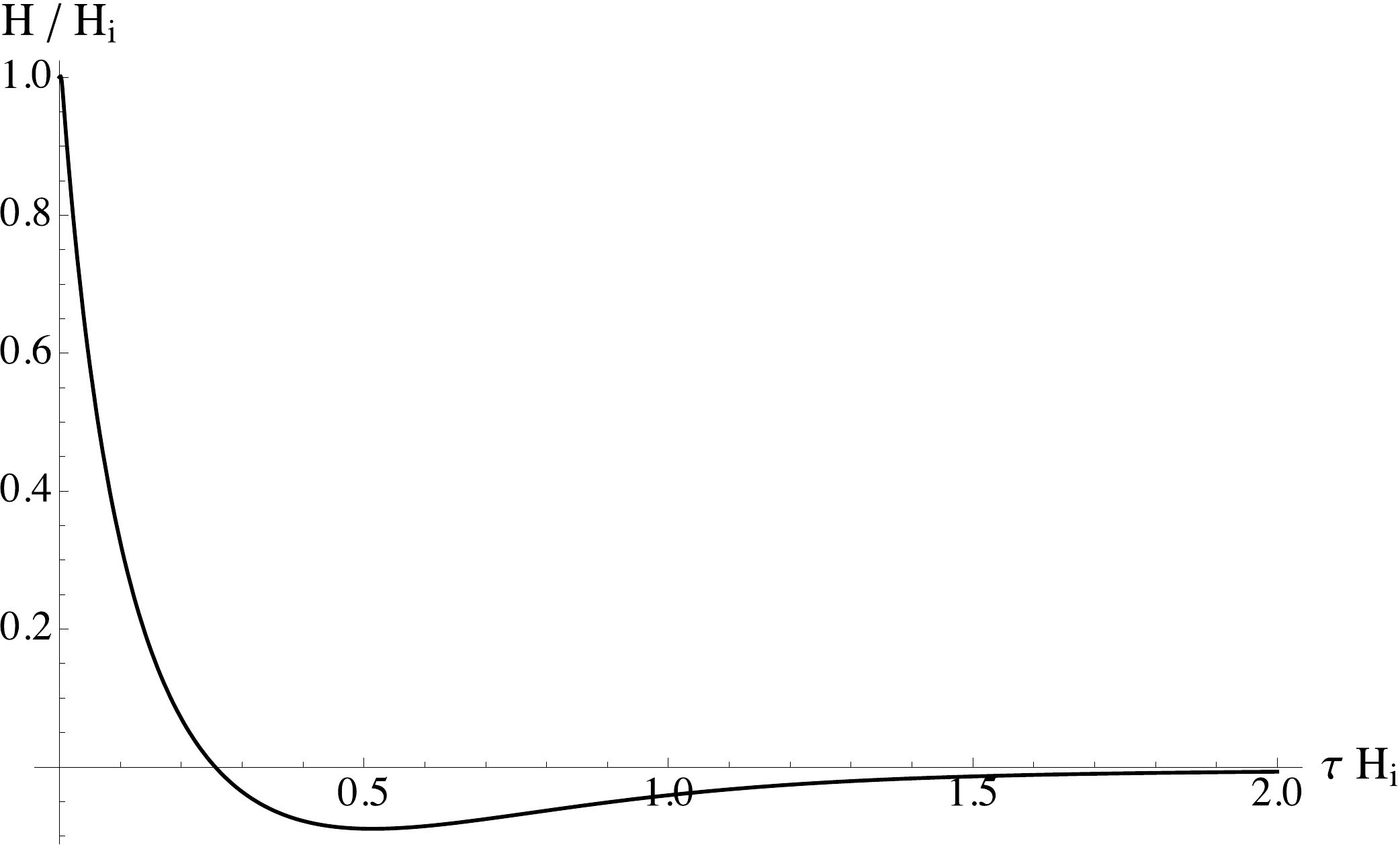}
		\label{fig:hubble_cc_degrav}
	}
	\hfill
	\subfloat[The radial profile for $ \alpha $ at different values of $ \tau $. The dots indicate the brane position as a function of time.]{
		\includegraphics[width=0.45\textwidth]{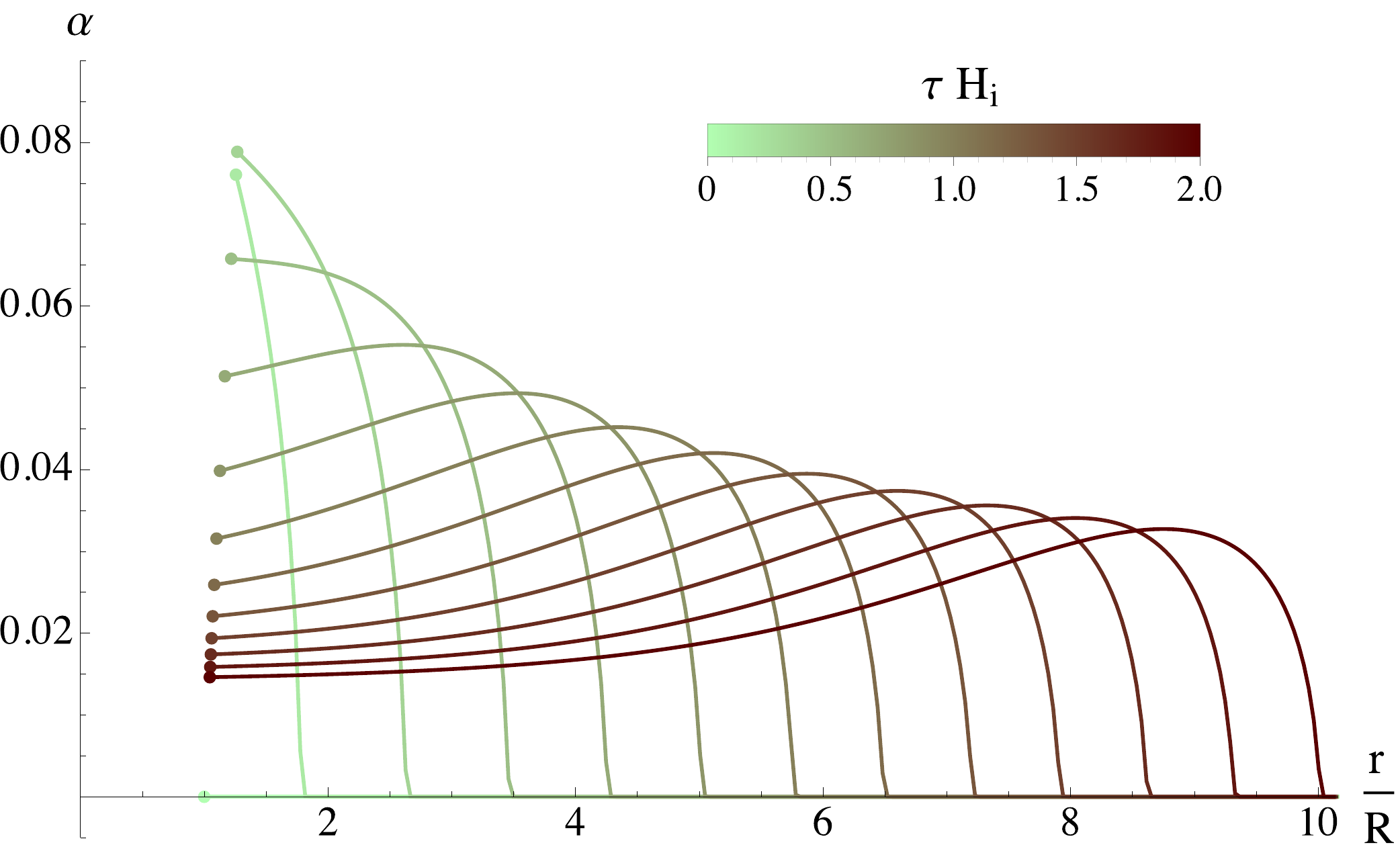}
		\label{fig:alpha_cc_degrav}
	}
	\\
	\subfloat[Equation of state of $ P_\phi $ that is needed to keep the brane circumference fixed for $ w=-1 $. It never falls below the value $-1$ corresponding to unphysical matter.]{
		\includegraphics[width=0.45\textwidth]{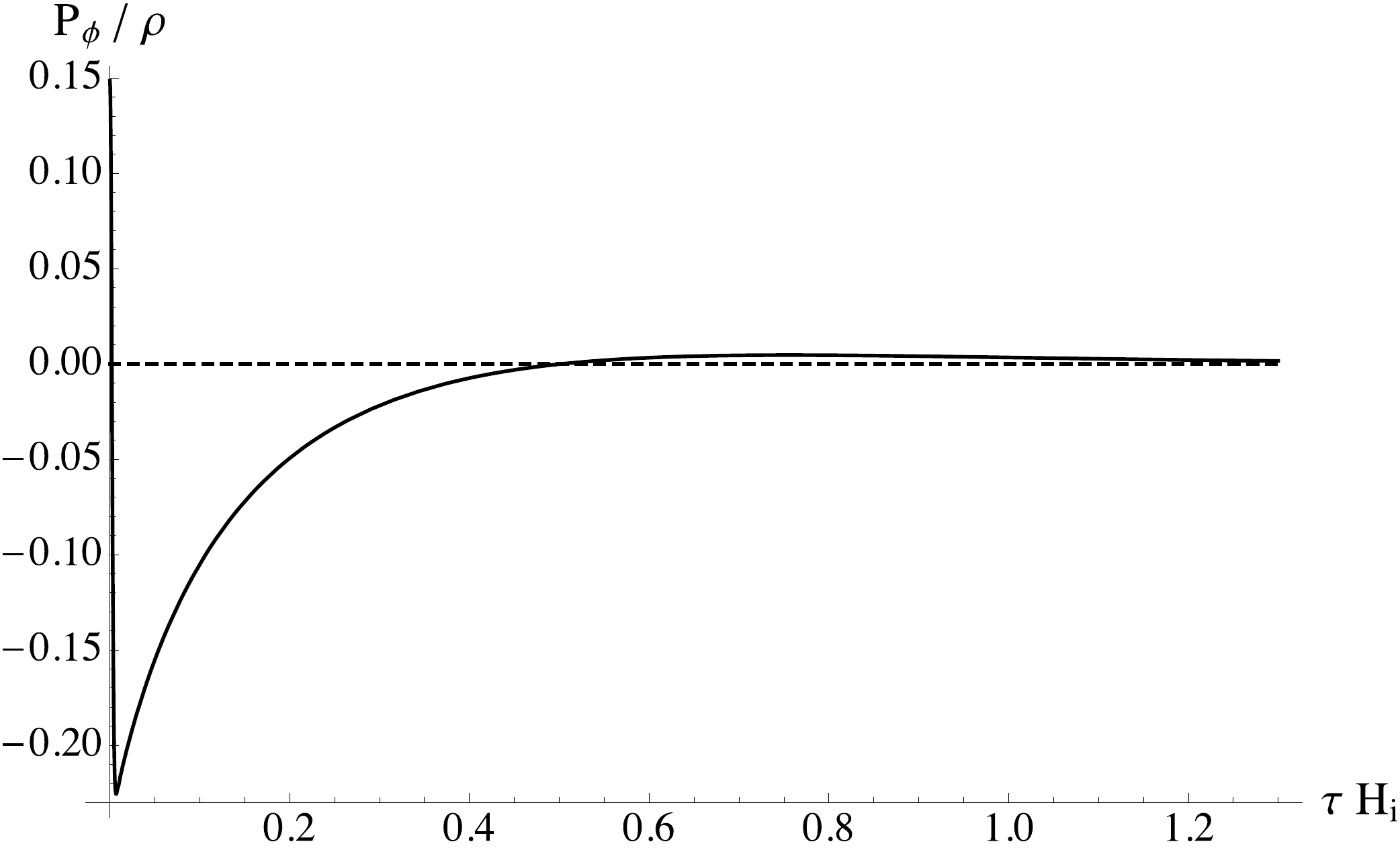}
		\label{fig:pPhi_cc_degrav}
	}
	\hfill
	\subfloat[The effective energy density, $\hat\rho\equiv \rho - 3M_4^2 H^2$, as ``seen'' by 6D GR. This approaches a positive value consistent with the static solution.]{
		\includegraphics[width=0.45\textwidth]{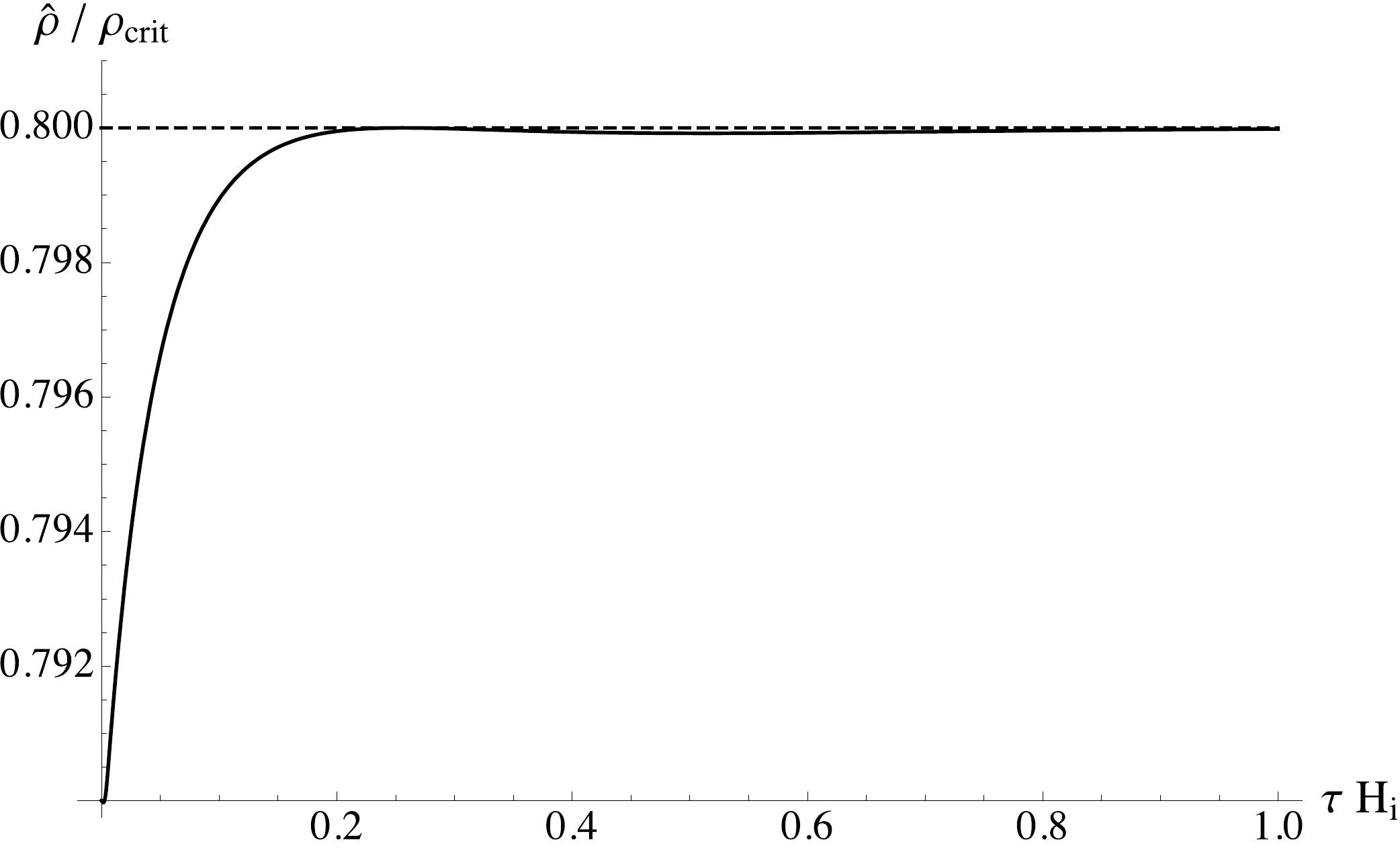}
		\label{fig:rhoHat_cc_degrav}
	}
	\caption{Example of a degravitating solution.} 
	\label{fig:cc_degrav}
\end{figure*}

\begin{figure*}[htb]
	\subfloat[The Hubble parameter on the brane grows in time, indicating super-acceleration.]{ %
		\includegraphics[width=0.45\textwidth]{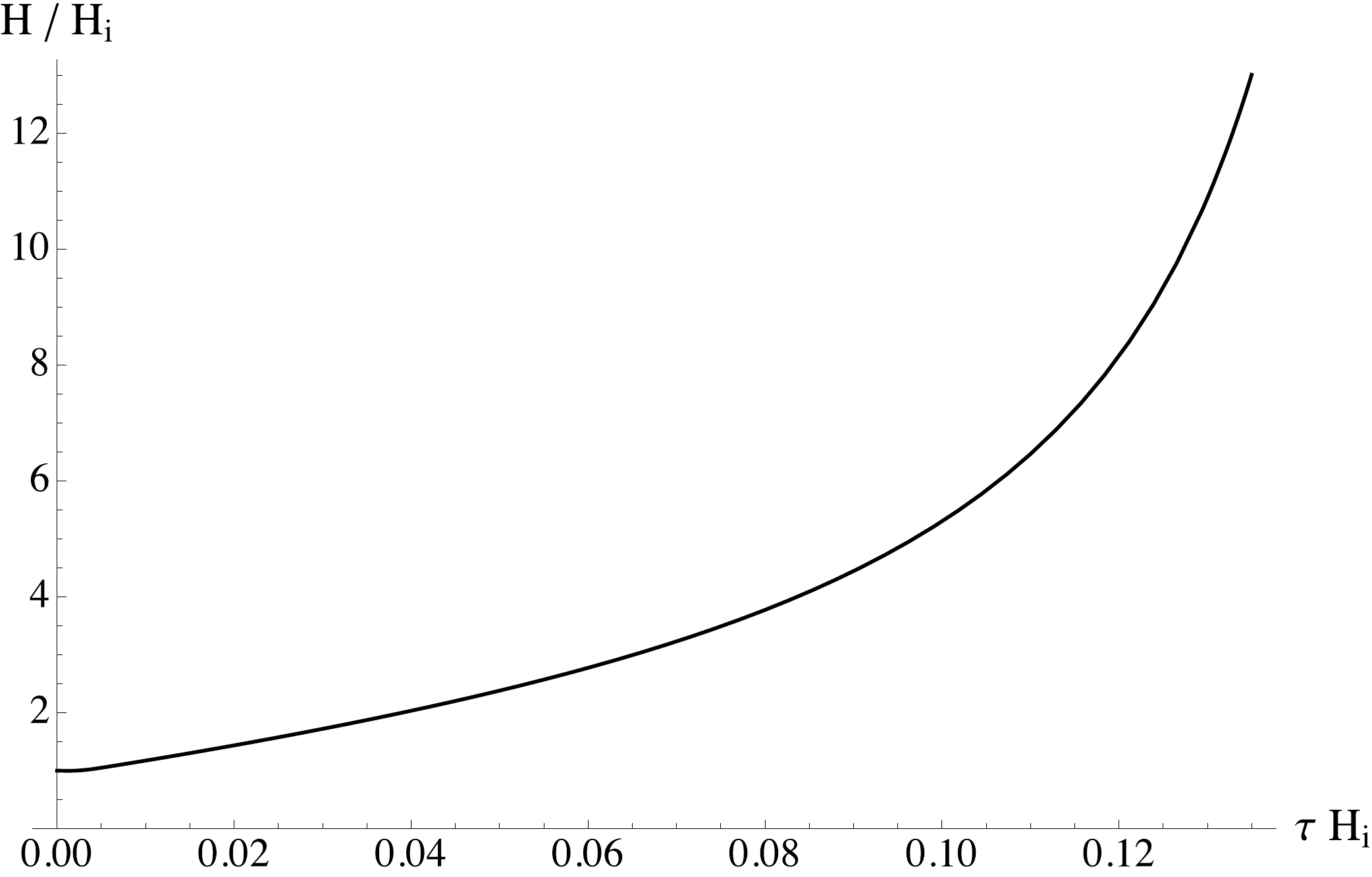} %
		\label{fig:hubble_cc_pathol} %
	} %
	\hfill %
	\subfloat[The radial profile of the function $ \alpha $ at different values of $ \tau $. At fixed $r$, $\alpha$ grows in time.]{ %
		\includegraphics[width=0.45\textwidth]{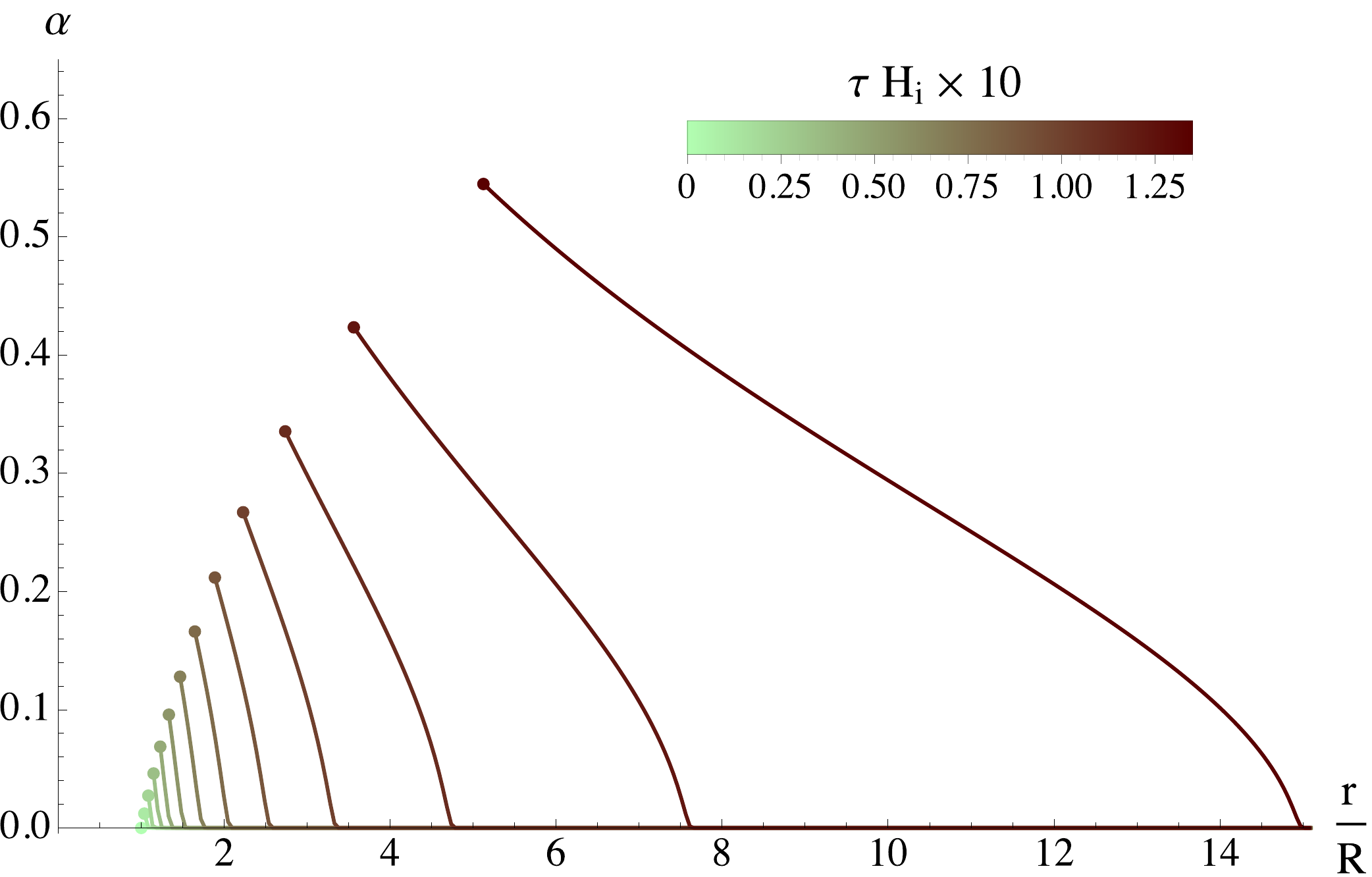} %
		\label{fig:alpha_cc_pathol} %
	} %
	\\
	\subfloat[Equation of state of $ P_\phi $ that is needed to keep the brane circumference fixed. It is negative and falls rapidly below $-1$.]{ %
		\includegraphics[width=0.45\textwidth]{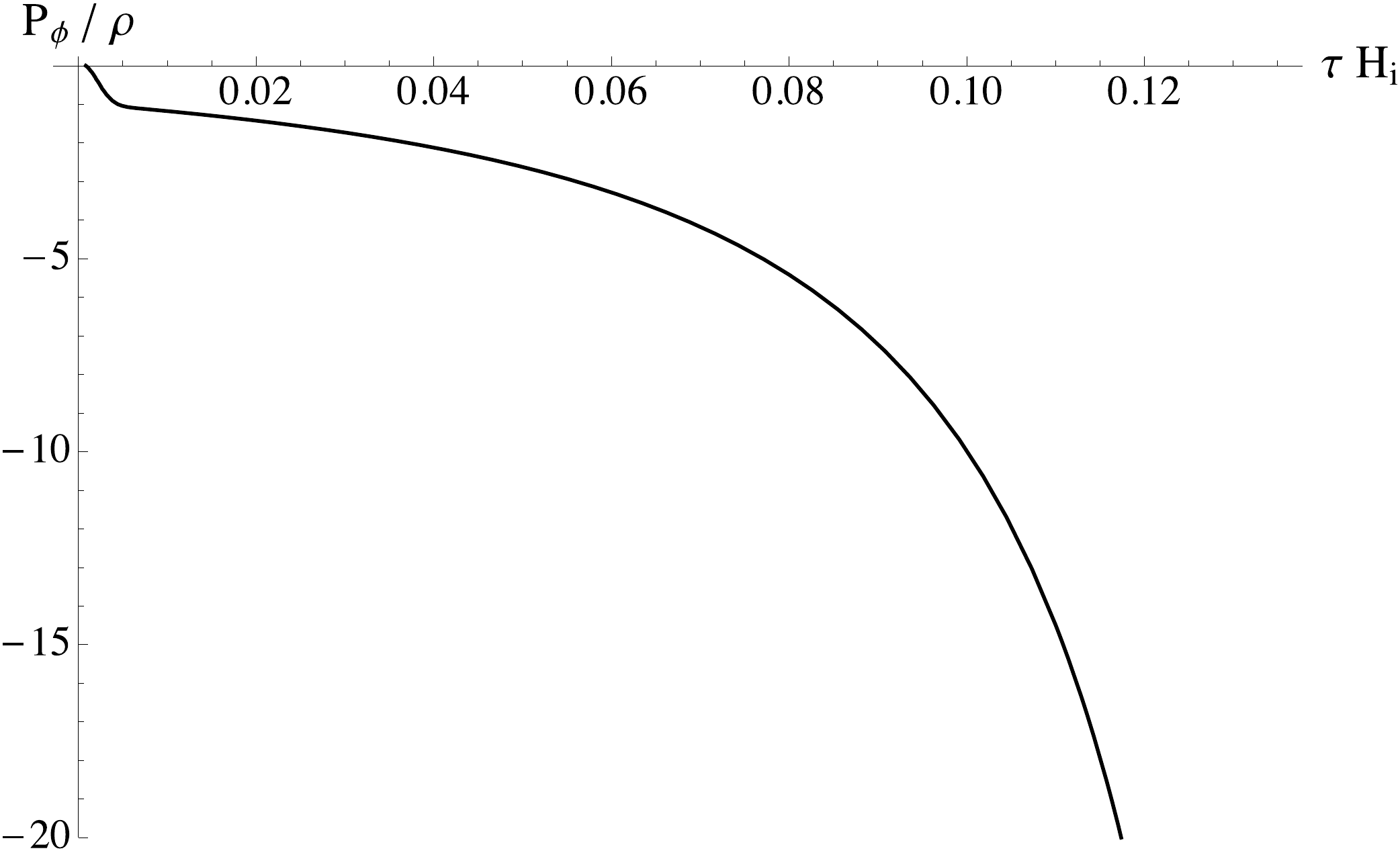} %
		\label{fig:pPhi_cc_pathol} %
	} %
	\hfill %
	\subfloat[The effective energy density, as ``seen'' by 6D GR, becomes negative. This is interpreted as the source of the physical instability.]{ %
		\includegraphics[width=0.45\textwidth]{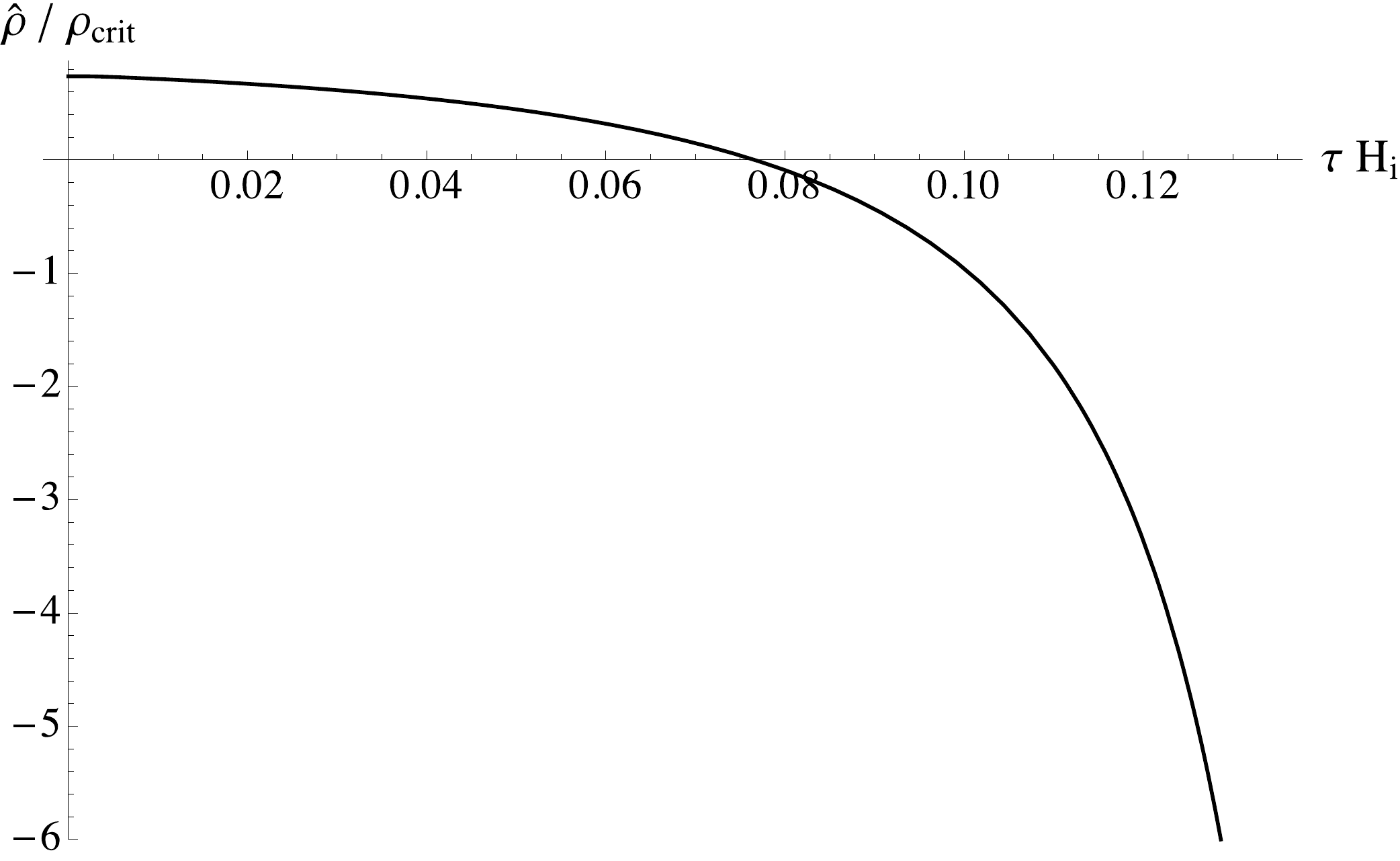} %
		\label{fig:rhoHat_cc_pathol} %
	} %
	\caption{Example of a  super-accelerating solution.}\label{fig:cc_pathol} %
\end{figure*}

\begin{figure*}
	\subfloat[Behavior of solutions for different choices of $r_c$ and $\rho_i$. The green region (Region (1)) shows stable solutions; the red region (Region (2)) shows unstable solutions. The solid line in between corresponds to $f = 0$. The gray region corresponds to super-critical solutions, which are not covered in our analysis.]{
		\includegraphics[width=0.42\textwidth]{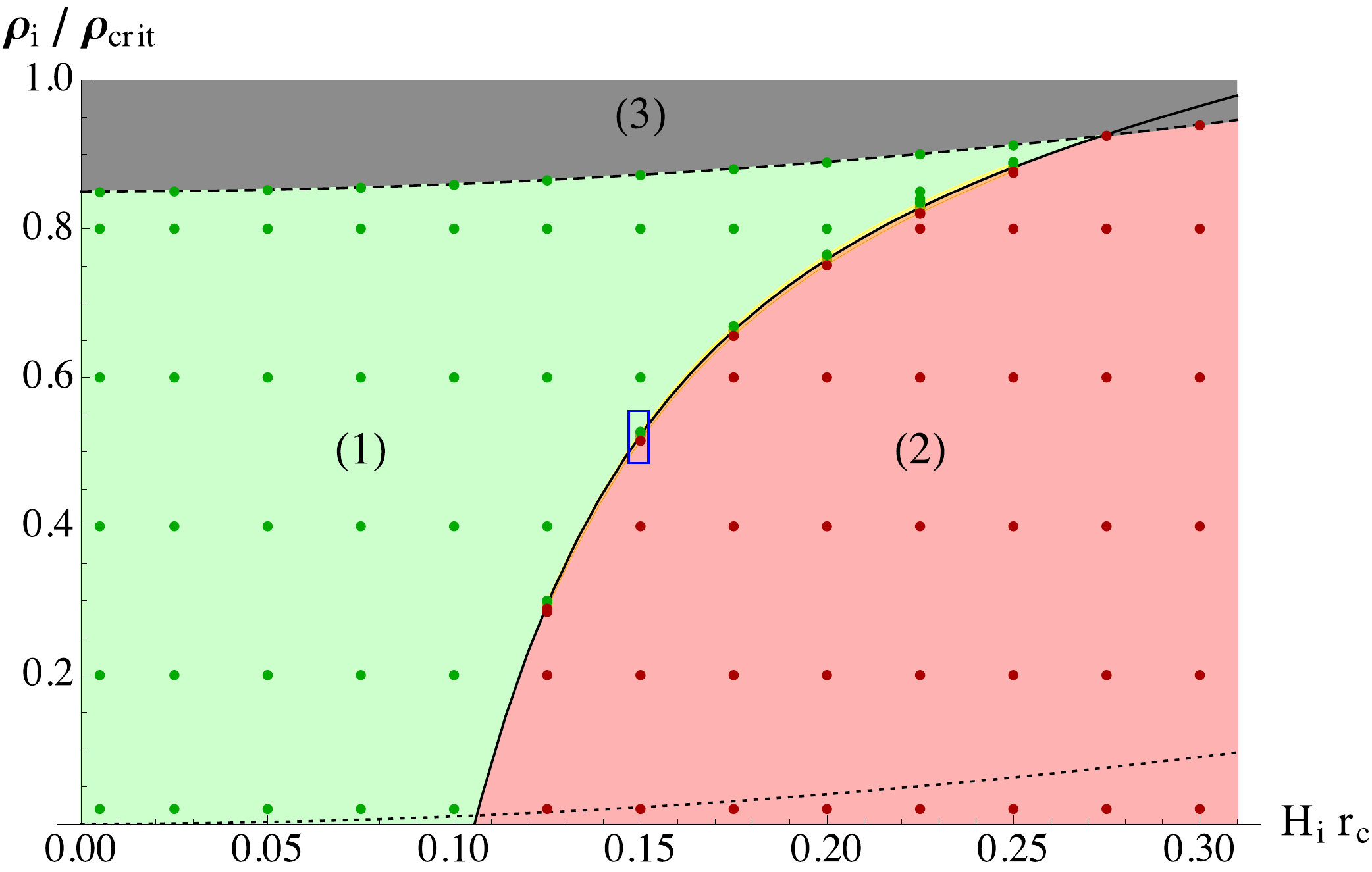}
		\label{fig:contour_plot_full}
	}
	\hfill
	\raisebox{0.1\height}{\includegraphics[width=0.11\textwidth]{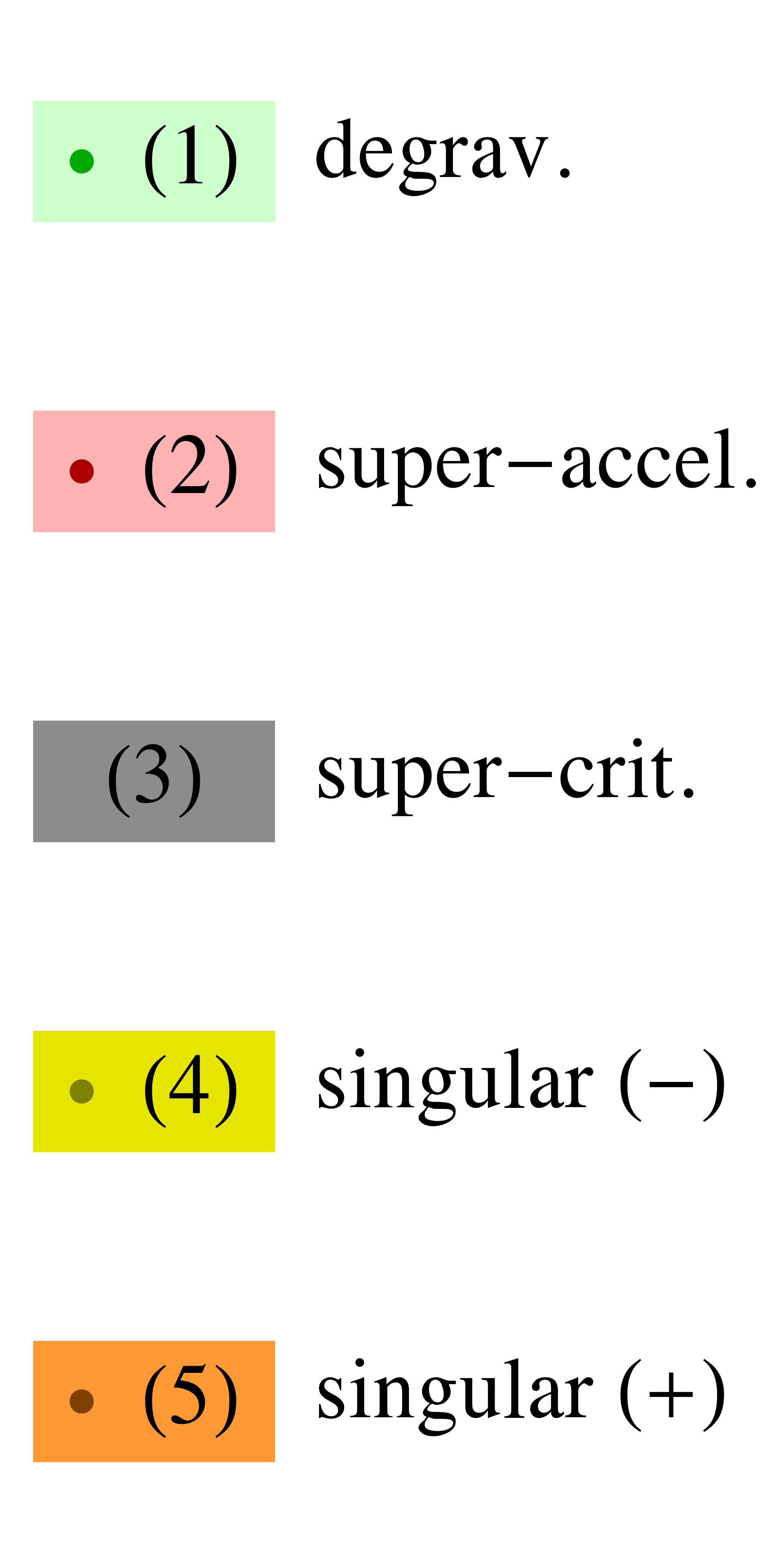}}
	\hfill
	\subfloat[Zoom into the small blue rectangle depicted in Fig.~\ref{fig:contour_plot_full}. The yellow/orange regions show solutions which hit the singularity at $f = 0$ in a finite time. The dashed lines have been inferred from the numerical results.]{
		\includegraphics[width=0.43\textwidth]{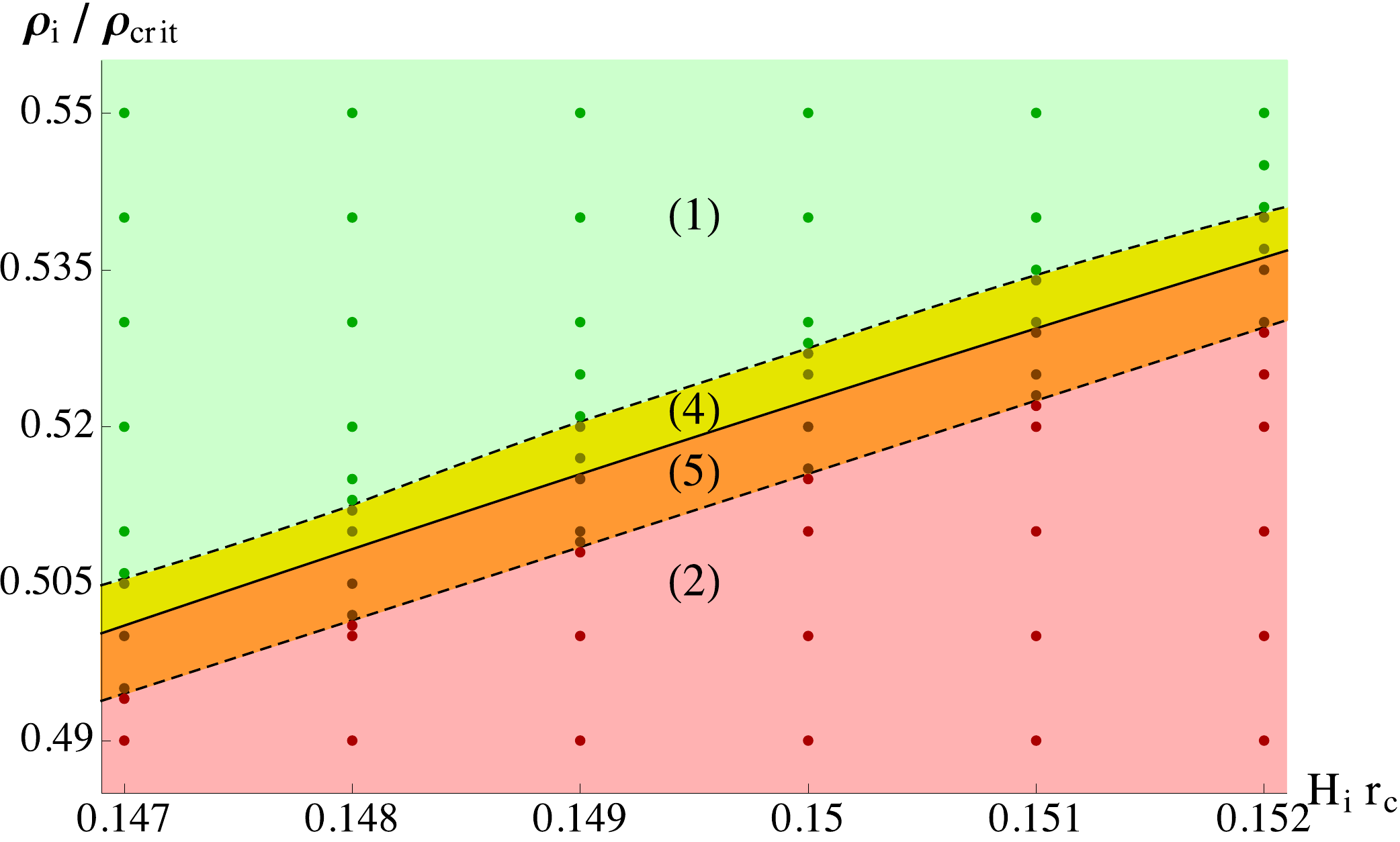}
		\label{fig:contour_plot_zoom}
	}\\
	\raisebox{0.75\height}{\includegraphics[width=0.15\textwidth]{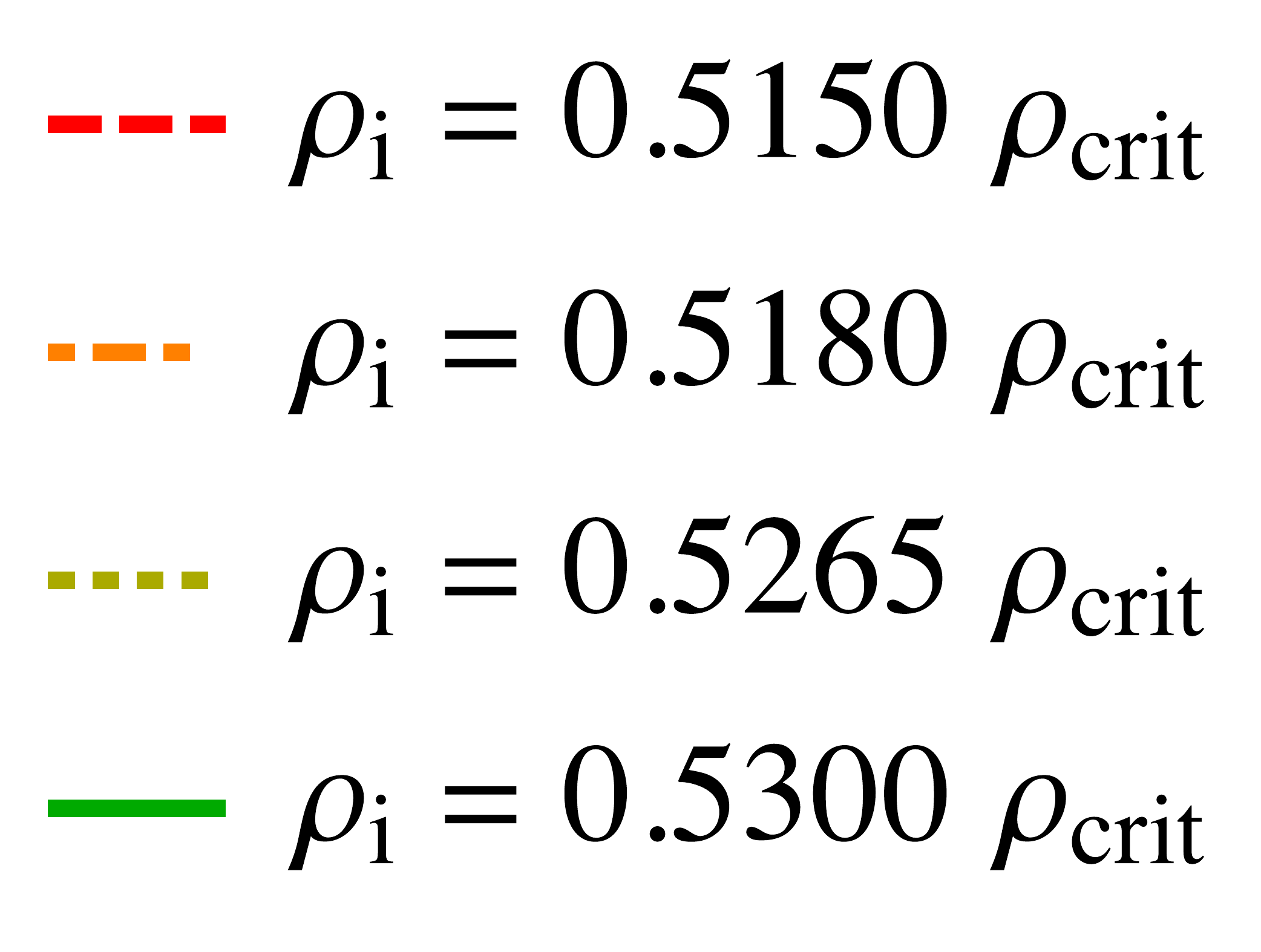}}
	\subfloat[The evolution of $f(\tau)  = 1 - \frac{9R^2}{2r_c^2\gamma}$ for $H_i r_c=0.15$ and different values of $\rho_i$. The color/numerical labels of the curves match those of Fig.~\ref{fig:contour_plot_zoom}. The yellow (4) and the orange (5) lines hit the singularity at $f=0$ in finite time, while the green (1) and red (2) curves avoid the singularity.]{
		\includegraphics[width=0.5\textwidth]{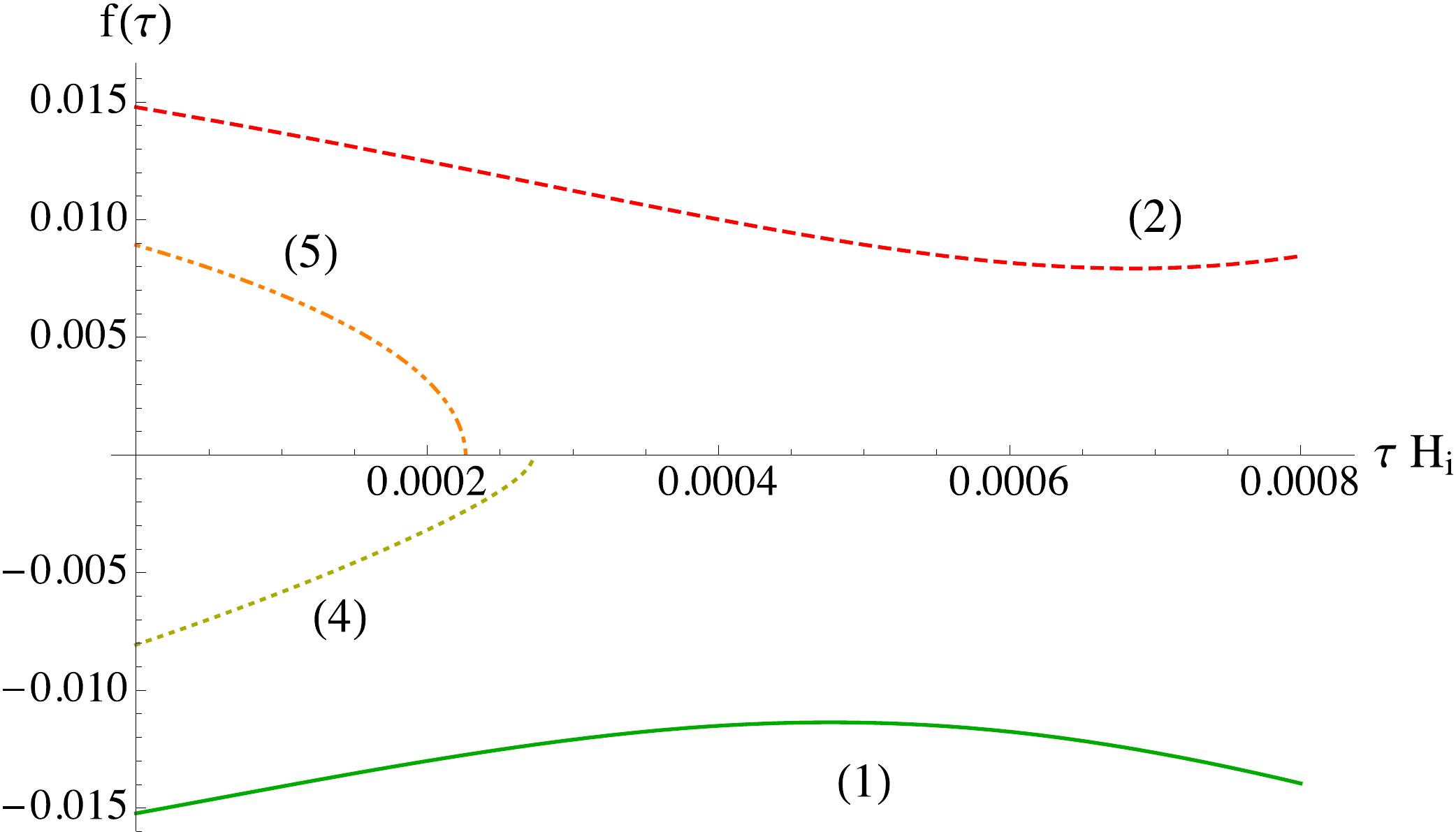}
		\label{fig:f_plot_zoom}
	}
	\caption{(color online) Results of the numerical stability analysis of the model.}
	\label{fig:contour_plot}
\end{figure*}

We have found two, qualitatively different classes of solutions, depending on the initial conditions.
The first class, called {\it degravitating} solutions, features a geometry which at late times approaches
the static profile. In particular, $H\rightarrow 0$ on the brane. The second class, called {\it super-accelerating}
solutions, features a run-away behavior for the Hubble parameter on the brane. The source for this apparent instability
is an effective energy density on the brane which violates the Null Energy Condition. 

After describing a fiducial degravitating (Sec.~\ref{sec:degrav_sol}) and super-accelerating (Sec.~\ref{sec:pathol_sol}) solution, 
we will discuss the regions of parameter space spanned by each class in Sec.~\ref{sec:contourPlot}.

\subsection{A degravitating solution}
\label{sec:degrav_sol}

As a first example, let us consider a 4D cosmological constant source ($ w=-1 $) with parameters\footnote{For completeness, the width of the initial Gaussian profile~\eqref{GaussianF} is set to $\sigma = R/50$, and the step size for integration~\eqref{stepsize} is $\epsilon = 2 \times 10^{-4} R$.}
\begin{equation}
		H_i r_c =  \frac{1}{10}\, ; ~~H_i R = \frac{1}{20} \, ; ~~\rho = \frac{4}{5} \rho_{\rm crit} \, .
		\label{eq:parameteres_dgrav}
\end{equation}
For this choice, the energy density lies in the sub-critical regime. Meanwhile, the cross-over scale $r_c$ is smaller than the initial Hubble radius, hence we expect a large modification to standard 4D gravity.
This can be seen directly from the Friedmann equation~\eqref{eq:rhoJunctCond}: The modification term $(\gamma-1)$ is controlled by $r_c$.

The results of the numerics are depicted in Fig.~\ref{fig:cc_degrav}. Fig.~\ref{fig:hubble_cc_degrav} shows the Hubble parameter on the brane as a function of time.
(The numerical error estimates for $ H $, discussed in Appendix~\ref{ap:numErrors}, are smaller than the line thickness.) We see that $H$ initially decreases to negative values, turns around and approaches zero at late times. This confirms that the static solutions of Sec.~\ref{sec:static_sol} have a finite basin of attraction. This is one of the central results of this work: it is the first example of dynamical {\it degravitation}, and demonstrates how the brane tension can be absorbed into extrinsic curvature while the intrinsic brane geometry tends to flat, Minkowski space. The evolution of the bulk geometry, characterized by $\alpha$, is shown in Fig.~\ref{fig:alpha_cc_degrav}. The initial configuration, as discussed in the last section, leads after a few time steps to a rather narrow Gaussian profile. As time evolves, we see that $\alpha$ describes a two dimensional gravitational wave that moves outwards, gets more and more diluted and asymptotically settles to a constant. 

It remains to check the physicality of the azimuthal pressure component $P_\phi$ required for stabilization. The equation of state corresponding to this pressure component is shown in Fig.~\ref{fig:pPhi_cc_degrav}. The equation of state satisfies the Null Energy Condition ($w_\phi \geq -1$), and is therefore physically reasonable. At late times, $P_\phi\rightarrow 0$, which is consistent
with the static solution for a 4D cosmological constant---see~\eqref{puretension}. Figure~\ref{fig:rhoHat_cc_degrav} shows the effective energy density (including the brane induced terms) that sources the 6D bulk gravity theory, $\hat\rho\equiv \rho - 3M_4^2 H^2$. This quantity remains positive at all times, which indicates a healthy source from the bulk perspective. At late times, $H\rightarrow 0$,
and $\hat\rho$ approaches $\frac{4}{5}\rho_{\rm crit}$, which is consistent with a static solution with brane density given by~\eqref{eq:parameteres_dgrav}.

We have repeated the analysis with a dust $(w=0)$ or radiation $(w=1/3)$ component on the brane and found similar behavior. The system approaches the corresponding static, deficit-angle solutions at late times. The azimuthal pressure $P_\phi$ and effective density $\hat\rho$ are healthy at all times.

\subsection{A super-accelerating solution}
\label{sec:pathol_sol}

Consider once again a 4D cosmological constant source ($ w=-1 $), with the same parameters as before except for a somewhat larger value of $r_c$:
\begin{equation}
		H_i r_c =  \frac{1}{4} \, .
	\label{eq:param_pathol}
\end{equation}
In this case we find completely different behavior. The Hubble parameter on the brane, shown in Fig.~\ref{fig:hubble_cc_pathol}, grows monotonically in time,
which indicates an effective violation of the Null Energy Condition. This growth propagates into the bulk, as can be seen from Fig.~\ref{fig:alpha_cc_pathol}: the wave
function $\alpha(\tau,r)$ grows in time at any $r$.

This pathological behavior is reflected in the azimuthal pressure $P_\phi$, whose equation of state (Fig.~\ref{fig:pPhi_cc_pathol}) becomes less than $-1$ and tends to $-\infty$.
Such an equation of state violates the Null Energy Condition and is rather unphysical. This suggests that no consistent stabilization mechanism exists for a super-accelerating solution. One might wonder 
whether this apparent instability is solely due to this strange azimuthal component required to fix the brane circumference. We found that this is not the case. In Appendix~\ref{sec:vanishingPPhi},
we show that fixing $ P_\phi = 0 $ by hand, and therefore allowing the circumference to evolve in time, still results in super-acceleration.

The instability can be clearly seen by looking at the effective energy density $\hat\rho\equiv \rho - 3M_4^2 H^2$ that sources 6D gravity. As shown in Fig.~\ref{fig:rhoHat_cc_pathol}, 
$\hat{\rho}$ starts out positive but eventually turns around and reaches negative values. This behavior bears resemblance to the DGP model, where the self-accelerating branch leads to a negative
effective energy density~\cite{Gregory:2007xy}. The self-accelerating branch is widely believed to contain a ghost in the spectrum~\cite{Luty:2003vm,Nicolis:2004qq,Koyama:2005tx,Charmousis:2006pn,Gregory:2007xy,Gorbunov:2005zk}. Although the study of perturbations is beyond the scope of this paper, we also expect that the super-accelerating solutions in 6D are likely to have ghosts. (The instability is even more severe in our case, since $\hat{\rho}$ decreases monotonically at late times whereas it is bounded below in DGP.) Note that this instability uncovered here is a nonlinear result which can only be inferred from the full Einstein equations. On a Minkowski background the linear 6D model is stable~\cite{Berkhahn:2012wg}.

\subsection{Contour plot}
\label{sec:contourPlot}

As the above examples show emphatically, our 6D model yields qualitatively very different solutions, depending on the choice of parameters. To study this more systematically,
we now perform a scan over $\rho_i$ and $r_c$, keeping $ H_i R = 0.05 $ fixed. This will allow us, in particular, to understand the border delineating degravitating
and super-accelerating solutions.

The results are shown in Fig.~\ref{fig:contour_plot_full}, where each dot corresponds to one set of parameters for which we ran the numerics. 
The green region (also labeled (1)) corresponds to degravitating solutions. As in the example of Sec.~\ref{sec:degrav_sol}, the brane Hubble parameter  $H$ tends to zero at late times,
and the effective energy density $ \hat\rho $ is always positive. The red region (also labeled (2)) indicates super-accelerating solutions. As in Sec.~\ref{sec:pathol_sol}, $H$ grows unbounded,
while $ \hat\rho $ eventually becomes negative, indicating a classical instability. Finally, the gray region (labeled (3)) corresponds to parameter choices for which the criticality
bound~\eqref{eq:criticalityBound} is violated. As explained earlier, our coordinate system is ill-defined in this case, and hence we cannot make any statements about solutions in this region. 

It turns out that the border between the stable and unstable regions matches perfectly the location in parameter space where 
\begin{equation}
f(\tau) \equiv 1- \frac{9R^2}{2r_c^2\gamma} \, , 
\label{eq:def_f_bis}
\end{equation}
first introduced in~\eqref{eq:def_f}, vanishes. This is drawn as a solid line in Fig.~\ref{fig:contour_plot_full}. In the degravitating regime, $ f $ is negative, and in the super-accelerating regime it is positive.
Since $ f $ appears in the denominator on the right-hand side of the $\dot{H}$ equation~\eqref{eq:pJunctCondRConst}, the evolution of $ H $ becomes ill-defined when $f$ vanishes. The system hits a (physical) singularity, where the numerics of course break down.

To better understand the boundary between the stable and unstable regions, Fig.~\ref{fig:contour_plot_zoom} zooms in on the boxed region of Fig.~\ref{fig:contour_plot_full}.
For parameters sufficiently close to the $f = 0$ line, $ f (\tau)$ dynamically approaches zero after a short time, and the system hits a singularity. 
The basin of attraction for the singularity corresponds to the yellow region (labeled (4)), in which case one starts in the ``healthy'' region,
and the orange region (labeled (5)), in which case one starts in the ``unstable'' region. This is shown in more detail in Fig.~\ref{fig:f_plot_zoom}. 
This yellow-orange attractor region of the singularity, which is hardly visible in Fig.~\ref{fig:contour_plot_full}, can be broadened by injecting
more energy into the bulk initially. This can be achieved by widening the initial Gaussian velocity profile.

We checked that these results are largely unchanged if one uses dust ($ w=0 $) or radiation ($ w=1/3 $) on the brane. Furthermore, we repeated the entire analysis for a different value of the circumference, namely \ $ R = 0.025 H_i^{-1} $, and found similar agreement. In particular, the border between the stable and unstable regimes again coincides with the $ f=0 $ line in parameter space.

\subsection{Interpretation}
\label{sec:interpretation}

The main lesson from the above analysis can be summarized as follows: For sub-critical energy densities, the model is stable if and only if the function $ f(\tau) <  0$.
Using the constraint~\eqref{eq:rhoJunctCond} to eliminate $ \gamma $, this stability condition can be cast into the form
\begin{equation}
\label{eq:stabilityBound}
	\frac{\rho}{\rho_{\mathrm{crit}}}  > r_c^2 H^2 + 1 -  \frac{9R^2}{2 r_c^2} \, .
\end{equation}
If this bound is violated, the model is unstable. The stable and unstable regions are separated by a physical singularity, so it is not possible to evolve dynamically from one region to the other. 

\begin{figure}
	\centering
		\includegraphics[width=0.45\textwidth]{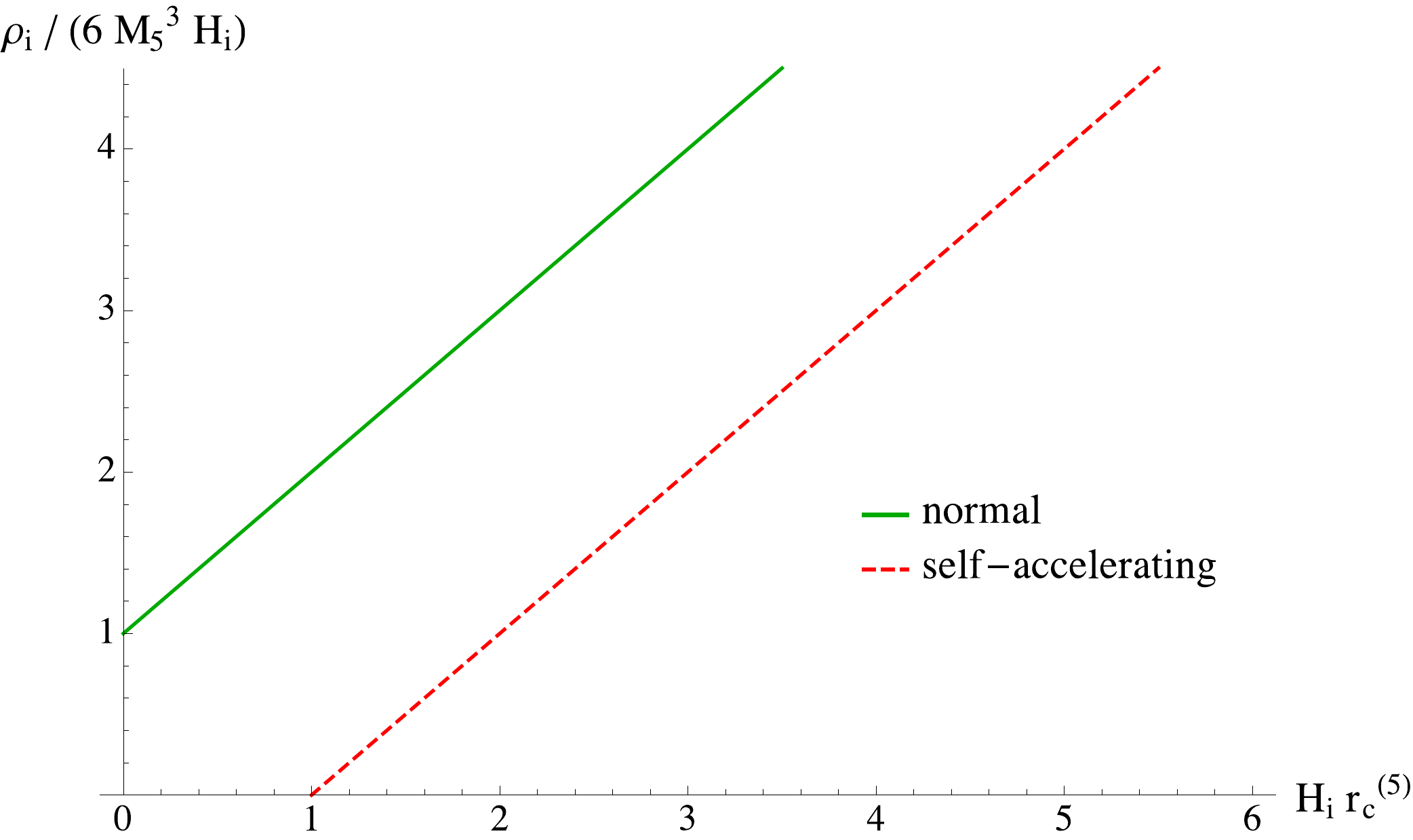}
		\caption{The ``contour'' plot for the DGP model consists of two disjoint lines. The green line is the normal branch, which is stable. The red line is the self-accelerated branch, which is unstable.}
		\label{fig:DGP_plot}
\end{figure}

It is instructive to compare this result with the analogous situation in the DGP model. In that case, the modified Friedmann equation reads~\cite{Deffayet:2000uy}
\begin{equation}
	H^2 = \frac{\rho}{3M_\mathrm{Pl}^2} \pm \frac{\left|H\right|}{r_c^{(5)}}\, ,
\end{equation}
where $r_c^{(5)} \equiv \frac{M_\mathrm{Pl}^2}{2M_5^3}$. The $ - $ sign corresponds to the ``normal'' branch and the $ + $ sign to the ``self-accelerated'' branch. At initial time, this can be rewritten as
\begin{equation}
	\frac{\rho_i}{6M_5^3 H_i} = H_i r_c^{(5)} \mp 1
\end{equation}
The ratio $\frac{\rho_i}{6M_5^3 H_i}$, which is the 5D analogue of $ \frac{\rho}{2\pi M_6^4}$, is fixed (up to the choice of branch) for a given crossover scale $r_c^{(5)}$.
Therefore, the DGP parameter space is only one-dimensional. This difference is due to the fact that in 6D there additional freedom in choosing the initial deficit angle.
The resulting DGP ``contour'' plot, shown in Fig.~\ref{fig:DGP_plot}, is remarkably similar to the 6D setup. The green line corresponds to the normal branch of DGP; this branch is stable, and
the effective density $\hat{\rho}$ is positive. The red line is the self-accelerated branch. On this branch, $H$ is always larger than $ H_{\text{self}} \equiv 1/r_c^{(5)} $, and $ \hat\rho $ is always negative.

Our results generalize this peculiarity of the DGP model to codimension-two. The main differences are: (i) the stable/unstable solutions lie on disconnected branches in the DGP model, whereas they are
separated by a physical singularity in 6D; (ii) there is no criticality bound on $\rho$ in DGP, hence no gray region. 
\section{Phenomenology}
\label{sec:phenomen}

The stable/degravitating (green) region of Fig.~\ref{fig:contour_plot_full} is bounded from above by the critical bound~\eqref{eq:criticalityBound}, and from below by the stability bound~\eqref{eq:stabilityBound}.
Since we have analytic expressions for both borders, we can discuss how this stable region depends on model parameters. Of particular interest is whether phenomenologically viable
points can lie inside this region.

\begin{figure*}[htb]

	\subfloat[$ H R=0.1 $]{
		\includegraphics[width=0.3\textwidth]{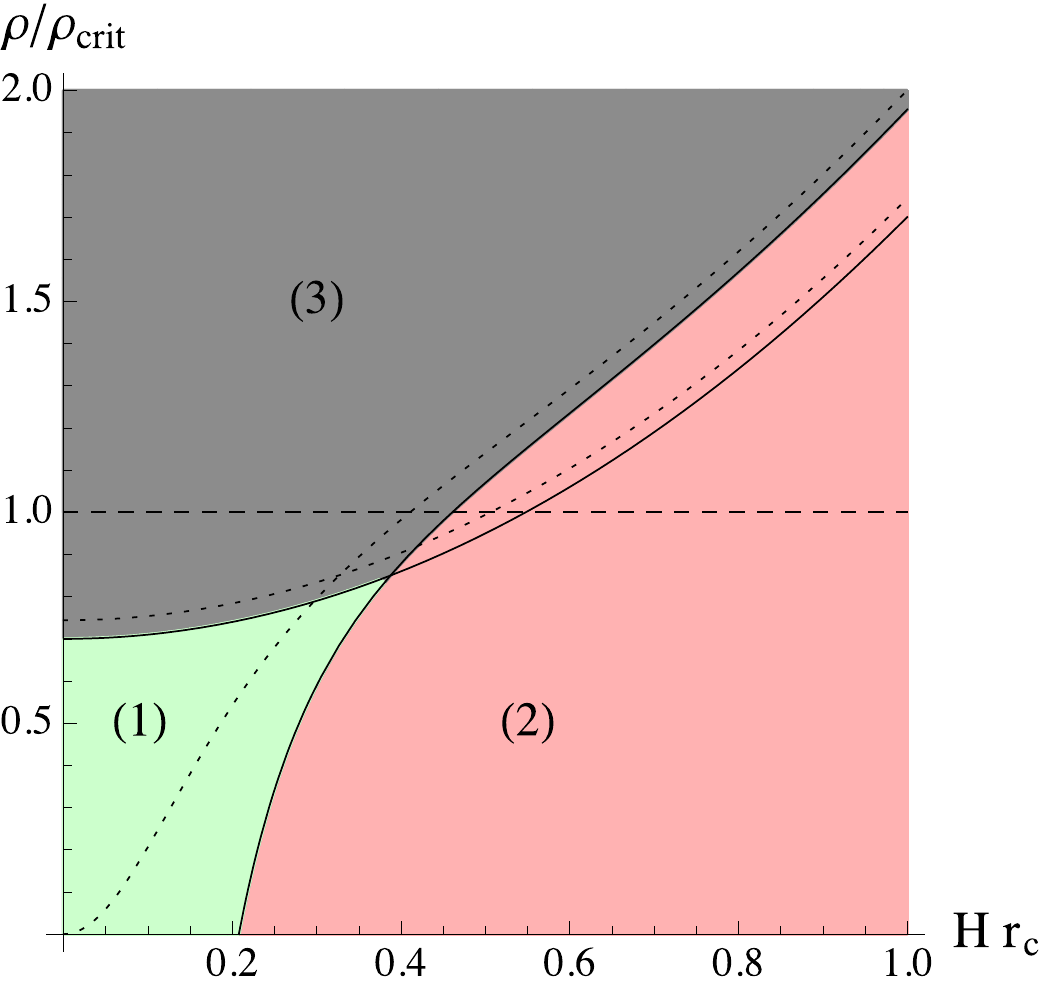}
	}%
	\hfill
	\subfloat[$ H R=0.05 $]{
		\includegraphics[width=0.3\textwidth]{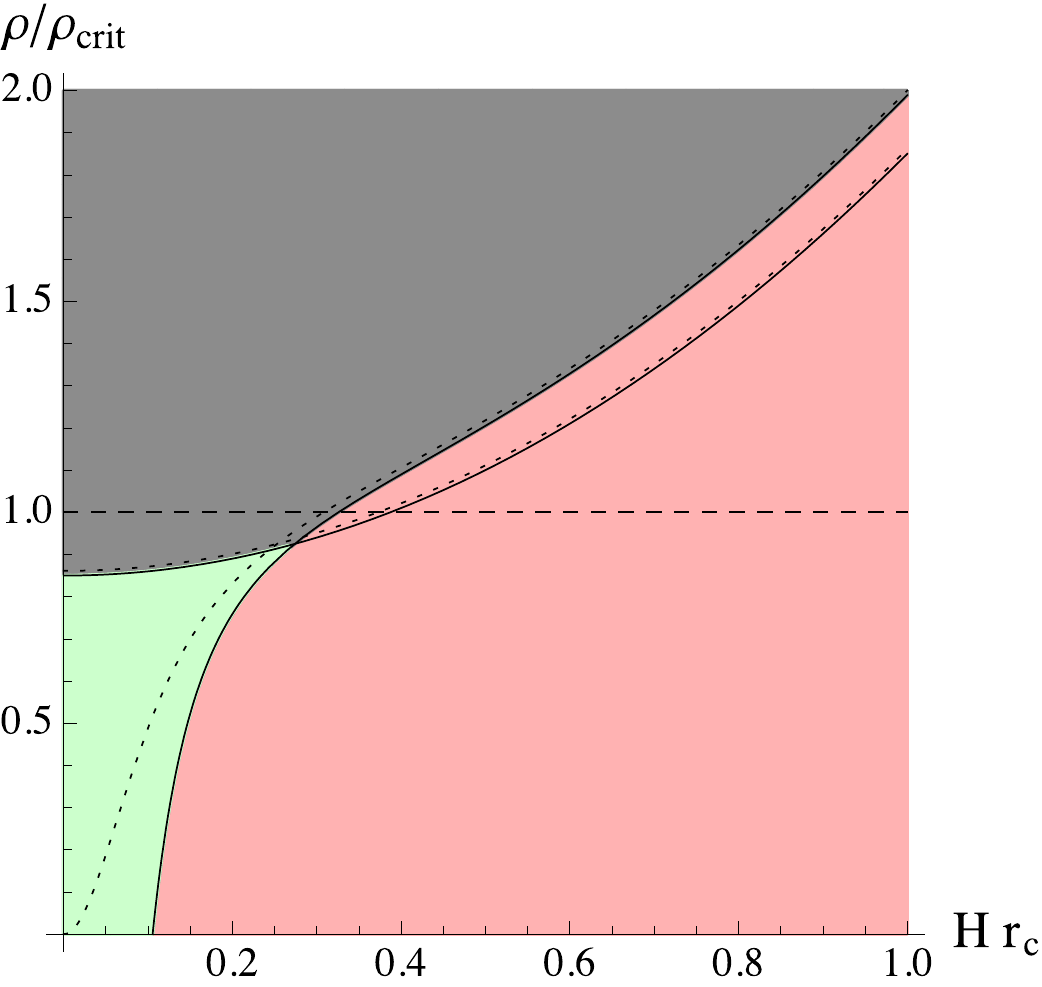}
	}%
	\hfill
	\subfloat[$ H R=0.01 $]{
		\includegraphics[width=0.3\textwidth]{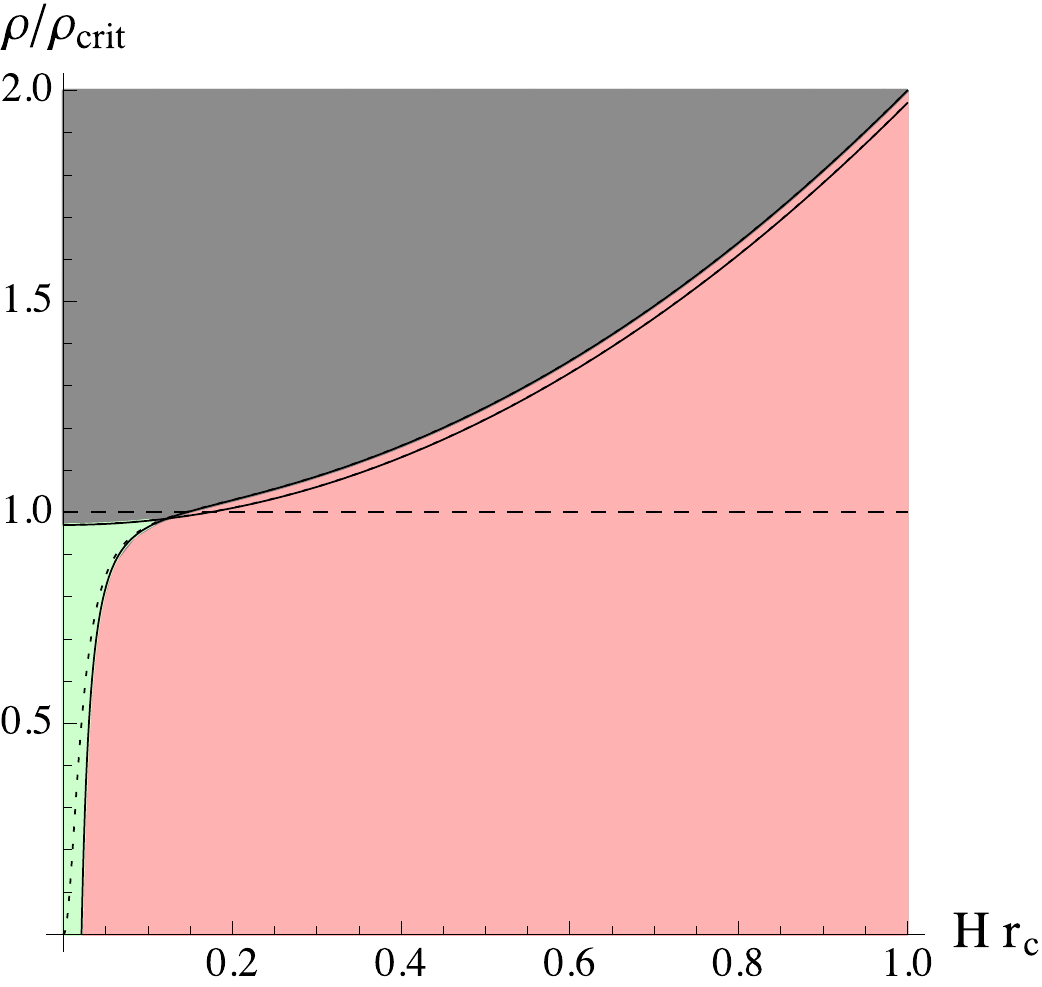}
	}%
	\caption{Contour plots for different values of $ H R $. The dotted lines correspond to the dynamical regularization discussed in Appendix \ref{ap:dynReg}.
	The color scheme is the same as in Fig.~\ref{fig:contour_plot}.}
	\label{fig:contour_plots_analytic}
\end{figure*}

Fig.~\ref{fig:contour_plots_analytic} shows three contour plots for different values of $H R $. In the limit $H R \to 0 $, the degravitating region gets squeezed towards the $H r_c=0$ axis, while approaching $ \rho = \rho_{\text{crit}} $ from below. The dotted lines are the corresponding boundaries for the dynamical regularization discussed in Appendix~\ref{ap:dynReg}. As $H R$ decreases, the dotted and solid lines approach each other, implying that the two regularization schemes agree in this limit, as expected.

The bounds~\eqref{eq:criticalityBound} and~\eqref{eq:stabilityBound} imply that sub-critical, stable solutions exists if and only if
\begin{equation}
\label{eq:maxCrossover}
	\left( H r_c\right)^2  < \frac{3}{2} \left|H\right| R \,.
\end{equation}
This bound can also be derived in the dynamical regularization, in which case it is only a necessary condition.

For phenomenological reasons, we need  $ H r_c \gg 1$ to reproduce standard 4D cosmological evolution on the brane, at least at early times. 
Indeed, if instead $ H r_c \lesssim 1 $, then the system will exhibit a 6D behavior. On the other hand, we must have $H R\ll 1$, as mentioned in~\eqref{UV_IR_Limit},
in order for brane physics to admit an effective 4D description. Clearly, these two requirements---$ H r_c \gg 1$ and $H R\ll 1$---are mutually incompatible,
given~\eqref{eq:maxCrossover}. In other words, {\it the model admits no (sub-critical) solutions that are both stable and phenomenologically viable}.

\begin{figure}
	\includegraphics[width=0.45\textwidth]{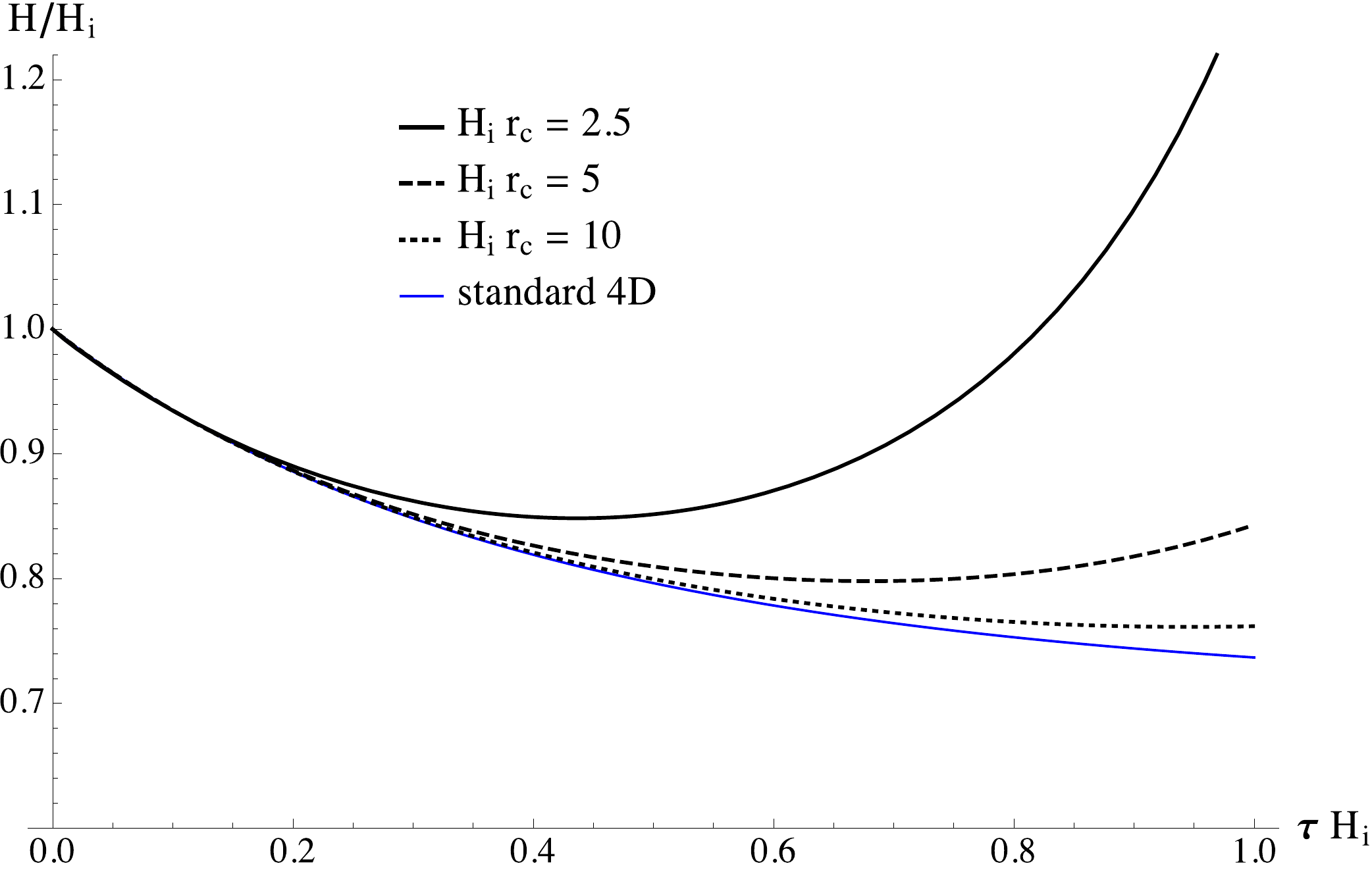}
	\caption{The Hubble evolution for different values of the cross-over scale $r_c$ (black curves), compared to the standard 4D evolution (blue curve). Since $ H_i r_c > 1 $, these curves all lie deep inside the super-accelerating/red region. As the value of $r_c$ is increased, the solution traces the 4D evolution for longer.}
	\label{fig:hubble_4D_regime}
\end{figure}

In the super-accelerating (red) region of Fig.~\ref{fig:contour_plot_full}, on the other hand, there is no problem with achieving arbitrarily large values of $Hr_c$.
Fig.~\ref{fig:hubble_4D_regime} shows the Hubble evolution for different values of $r_c$ (black curves), compared to the standard 4D evolution (blue curve). 
The matter consists of dust and cosmological constant, with
\begin{equation}
\rho_i^{\text{cc}} = \rho_i^{\text{dust}} = \frac{1}{2} \left( H_i^2 r_c^2 + 0.8 \right) \rho_{\text{crit}} \,.
\end{equation}
As expected, the larger the $r_c$ value, the longer the standard evolution is traced. Once the modification kicks in, however, the evolution becomes unstable and super-accelerating.
This instability, accompanied by a negative effective energy density, should be regarded as strong indications against the physical relevance of those solutions. We expect fluctuations
around such backgrounds to exhibit ghost instabilities, analogous to the DGP model. It would of course be worthwhile to verify this expectation through explicit calculation.
While it would be desirable to further verify this last claim, we think that our current results already suggests that the super-accelerating solutions should not be regarded as consistent alternative cosmologies.

\section{Conclusion}
\label{sec:conclusion}

In this work, the cosmology of the brane induced gravity model in $ 6 $ dimensions has been investigated. The existence of bulk gravitational waves, and the fact that a (nontrivial) FRW codimension-two brane cannot be embedded in a Minkowski bulk, makes it impossible to derive a local on-brane Friedmann equation as in the DGP case. Therefore, we solved the full (nonlinear) system of bulk-brane equations numerically.

We found that the model can show two qualitatively different behaviors: either the solutions degravitate, \textit{i.e.}, they dynamically approach the static deficit angle solution, or they super-accelerate, \textit{i.e.}, the Hubble parameter grows unbounded. This instability originates from the effective energy density $ \hat\rho $, which sources six-dimensional GR, becoming negative in those cases. It is very likely---though we have not shown this in the present work---that perturbations around those solutions would allow for ghosts, on top of the classical instability of the background itself. It would certainly be desirable to verify this claim; one strong indication for it is that this is exactly what happens in the DGP case: ghosts are present in fluctuations around the self-accelerated branch, which also has $ \hat\rho < 0 $. But in 6D the instability already shows up in the background solution, which is why we already consider them physically irrelevant.

Whether a solution degravitates or super-accelerates depends on the three independent (dimensionless) parameters $ H R $, $ H r_c $ and $ \rho / \rho_{\mathrm{crit}} $. We were able to derive an analytic expression that determines the border between the two regimes and showed that it corresponds to a physical singularity. Thus, a solution can never dynamically evolve from one regime to the other.

Unfortunately, it turned out that the stable, degravitating solutions are not phenomenologically viable because they never lead to an almost 4D behavior, and thus could never match the past history of our universe which is very well described by the standard FRW evolution. On the other hand, phenomenologically interesting parameters $ H r_c \gg 1 $, $ H R \ll 1 $ which are indeed able to mimic a 4D evolution, always lead to an instable behavior once the modification sets in. Unless there is some way to make sense of those instable solutions---which seems very unlikely---we conclude that the BIG model in $ d=6 $ is ruled out (for sub-critical energy densities).

It should be noted that we have not investigated super-critical energy densities. An effective field theory (EFT) analysis in Appendix~\ref{ap:EFT}  shows that for large enough values of the regularization scale ($R> M_6^{-1}$) this constitutes the remaining window in parameter space which could allow for a phenomenologically interesting solution.   %
Finally, we have not considered a cosmological constant in the bulk. It might be interesting to check how relaxing this assumption would change the size of the healthy region in parameter space.

\begin{acknowledgments}
We thank Felix Berkhahn, Gia Dvali and Michael Kopp for helpful discussions.
FN and RS would like to thank the Department of Physics and Astronomy at the University of Pennsylvania for its hospitality in the course of this work. The work of SH was supported by the DFG cluster of excellence `Origin and Structure of the Universe' and by TRR 33 `The Dark Universe'. The work of FN and RS was supported by the DFG cluster of excellence `Origin and Structure of the Universe'. JK is supported in part by NSF CAREER Award PHY-1145525 and NASA ATP grant NNX11AI95G. 
\end{acknowledgments}
\appendix
\section{Dynamical regularization}
\label{ap:dynReg}

In the dynamical regularization, the space-time geometry in the interior of the cylinder is resolved and its dynamical impact on the brane evolution is properly taken into account. This has the advantage that the regularity condition at the axis can be implemented and thus one obtains a fully self-consistent and non-singular solution of the (modified) Einstein equations in the whole space-time. Note that in the static regularization, the brane dynamics was not influenced by an interior geometry because the system (brane + exterior bulk) was closed by defining the brane as the boundary of space-time \eqref{israel2} or, equivalently, setting the extrinsic curvature in the interior to its static value \eqref{eq:RegII}. The geometrically more consistent boundary condition is the one that ensures regularity at the axis. However, this has the drawback that one has to specify more initial data, and that the solutions will become more sensitive to those initial conditions, because gravitational waves that are reflected at the axis can influence the on-brane evolution. However, it turns out that the solutions obtained in the two regularizations agree very well, up to small oscillations in the dynamical case which are caused by the initial conditions. This result shows that the static regularization is indeed an efficient way to get rid of the dependency on the interior geometry, but without affecting the evolution on the time-scales we are actually interested in.

In this section, we give the details of the dynamical regularization. We will consider the case in which the brane circumference is fixed (as in the main text), but also the case $ P_\phi = 0 $ in which the brane circumference becomes time-dependent. The latter case serves as a proof that super-acceleration in the stabilized scenario is not caused by the (unphysical) equation of state of $ P_\phi $. We present the numerical results in Sec.~\ref{ap:num_sol} and compare them to the ones obtained in the static regularization, which were shown in the main body of the paper.

\subsection{Brane bulk dynamics}

As discussed in Appendix~\ref{ERcoords}, the Einstein-Rosen coordinates \eqref{eq:met_cyl_symm_2} can only be introduced in vacuum regions of space-time.
Since the interior of the cylinder is also source-free, we can use the same metric ansatz there.
However, the energy-momentum tensor that is localized on the brane then implies that the interior and exterior coordinate patches will not be continuously connected.
To distinguish them, we will put tildes on all coordinates and functions that live in the interior, so the line element inside is
\begin{equation}\label{eq:met_int}
	\rd \tilde s^2 = \re^{2(\tilde \eta - 3\tilde \alpha)} \left( -\rd \tilde t^2 \!+ \rd \tilde r^2 \right) + \re^{2\tilde \alpha} \rd  \vec{x}^2 + \re^{-6\tilde\alpha} \tilde r^2 \rd\phi^2 \; ,
\end{equation}
with $ \tilde \alpha $ and $ \tilde \eta $ being functions of $ (\tilde t,\tilde r) $. (The coordinates $ \vec{x} $ and $ \phi $ are continuous, so there is no need for tildes on them.)
Einstein's field equations inside the cylinder take of course the same form as outside, equation \eqref{eq:einstein_vacuum} with the replacement $(t,r,\alpha,\eta)\rightarrow (\tilde t,\tilde r,\tilde \alpha,\tilde \eta)$.

Regularity at the axis implies the condition
\begin{equation}\label{eq:cond_axis_alpha}
	\lim_{\tilde r \to 0} \partial_{\tilde r} \tilde \alpha = 0
\end{equation}
and elementary flatness, \textit{i.e.}, the absence of a conical singularity, requires 
\begin{equation}\label{eq:cond_axis_eta}
	\qquad \lim_{\tilde r \to 0} \tilde \eta = 0 \;.
\end{equation}

Denoting with $ \tilde r_0(\tilde t)$ the brane position in the interior coordinate patch, and defining $\tilde \gamma$ analogously to $ \gamma $ \eqref{eq:defGamma},
continuity of the metric at the position of the brane yields
\begin{subequations}
	\begin{align}
		\alpha_0(t)		&=\tilde \alpha_0(\tilde t)\;, \label{eq:cont_alpha}\\
		r_0(t)			&=\tilde r_0(\tilde t)\;,\\
		\frac{\rd t}{\gamma}	&=\frac{\rd \tilde t}{\tilde \gamma}\;. \label{eq:timeRelation}
	\end{align}
\end{subequations}

The extrinsic curvature at the exterior and interior boundary of the cylinder are calculated using the outward pointing normal vectors
\begin{align}\label{eq:normal_vec_int}
	n^A & = \gamma \re^{3\alpha_0} \left( \frac{\rd r_0}{\rd t}, 1, 0,0,0,0 \right)\,, \\
	\tilde n^A & = \tilde\gamma \re^{3\tilde\alpha_0} \left( \frac{\rd \tilde r_0}{\rd \tilde t}, 1, 0,0,0,0 \right)\,,
\end{align}
respectively.
Using this, Israel's junction conditions \eqref{israel2} become:
\begin{widetext}
\begin{subequations}
\label{eq:junctCond_dynReg}
	\begin{align}
		- \frac{\rho}{3 M_{\rm Pl}^2} + \biggl ( H^2 + H H_R \biggr ) & = \frac{1}{r_c^2} \left( \gamma - \tilde\gamma \right)  \label{eq:rhoJunctCond_dynReg} \\
		 \frac{P}{M_{\rm Pl}^2} +  \left( 2\frac{\rd H}{\rd \tau} + \frac{\rd H_R}{\rd \tau} + 3 H^2 + H_R^2 + 2 H H_R \right)  & = 
		 \frac{3 }{r_c^2}  \left[ \gamma \left( 1 + \frac{r_0 \frac{\rd^2 r_0}{\rd t^2}}{1 - \left( \frac{\rd r_0}{\rd t} \right)^2} \right) + R \, n^A\partial_A \left( \eta - 4 \alpha\right) |_0 - \text{``tilde''} \right]  \label{eq:pJunctCond_dynReg} \\
		\frac{P_\phi}{3M_{\rm Pl}^2} +  \left ( \frac{\rd H}{\rd \tau} + 2 H^2 \right ) & = \frac{1}{r_c^2}  \left[\gamma  \frac{r_0 \frac{\rd^2 r_0}{\rd t^2}}{1 - \left( \frac{\rd r_0}{\rd t} \right)^2} + R\, n^A\partial_A \eta|_0 - \text{``tilde''}  \right]   \label{eq:pPhiJunctCond_dynReg}
	\end{align}
\end{subequations}
\end{widetext}
Here ``tilde'' is shorthand for repeating all the terms in the square brackets, but with tildes on all functions and variables.
Note that we have not assumed $ R = \mathrm{constant} $, and so the brane induced gravity terms $ \propto G^{(5)a}_{\hphantom{(5)a}b} $ on the left-hand side of \eqref{eq:junctCond_dynReg} receive contributions not only from $ H $, but also from $ H_R \equiv \dot R / R $. Furthermore, the energy conservation equation now reads
\begin{equation}
\label{eq:enCons5D}
	\frac{\rd \rho^{(5)}}{\rd \tau} + 3H \left( \rho^{(5)} + P^{(5)} \right) + H_R \left( \rho^{(5)} + P_\phi^{(5)} \right) = 0 \, ,
\end{equation}
where the five-dimensional source terms are related to the four dimensional ones by $ T^{(5)a}_{\hphantom{(5)a}b} = T^{a}_{\hphantom{a}b} / (2\pi R) $.
As a consistency check, one can verify that this conservation equation follows from the junction conditions \eqref{eq:junctCond_dynReg}, together with the vacuum Einstein equations \eqref{eq:einstein_vacuum}.

Finally, it will be convenient to work with $ R $ instead of $ r_0 $ and $ \tilde r_0 $ in \eqref{eq:junctCond_dynReg}. To this end, note that the definition of $R$ in \eqref{eq:defR} implies $ \dot r_0 = \left(3H + H_R \right) r_0 $ and so \eqref{eq:defGamma} can be written as
\begin{equation}\label{eq:gamma_gen}
	\gamma = \sqrt{\re^{-2\eta_0} + \left( 3H + H_R \right)^2 R^2} \, .
\end{equation}
Another straightforward calculation gives
\begin{multline}\label{eq:dotdot_r0_gen}
	\frac{r_0 \frac{\rd^2 r_0}{\rd t^2}}{1 - \left( \frac{\rd r_0}{\rd t} \right)^2} = \frac{R^2}{\gamma^2} \Bigl[ 3\dot H + \dot H_R \\
	+ \left( 3H + H_R \right) \left( H_R + \dot \eta_0 \right) \Bigr].
\end{multline}
Equations \eqref{eq:gamma_gen} and \eqref{eq:dotdot_r0_gen} similarly hold for the tilde quantities, \textit{i.e.}, for $ (\gamma, \eta_0, r_0) \to (\tilde\gamma, \tilde \eta_0, \tilde r_0) $.

As before, the equations of motion are only closed after specifying an equation of state for both of the two pressure components $ P $ and $ P_\phi $. For the former we will again assume a fixed (but arbitrary) linear equation of state $ P = w \rho $. For the latter, we will consider two different possibilities: $(a)$ $ P_\phi $ is chosen to stabilize the brane circumference, exactly as it was done in the main part of this work; $(b)$ $ P_\phi = 0 $. Let us now further discuss the two cases separately.

\subsubsection{Fixed brane width}

As in the main text, we set $ H_R = 0 $ and use the junction condition \eqref{eq:pPhiJunctCond_dynReg} only to infer the value of $ P_\phi $ that is needed to stabilize the brane. The remaining junction conditions take the form:
\begin{subequations}
\label{eq:modFried_dynReg}
	\begin{align}
		H^2 &= \frac{\rho}{3 M_{\rm Pl}^2} + \frac{1}{r_c^2} \left( \gamma - \tilde{\gamma} \right) \label{eq:modFried1_dynReg} \, , \\
		\dot H &= \frac{-3}{2\hat f(\tau)}\left[ \frac{P}{3 M_{\rm Pl}^2} + H^2 - \frac{1}{r_c^2} \left( \gamma\, g(\xi, \chi) -\tilde \gamma\, g(\tilde \xi, \tilde \chi) \right) \right]\,, \label{eq:modFried2_dynReg}
	\end{align}
\end{subequations}
with
\begin{align}
	\gamma = \sqrt{\re^{-2\eta_0} + 9 H^2 R^2 } \, , &&
	\tilde\gamma = \sqrt{\re^{-2\tilde\eta_0} + 9 H^2 R^2 } \, ,
\end{align}
the function $ g $ is the one defined in \eqref{eq:def_g} and
\begin{equation}
	\hat f(\tau) \equiv 1- \frac{9R^2}{2 r_c^2}\left(\frac{1}{\gamma}-\frac{1}{\tilde\gamma}\right) \, . \label{eq:def_fHat}
\end{equation}


The modified Friedmann equations \eqref{eq:modFried_dynReg} are very similar to the ones of the static regularization, \eqref{eq:rhoJunctCond} and \eqref{eq:pJunctCondRConst}, with the crucial difference that now the quantities $ \tilde\eta_0 $ and $ \partial_{\tilde r} \tilde \alpha_0 $ enter, which are determined by the interior bulk evolution. In this way, the brane evolution is now influenced by the space-time dynamics inside the cylinder. Furthermore, note that the function $ f $ is now slightly modified to $ \hat f $.

\subsubsection{Vanishing azimuthal pressure}

For $ P_\phi = 0 $ the brane circumference $ R $ will in general be time-dependent, and so the energy conservation equation \eqref{eq:enCons5D} now implies
\begin{equation}
	\rho^{(5)}(\tau) \propto \frac{1}{R(\tau)} \re^{-3(1+w)\alpha_0(\tau)} \, .
\end{equation}
As a consequence, the dimensionally reduced quantity $ \rho \equiv 2\pi R \, \rho^{(5)} $ scales exactly as before. 

We can still formally introduce the four dimensional Planck-scale and the crossover-scale as in \eqref{M4} and \eqref{eq:defCrossover} respectively, but one has to keep in mind that they will now be functions of time as well. Specifically, they scale with $R$ as $ M_{\rm Pl}(\tau), r_c(\tau) \propto \sqrt{R(\tau)} $.

The junction conditions then become
\begin{subequations}
\label{eq:modFried_pPhi_0}
\begin{align}
	H^2 + H H_R &= \frac{\rho}{3 M_{\rm Pl}^2} + \frac{1}{r_c^2} \left( \gamma - \tilde{\gamma} \right) \label{eq:modFried1_pPhi_0} \, , \\
	\dot H &= \frac{A \delta + B}{1 - 4 \delta} \, , \label{eq:modFried2_pPhi_0}\\
	\dot H_R &= \frac{A \left( 1 - 3\delta \right) + B}{1 - 4 \delta} \, , \label{eq:modFried3_pPhi_0}
\end{align}
\end{subequations}
with the following definitions:
\begin{widetext}
\begin{subequations}
\begin{align}
	A &\equiv -\frac{P}{M_{\rm Pl}^2} + 3H^2 - 2 H H_R - H_R^2 + \frac{3}{r_c^2} \Bigl \{ \gamma \left [ 1 - 4 \left ( \xi + \frac{\rd r_0}{\rd t} \psi \right ) \right ] -  \text{``tilde''} \Bigr \} \, ,\\
	B &\equiv - 2 H^2 + H_R \left ( 3H + H_R \right ) \delta + \frac{6}{r_c^2} \Bigl \{ \gamma \left [ 4 \frac{\rd r_0}{\rd t} \xi \psi + \left (1 + \left( \frac{\rd r_0}{\rd t} \right)^2 \right ) \left ( \xi^2 + \psi^2 \right ) \right ] -  \text{``tilde''} \Bigr \} \, ,\\
	\delta &\equiv \frac{R^2}{r_c^2} \left( \frac{1}{\gamma} - \frac{1}{\tilde{\gamma}} \right) \, ,\\
	\psi &\equiv r_0 \partial_t \alpha_0 = \frac{R}{\gamma} \left [ H - \xi \left ( 3H + H_R \right ) \right ] \quad\quad \text{(and similarly for $\tilde\psi$)} \, .
\end{align}
\end{subequations}
\end{widetext}
This time there are two dynamical equations of motion, equations \eqref{eq:modFried2_pPhi_0} and \eqref{eq:modFried3_pPhi_0}, which will be used to numerically determine $ H $ and $ H_R $, respectively. The constraint equation \eqref{eq:modFried1_pPhi_0} again serves as a non-trivial consistency check for the numerics.

\subsection{Numerical implementation and initial data}
\label{ap:num_impl}
The numerical scheme is the same as the one used for the solutions in the static regularization, with two slight modifications: First, we now also have to specify initial data in the interior that have to be compatible with the boundary conditions \eqref{eq:cond_axis_alpha} and \eqref{eq:cond_axis_eta}. Second, since we have two discontinuous time coordinates for the interior and exterior region, the corresponding temporal grid points will in general not agree at the brane position. We again deal with this problem by some suitable interpolation scheme. The details of the numerical implementation can be found in Appendix \ref{ap:numImpl}.

The initial data in the exterior is chosen in the same way as before, \textit{i.e.}\ as discussed in Sec.~\ref{sec:num_impl}.
But now we also have to specify the initial radial profile $ \tilde\alpha_i(\tilde r) $ and its time derivative $ \partial_{\tilde t} \tilde \alpha_i(\tilde r) $ for $ \tilde r \in [0, R] $.
As for the exterior, we choose the profile of the static solution discussed in Sec.~\ref{sec:static_sol}, which is simply
\begin{equation}
\label{eq:initProfileAlphaTilde}
	\tilde\alpha_i(\tilde r) = 0 \, .
\end{equation}
For the velocity profile, regularity at the axis \eqref{eq:cond_axis_alpha} implies
\begin{equation}
	\partial_{\tilde r}\partial_{\tilde t}\tilde\alpha_i(0) = 0.
\end{equation}
At the brane position, it is related to the initial Hubble parameter $ H_i $ via
\begin{equation}
\label{eq:alpha0TildeDot_init_dyn}
	\partial_{\tilde t} \tilde \alpha_{0i} = \frac{\rd \tilde \alpha_{0i} }{\rd \tilde t} = \frac{H_i}{\tilde\gamma_i}
\end{equation}
where the first equality uses $ \partial_{\tilde r}\tilde\alpha_i = 0 $ which is satisfied for our choice \eqref{eq:initProfileAlphaTilde}. We can thus write
\begin{equation}
	\partial_{\tilde t} \tilde \alpha_i(\tilde r) = \frac{H_i}{\tilde\gamma_i} \tilde F(\tilde r)
\end{equation}
with some profile function $ \tilde F(\tilde r) $ satisfying the boundary conditions $ \tilde F'(0) = 0 $ and $ \tilde F(R) = 1 $.
For definiteness, we will choose the flat profile
\begin{align}
	\tilde F(\tilde r) \equiv 1 \, .
\end{align}
This choice is motivated by the observation that for $ R $ small enough, the regularity condition at the axis implies that $ \partial_{\tilde r}\tilde\alpha \approx 0 $. We again expect the on-brane evolution to become insensitive to the initial conditions for late times.

For the case $ P_\phi = 0 $ we also have to specify an initial value for $ H_R $, which we will (for simplicity) set to zero:
\begin{equation}
	(H_{R})_{ i} = 0 \,.
\end{equation}
In particular, this implies that the initial constraint \eqref{eq:modFried1_pPhi_0} is identically to the case of fixed $ R $, \eqref{eq:modFried1_dynReg}.
Furthermore, the values for $ R $ and $ r_c $---which are constant for $ H_R = 0 $---will be used as the initial values $ R_i $ and $ (r_{c})_{i} $ in the case $ P_\phi = 0 $, when comparing the corresponding solutions.

This completes the specification of initial data. Indeed, the remaining variables $ \tilde\eta_{0i} $ and $ \eta_{0i} $ are determined by the regularity condition \eqref{eq:cond_axis_eta} together with the constraints \eqref{eq:etaPrime_vac} and \eqref{eq:modFried1_dynReg}\footnote{The full radial profile of $ \tilde\eta, \eta $ can be calculated from \eqref{eq:etaPrime_vac}, but is actually not needed for the evolution of $ \alpha $. Only $ \tilde\eta_0, \eta_0 $ enter through the junction conditions, and those can be calculated at later times from their initial values using \eqref{eq:etaPrime_vac}, \eqref{eq:etaDot_vac} only locally at the brane position.}.
Specifically,
\begin{subequations}
\label{eq:eta0Tilde_init}
\begin{align}
	\tilde \eta_{0i} & = 6 \int_0^{R} \!\rd \tilde r \, \tilde r \left[ \left(\partial_{\tilde r} \tilde\alpha_i \right)^2 + \left(\partial_{\tilde t} \tilde\alpha_i \right)^2 \right] \label{eq:eta0Tilde_init_a}\\
	& = \frac{6 H_i^2}{\gamma_i^2} \int_0^{R} \!\rd \tilde r \, \tilde r \tilde F^2 \\
	& = \frac{3 H_i^2 R^2}{\re^{-2\tilde\eta_{0i}} + 9 H_i^2 R^2} \, ,
\end{align}	
\end{subequations}
an implicit equation for $ \tilde\eta_{0i} $ which can be solved numerically. (Note that for any value of $ H_i R $ there exists a unique real solution for $ \tilde\eta_{0i} $ to this equation.) It is also interesting that $ \tilde\eta_{0i} $ is a direct measure of the gravitational energy stored inside the cylinder initially, which is suggested by \eqref{eq:eta0Tilde_init_a}. In fact, it is (up to a constant factor) nothing but the so called {\it C-energy} introduced by Thorne \cite{Thorne:1965}, generalized to 6 dimensions. We will come back to this point in Sec.\ \ref{sec:contourPlot_dynReg}, when discussing the dependence on the interior initial data.

The exterior $ \eta_{0i} $ is finally obtained from \eqref{eq:modFried1_dynReg}, which can be rewritten as
\begin{equation}
	\frac{\rho_i}{\rho_{\rm crit}} = r_c^2 H_i^2 + \tilde\gamma_i - \sqrt{ \re^{-2\eta_{0i}} + 9 H_i^2 R^2 } \, .
\end{equation}
The existence of a real solution for $ \eta_{0i} $ (and in fact for $ \eta_0 $ at any other time) again places an upper bound on the energy density:
\begin{equation}
\label{eq:criticalityBound_dynReg}
	\frac{\rho}{\rho_{\rm crit}} < r_c^2 H^2 + \tilde\gamma - 3 \left|H\right| R \, .
\end{equation}
This should be compared to the criticality bound \eqref{eq:criticalityBound} in the static regularization, which is formally obtained from \eqref{eq:criticalityBound_dynReg} by the replacement $ \tilde\gamma \to 1 $. As soon as this bound is violated, the initial constraint cannot be fulfilled, and we are in the super-critical regime which is excluded from our analysis.

Finally, one also has to specify a grid spacing for the interior domain. As for the exterior, we will take the temporal and radial grid spacing to be equal and constant, but it can be different from the spacing outside:
\begin{equation}
	\Delta \tilde t = \Delta \tilde r \equiv \tilde \epsilon
\end{equation}
In what follows we will present the results.

\subsection{Numerical solutions}
\label{ap:num_sol}

\begin{figure*}[bht]
	\centering
	\subfloat[The radial profile of $ \alpha $ at different values of $ \tau $. The dots indicate the brane position, left of which the plotted function is the interior $ \tilde\alpha(\tilde r) $.]{
		\includegraphics[width=0.45\textwidth]{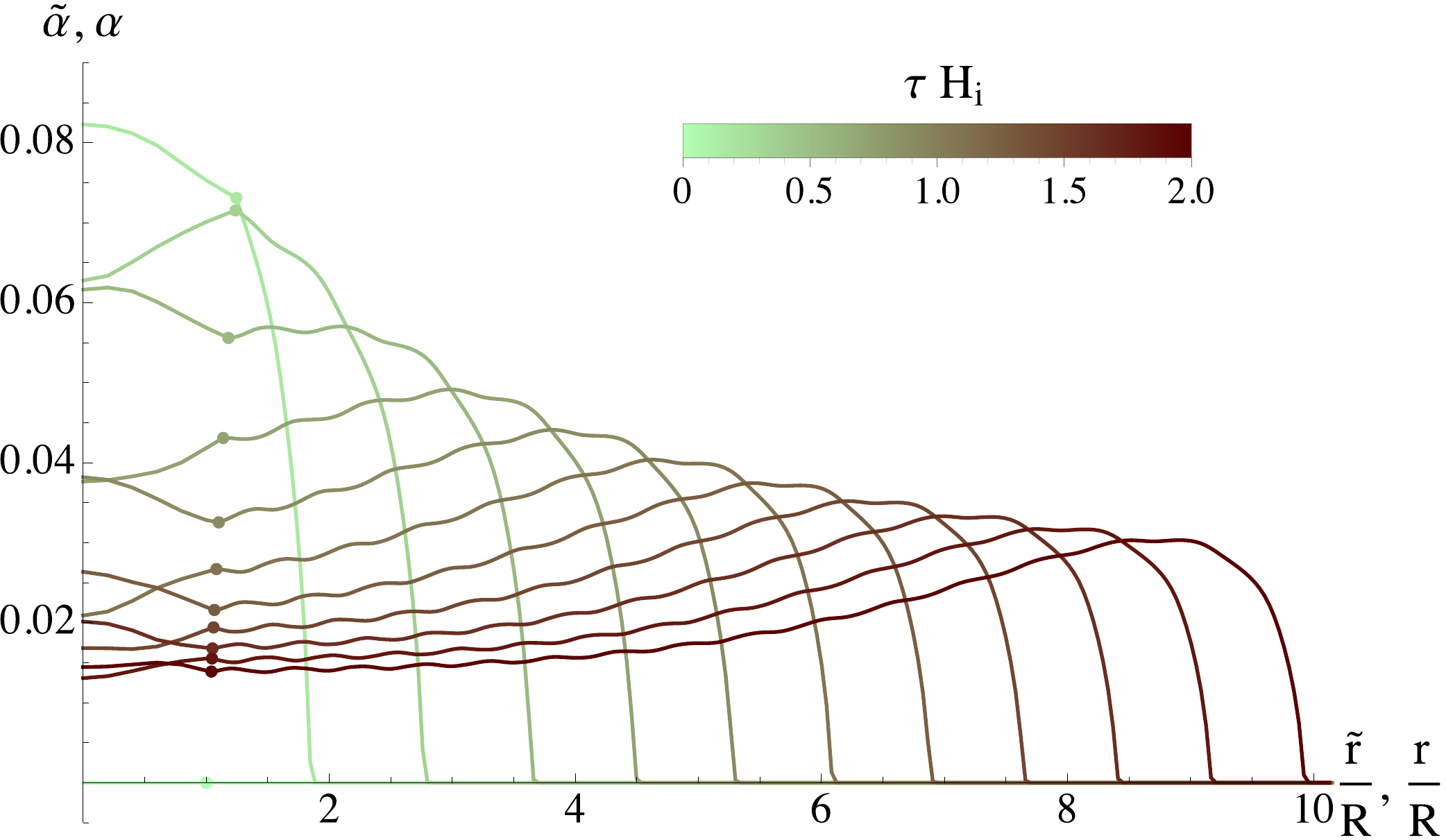}
		\label{fig:alpha_cc_degrav_dynReg}
	}
	\hfill
	\subfloat[The Hubble parameter in the static regularization smoothly traces the mean of the one in the dynamical regularization.]{
		\includegraphics[width=0.45\textwidth]{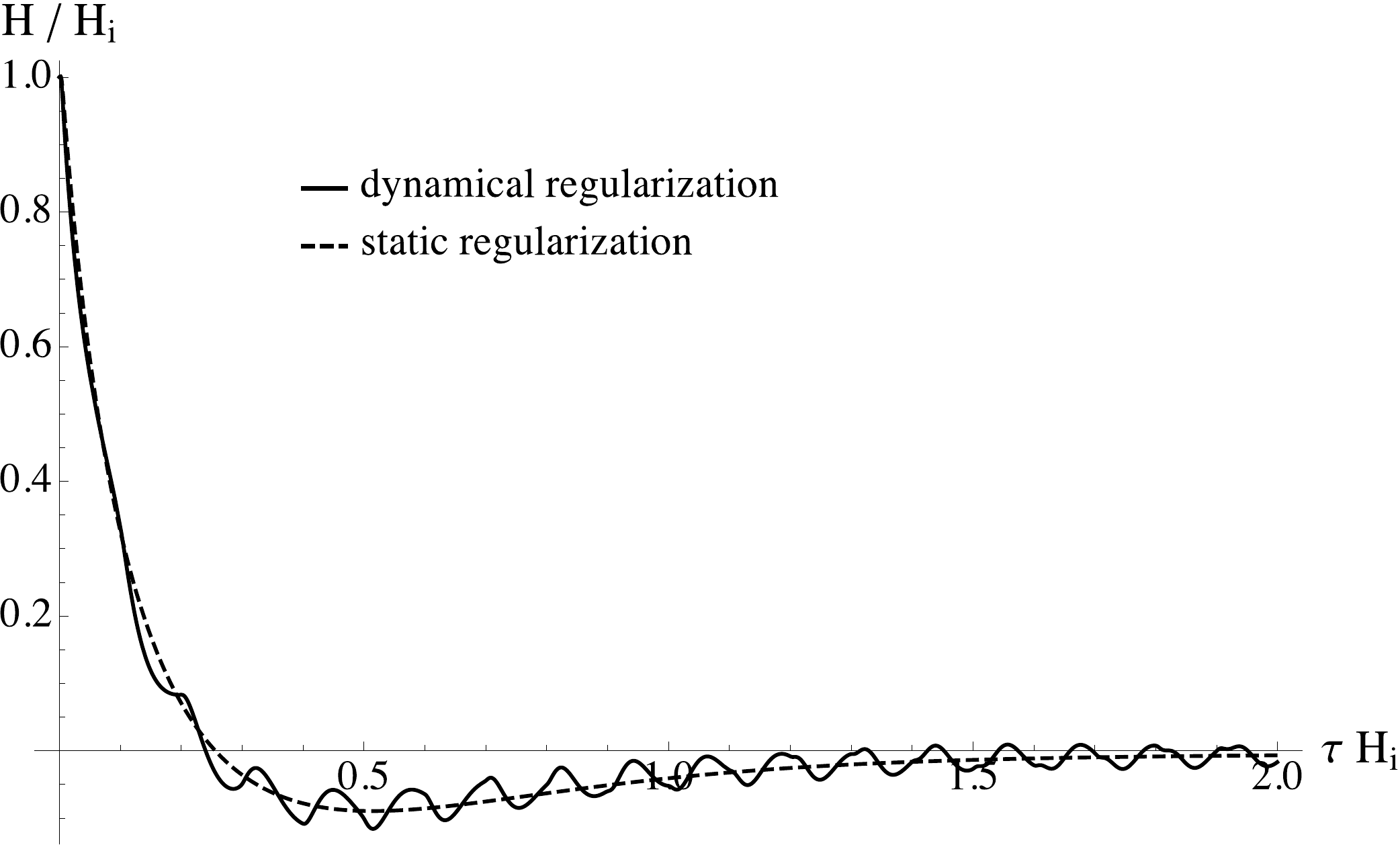}
		\label{fig:hubble_cc_degrav_dynReg}
	}
	\caption{The degravitating solution of Fig.~\ref{fig:cc_degrav}, but in the dynamical regularization.}\label{fig:cc_degrav_dynReg}
\end{figure*}

In Secs.~\ref{sec:degrav_sol_dynReg} and \ref{sec:patho_sol_dynReg}, respectively, we re-investigate the two fiducial degravitating and super-accelerating solutions from the main text in the dynamical regularization. We thereby find that our results are independent of the choice of regularization scheme. In Sec.~\ref{sec:vol_stabil}, we show that the interior volume of the cylinder is sufficiently stabilized in the dynamical regularization, and Sec.~\ref{sec:vanishingPPhi} excludes the unphysical behavior of $ P_\phi $ in the stabilized scenario as the source of super-acceleration by considering $ P_\phi = 0 $. Finally, Sec.\ \ref{sec:contourPlot_dynReg} investigates the regimes in parameter space corresponding to the stable/unstable solutions in the dynamical regularization.

\subsubsection{Degravitating solution}
\label{sec:degrav_sol_dynReg}

Let us first consider the case $ H_R = 0 $, for the parameters \eqref{eq:parameteres_dgrav}, which led to a degravitating solution in the static regularization (cf.\ Sec.~\ref{sec:degrav_sol}).

The numerical results are shown in Fig.~\ref{fig:cc_degrav_dynReg}. Unlike in the static regularization, now the geometry inside the cylinder---characterized by $ \tilde\alpha $---is obtained as well, which is plotted in Fig.~\ref{fig:alpha_cc_degrav_dynReg} together with $ \alpha $. It is evident from this plot that dynamically resolving the interior indeed allows for gravitational waves moving back and forth between the axis $ \tilde r = 0 $ and the brane, where they are partially transmitted to the exterior. As a consequence, the $ r $-profile of $ \alpha $ is not as smooth as before (cf.\ Fig.~\ref{fig:alpha_cc_degrav}), but is slightly distorted by those waves. But apart from this, the two solutions are practically identical.

This becomes even more obvious when comparing the time evolution of Hubble, shown in Fig.~\ref{fig:hubble_cc_degrav_dynReg}. In the dynamical regularization (solid line), the gravitational waves in the interior region produce small oscillations (with frequency $ \sim 1/R $) in the on-brane evolution. However, this is exactly the part which is sensitive to the initial conditions that are chosen in the bulk, so we would not trust them anyway. But now we see that the dashed line (static regularization) perfectly follows the mean of this oscillatory behavior. (Note that the same holds true for all other observables like $ \hat\rho $ or $ P_\phi $.) This confirms that the static regularization is indeed an efficient way to get rid of the dependency on the interior geometry, but in such a way that the long-time evolution (on time scales $ \sim 1/H_i $) is not affected. Furthermore, it shows that the predicted Hubble evolution on the time-scales of interest $ \Delta t \sim 1/H_i \gg R $ is completely insensitive to what is going on inside the cylinder, and is in that sense regularization independent.

We also checked that this remarkable agreement between the two regularizations is not altered when considering dust $(w=0)$ or radiation $(w=1/3)$.

\begin{figure*}[htb]
	\centering
	\subfloat[The radial profile of $ \alpha $ at different values of $ \tau $. The dots indicate the brane position, left of which the plotted function is the interior $ \tilde\alpha(\tilde r) $.]{
		\includegraphics[width=0.45\textwidth]{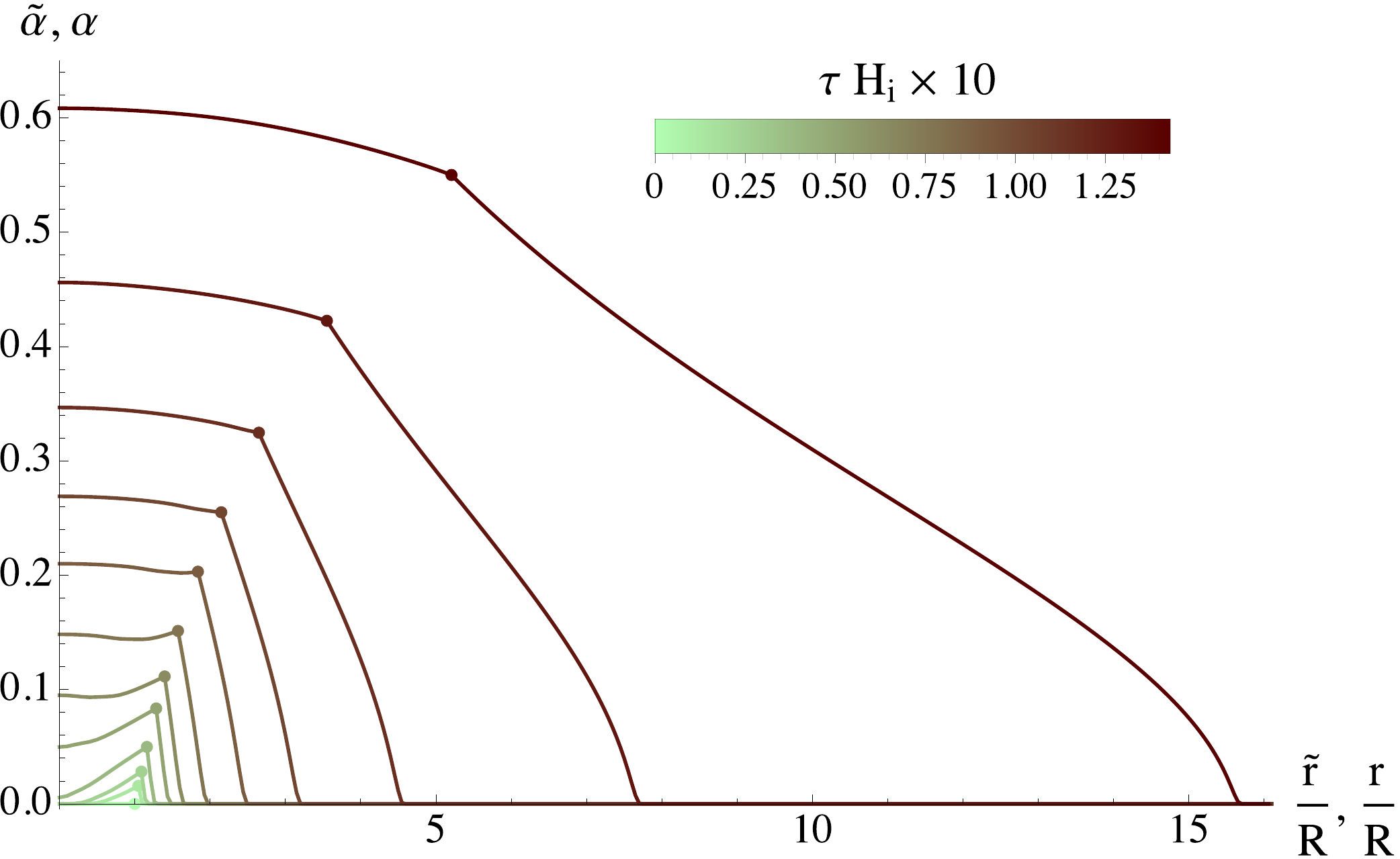}
		\label{fig:alpha_cc_pathol_dynReg}
	}
	\hfill
	\subfloat[The Hubble parameter shows qualitatively the same super-accelerating behavior in both regularizations.]{
		\includegraphics[width=0.45\textwidth]{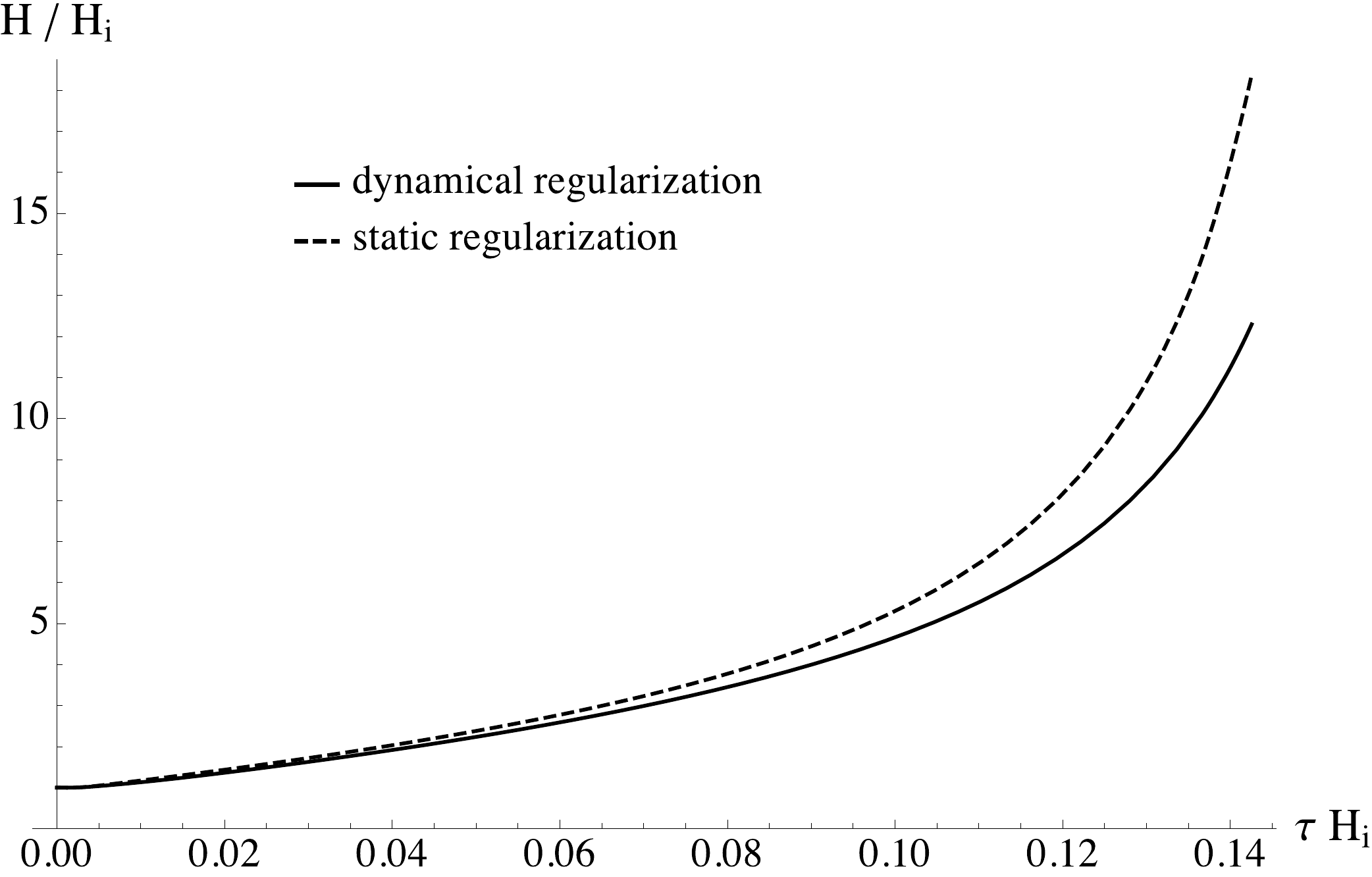}
		\label{fig:hubble_cc_pathol_dynReg}
	}%
	\caption{The super-accelerating solution of Fig.~\ref{fig:cc_pathol}, but in the dynamical regularization.}\label{fig:cc_pathol_dynReg}
\end{figure*}

\begin{figure*}[htb]
	\centering
	\subfloat[The degravitating solution.]{
		\includegraphics[width=0.45\textwidth]{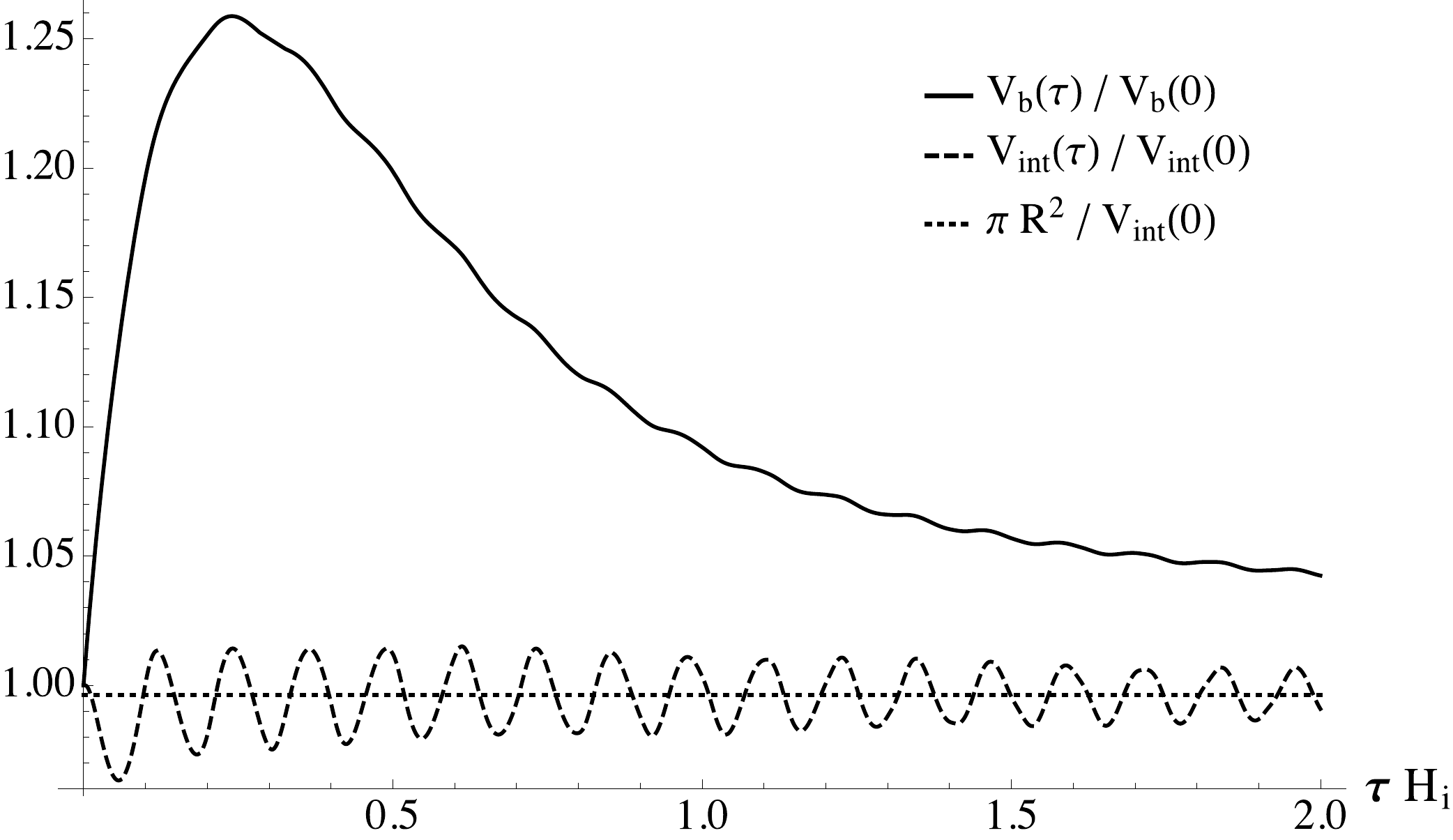}
		\label{fig:vol_degrav}
	}
	\hfill
	\subfloat[The super-accelerating solution.]{
		\includegraphics[width=0.45\textwidth]{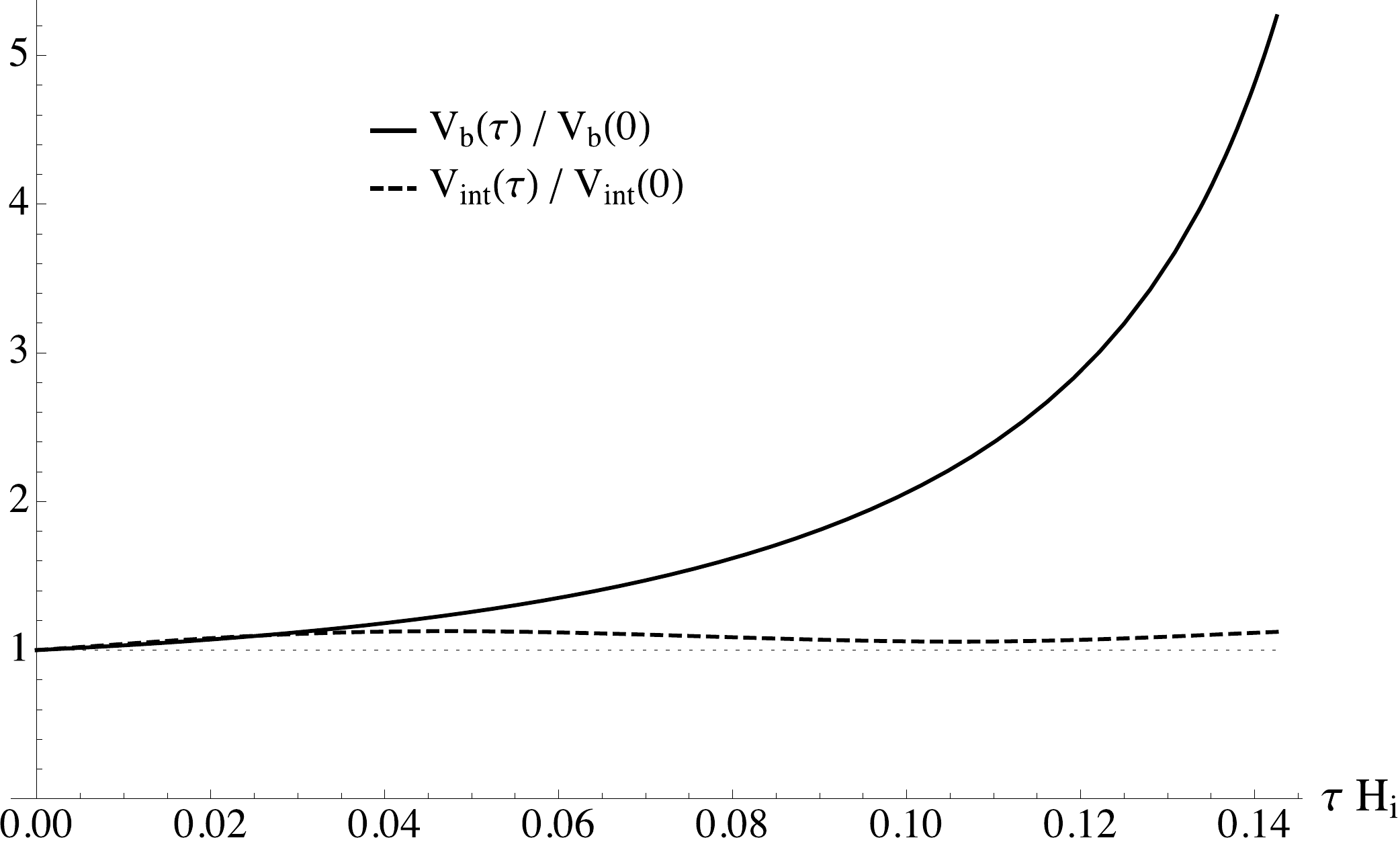}
		\label{fig:vol_pathol}
	}
	\caption{The 2D volume of the interior of the cylinder $ V_{\mathrm{int}} $ \eqref{eq:v_int} is approximately constant as compared to the 3D brane volume $ V_{\rm b} \propto \re^{3\alpha_0} $, confirming a successful stabilization in the dynamical regularization.}\label{fig:vol_dynReg}
\end{figure*}

\subsubsection{Super-accelerating solution}
\label{sec:patho_sol_dynReg}

Next, let us investigate the super-accelerating solution presented in Sec.~\ref{sec:pathol_sol} in the dynamical regularization. That is, we still keep $ R $ fixed and use the parameters~\eqref{eq:param_pathol} (and grid spacings  $ \epsilon = 5 \times 10^{-4} R  $, $ \tilde \epsilon = 10^{-3} R  $).

The results are shown in Fig.~\ref{fig:cc_pathol_dynReg}. The evolution of $ \alpha $ is qualitatively the same as in the static regularization, cf.\ Fig.~\ref{fig:alpha_cc_pathol}. Moreover, there are no visible small oscillations as in the degravitating solution, because the dynamics is completely dominated by the overall super-acceleration. In the Hubble plot, Fig.~\ref{fig:hubble_cc_pathol_dynReg}, the dashed line again corresponds to the static regularization. While there is no perfect agreement in this case, the qualitative behavior is not altered. Moreover, we checked that the curves approach each other as the regularization size $ R $ is decreased. Note that the value $ R = 0.05 H_i^{-1} $ is still vastly larger than a phenomenologically realistic value, \textit{e.g.}, $ R = 10^{-36} H_i^{-1} $ for $ R \sim 10^{-3} \eV $ and $ H_i \sim H_{\mathrm{today}} \sim 10^{-33} \eV $. The faster growth in the static regularization is due to the fact that in this case the parameters \eqref{eq:param_pathol} are closer to the stability bound, cf.\ Fig.~\ref{fig:contour_plots_analytic}.

As with Hubble, the effective equation of state for $ P_\phi $ is qualitatively the same as in the static regularization. It becomes smaller than $ -1 $ and tends to $ -\infty $. As already discussed in the main text, this unphysical source might in principle happen to be the reason for the super-acceleration. To exclude this possibility, we accordingly present the solution for the case $ P_\phi = 0 $ in Sec.~\ref{sec:vanishingPPhi}.

\begin{figure*}[htb]
	\subfloat[The two Hubble parameters as functions of $ \tau $, showing a super-accelerated behavior.]{
		\includegraphics[width=0.45\textwidth]{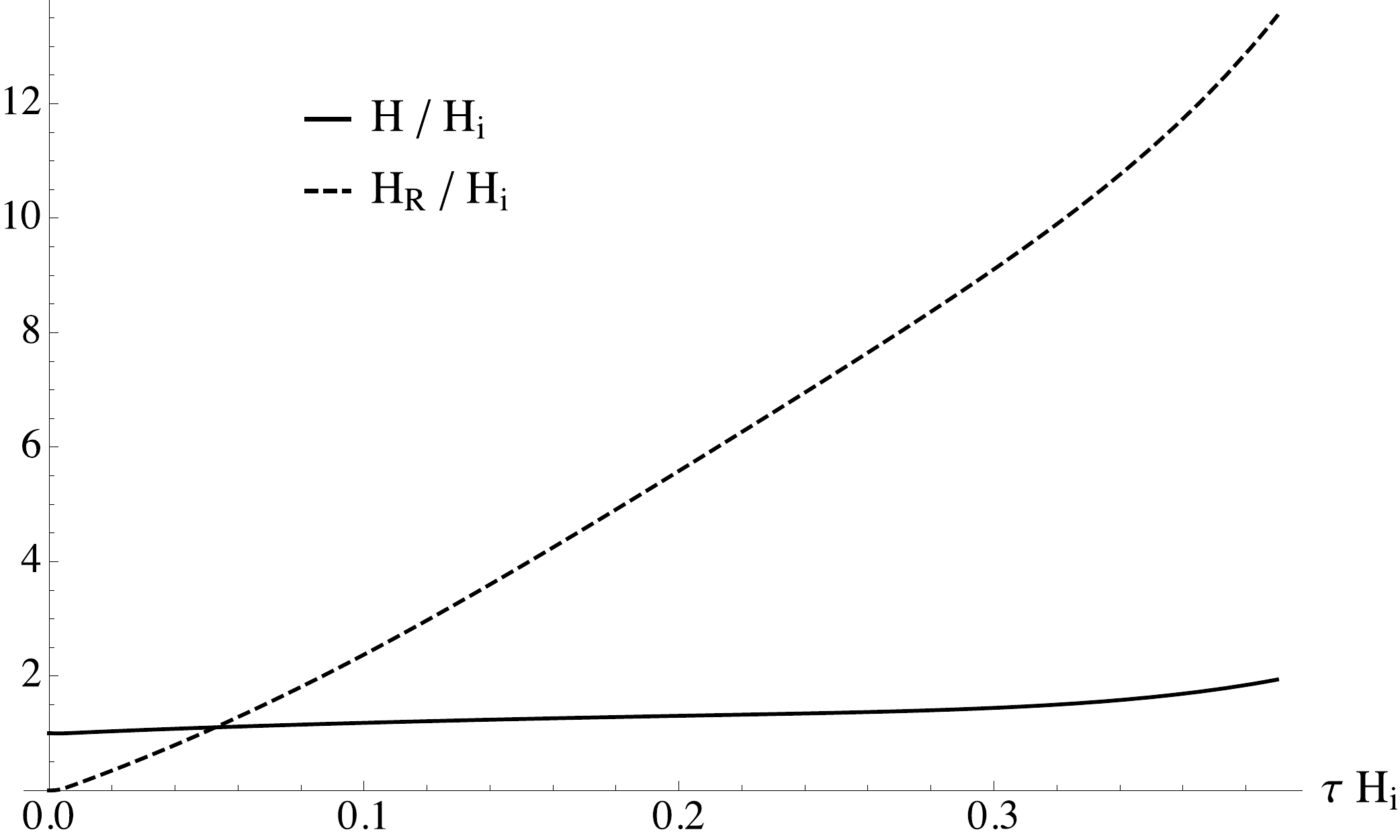}
		\label{fig:hubble_cc_pathol_pPhi_0}
	}%
	\hfill
	\subfloat[The effective energy density becomes negative and tends to $ -\infty $.]{
		\includegraphics[width=0.45\textwidth]{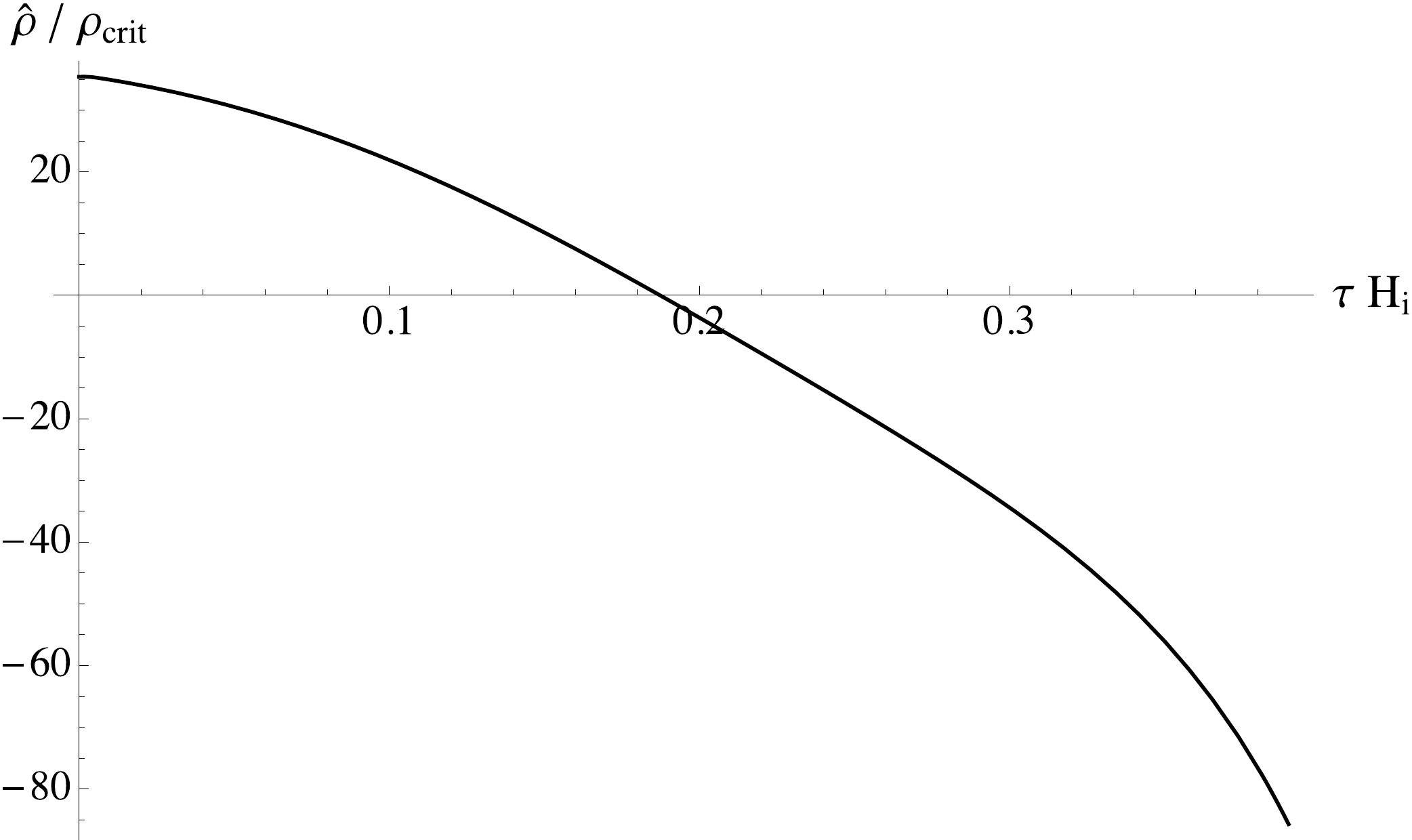}
		\label{fig:rho_hat_cc_pathol_pPhi_0}
	}
	\caption{Plots of the numerical results for the super-accelerated solution in the case $ P_\phi = 0 $. The instable behavior encountered in the case $ H_R = 0 $ is not cured by setting $ P_\phi = 0 $.}\label{fig:cc_pathol_pPhi_0}
\end{figure*}

\subsubsection{Volume stabilization}
\label{sec:vol_stabil}

In the dynamical regularization we can also address the question whether the extra space volume inside the cylinder is approximately constant and, in particular, vanishes for $R \rightarrow 0$, as is required by a consistent regularization. \textit{A priori}, it is not clear whether this condition is fulfilled, since we only fixed the circumference $R$. From \eqref{eq:met_int} we derive for the volume in the interior
\begin{equation}
\label{eq:v_int}
	V_{\rm int}(\tau)=2\pi\int_0^{r_0(\tau)} {\rm d}r\; r\, \re^{\tilde \eta-6\tilde \alpha }\;,
\end{equation} 
which can be integrated numerically, and is depicted by the dashed lines in Fig.~\ref{fig:vol_dynReg}. We find that the interior volume oscillates with a frequency of order $R^{-1}$. The oscillations are again due to small wave excitations in the interior of the cylinder and are thus an artifact of the initial conditions. The closer we approach the attractor solution in the degravitating case, the more they are washed out, and $ V_{\rm int} $ approaches the flat space value $V_{\rm flat} = \pi R^2 $ (dotted line) which lies slightly below the initial volume $V_{\rm int}(0)$.

The solid curves describe the evolution of a certain initial 3D volume $ V_{\rm b} \propto \re^{3\alpha_0} $ intrinsic to the brane. Evidently, the interior volume can be regarded as approximately constant as compared to the brane volume, in both the degravitating and super-accelerating solutions.
We consequently conclude that by fixing the circumference, the volume of the cylinder is sufficiently stabilized in the dynamical regularization.
Furthermore, this volume vanishes for $R \rightarrow 0$ as demanded by a consistent regularization.

Regarding the super-accelerating solution, note that in particular the volume inside the cylinder does not collapse, which could have been a potential source of energy for the super-acceleration in the brane-direction. Instead, the energy for this expansion is provided by the brane induced gravity terms, as already discussed. This conclusion can also be drawn from the fact that we find the same Hubble evolution for the static regularization, where the system is not influenced by an interior geometry. In summary, the interior of the cylinder has to be discarded as a potential source for the instability.

\subsubsection{Vanishing azimuthal pressure}
\label{sec:vanishingPPhi}

We now set $ P_\phi = 0 $, and choose the same parameters as for the super-accelerating solution before (but with grid-spacings $ \tilde\epsilon = 10^{-3} \times R_i $ and $ \epsilon  = 2 \times 10^{-3} R_i $). The results are shown in Fig.~\ref{fig:cc_pathol_pPhi_0}. The estimated numerical error-bars are again smaller than the line widths.

The two Hubble parameters $ H $ and $ H_R $ are plotted in Fig.~\ref{fig:hubble_cc_pathol_pPhi_0}. They both increase, implying a super-accelerated expansion. 
Fig.~\ref{fig:rho_hat_cc_pathol_pPhi_0} shows the effective energy density from a 6D perspective, which for $ H_R \neq 0 $ is given by
\begin{equation}
\label{eq:rhoHat_gen}
	\hat\rho \equiv \rho - 3M_{\rm Pl}^2 \left ( H^2 + H H _R \right ) \, .
\end{equation}
Again, it becomes negative and tends towards $ -\infty $.
This shows that the instability is not due to the unphysical pressure $ P_\phi $ encountered in the $ H_R = 0 $ scenario. On the contrary, the unphysical behavior of $ P_\phi $ is a consequence of the instability and the requirement of stabilizing the brane width $ R $. This can also be understood from Fig.~\ref{fig:hubble_cc_pathol_pPhi_0} which shows that without any stabilization the acceleration is dominantly in $ \phi $-direction.

\subsubsection{Contour plot}
\label{sec:contourPlot_dynReg}

\begin{figure}
	\includegraphics[width=0.45\textwidth]{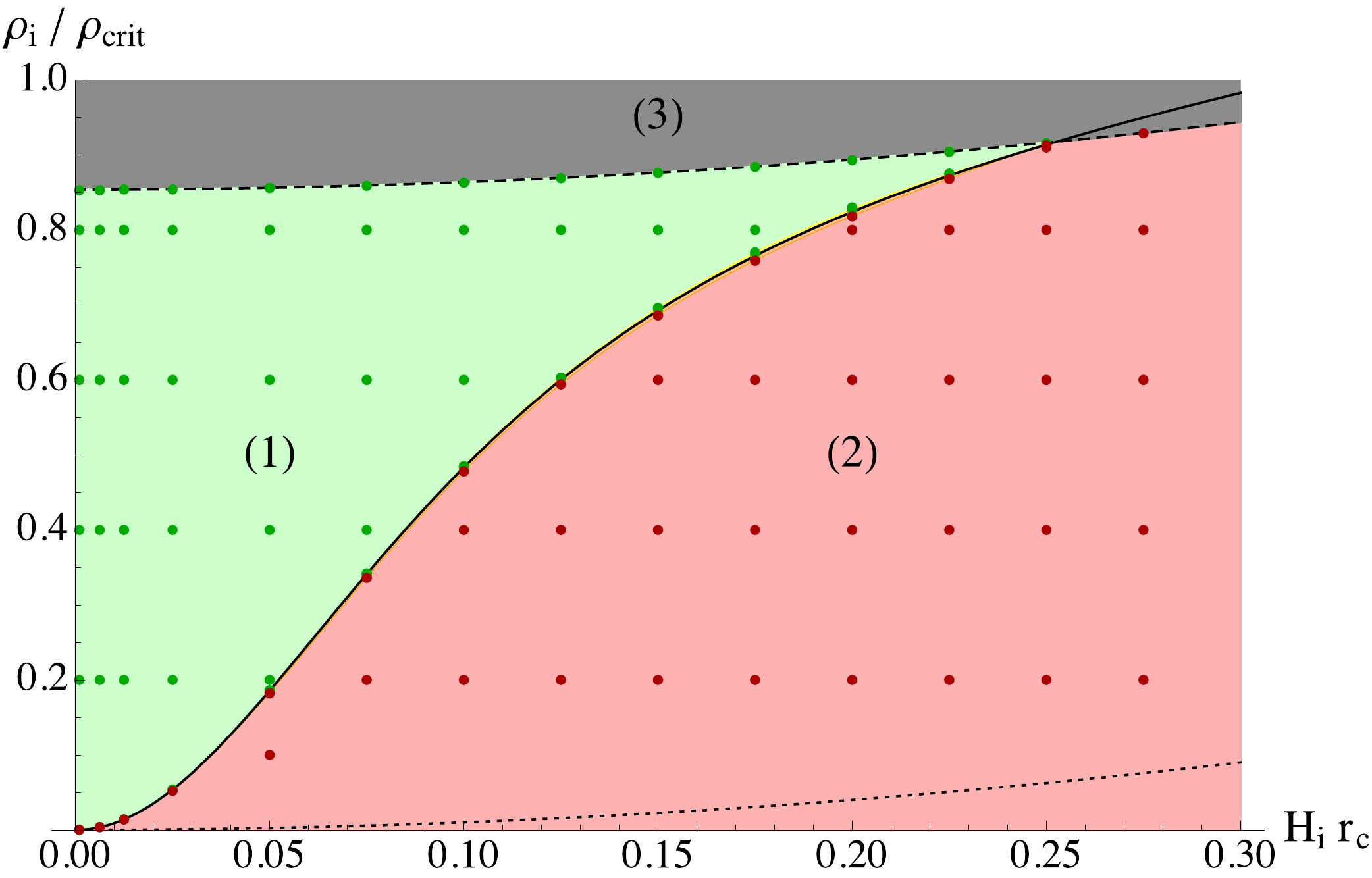}
	\caption{The same contour plot as in Fig.~\ref{fig:contour_plot}, but for the dynamical regularization. The dashed line corresponds to the criticality bound \eqref{eq:criticalityBound_dynReg}, the solid line to the stability bound \eqref{eq:stabilityBound_dynReg}, and the dotted line to $ \hat\rho_i = 0 \Leftrightarrow \rho_i = 3M_{\rm Pl}^2H_i^2$, \textit{i.e.}, the standard 4D constraint.}
	\label{fig:contour_plot_dynReg}
\end{figure}

\begin{figure*}[htb]
	\subfloat[$ q=1 $]{
		\includegraphics[width=0.3\textwidth]{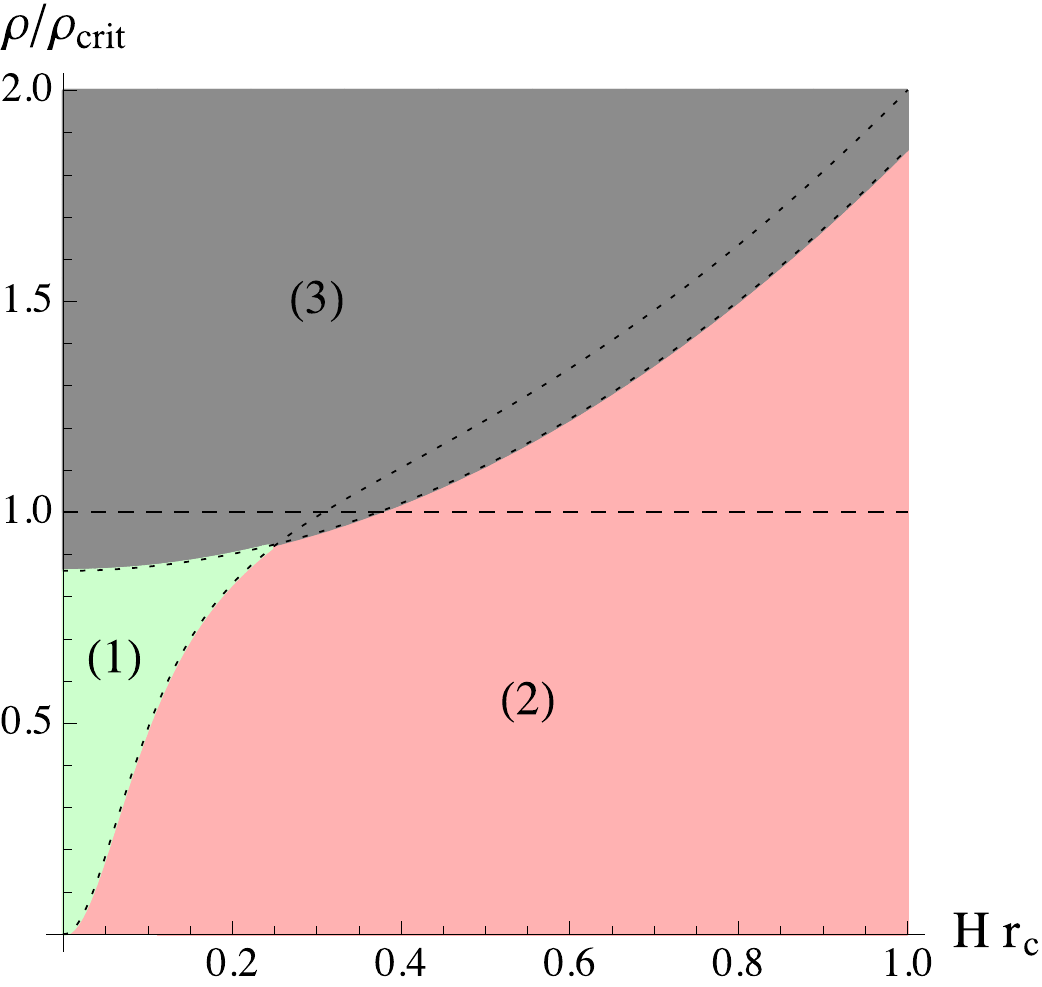}
	}%
	\hfill
	\subfloat[$ q=0.5 $]{
		\includegraphics[width=0.3\textwidth]{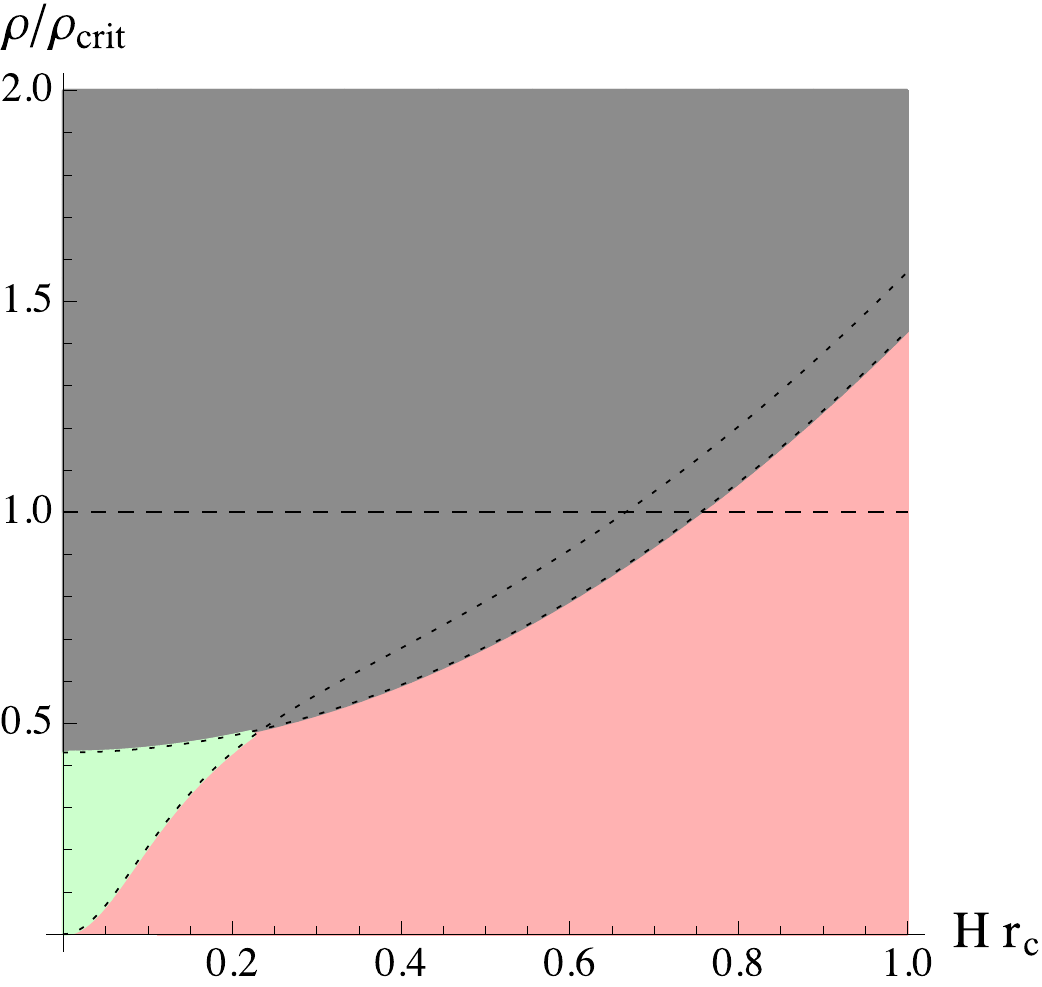}
	}%
	\hfill
	\subfloat[$ q=0 $]{
		\includegraphics[width=0.3\textwidth]{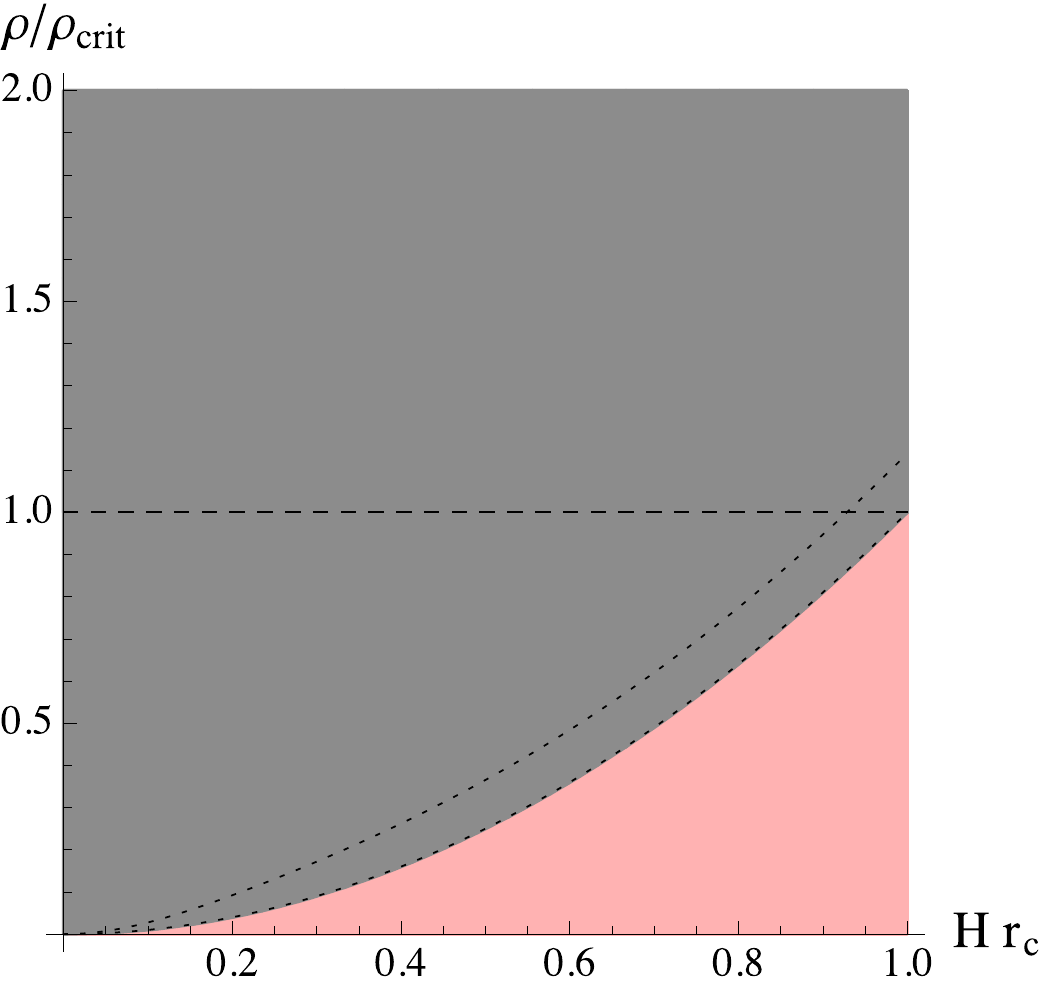}
	}%
	\caption{Contour plots in the dynamical regularization for different values of $ q $ as defined in \eqref{eq:q}, with $ H R=0.05 $.}
	\label{fig:contour_plots_analytic_q}
\end{figure*}

Fig.~\ref{fig:contour_plot_dynReg} shows the result of the classification of parameter space in the dynamical regularization with fixed brane width $ R = 0.05 H_i^{-1}$. The coloring (labeling) is the same as in Fig.~\ref{fig:contour_plot}.

Similarly to the static regularization, the degravitating and super-accelerating regimes are separated by the region in parameter space where the function $ \hat f $, introduced in equation \eqref{eq:def_fHat}, vanishes. It is negative in the degravitating regime, and positive in the super-accelerating regime.
After eliminating $ \gamma $ by using the constraint \eqref{eq:modFried1_dynReg}, the stability bound $ \hat f < 0 $ can be cast in the form
\begin{equation}
\label{eq:stabilityBound_dynReg}
	\frac{\rho}{\rho_{\mathrm{crit}}} > r_c^2 \left( H^2 + \frac{2\tilde\gamma^2}{9R^2 + 2r_c^2 \tilde\gamma} \right) \, .
\end{equation}
If it is violated, the model is unstable. The two regions are again separated by a physical singularity, so it is not possible to evolve dynamically from one region to the other. Equation \eqref{eq:stabilityBound_dynReg} also immediately shows that $ \hat\rho \equiv \rho - \rho_{\mathrm{crit}} r_c^2 H^2 $ is always positive when the bound is satisfied. On the other hand, the violation of this bound does not immediately imply $ \hat\rho < 0 $. (In Fig.~\ref{fig:contour_plot_dynReg} $ \hat\rho_i $ is zero along the dotted line and positive above.) But the numerical results show that whenever the bound is violated, $ \hat \rho $ dynamically becomes negative at some (later) time.

One disturbing fact with the criticality bound \eqref{eq:criticalityBound_dynReg} and the stability bound \eqref{eq:stabilityBound_dynReg} in the dynamical regularization is their dependence on the interior bulk geometry, through the appearance of $ \tilde\gamma $. However, $ \tilde\gamma $ can only take values in the interval
\begin{equation}
	3 \left| H \right| R < \tilde\gamma < \sqrt{1 + 9 H^2 R^2} \, .
\end{equation}
Let us, for convenience, introduce the parameter
\begin{equation}\label{eq:q}
	q := \frac{\tilde\gamma - 3 \left| H \right| R}{\sqrt{1 + 9 H^2 R^2} - 3 \left| H \right| R} \in (0, 1) \, .
\end{equation}
The maximum value $ q=1 $ corresponds to $ \tilde\eta_{0} = 0 $, which by inspection of equation \eqref{eq:eta0Tilde_init_a} is equivalent to $ \partial_{\tilde r}\tilde\alpha = \partial_{\tilde t}\tilde\alpha = 0 $, or in other words, to zero gravitational energy inside the cylinder. Initially, for $ H_i \neq 0 $, this can never be achieved exactly, because of the constraint \eqref{eq:alpha0TildeDot_init}. However, by making the profile function $ \tilde F $ sharply localized, it could be approached asymptotically. On the other hand, the minimum value $ q=0 $ would correspond to $ \tilde\eta_{0} = \infty $, \textit{i.e.}, an infinite amount of gravitational energy inside the cylinder. This is clearly not what we want, so we are mainly interested in values of $ q $ close to 1. In particular, for the flat initial conditions that we used for our numerics and $ H_i R = 0.05 $ (which was chosen in Fig.~\ref{fig:contour_plot_dynReg}) one finds $ q = 0.9915 $.

Fig.~\ref{fig:contour_plots_analytic_q} shows how the contour plots depend on this parameter $ q $: As $ q $ decreases, the green region (1) becomes smaller and is replaced by the gray region (3). This is due to the fact that as $ q \to 0 $, we are putting more and more energy into the gravitational field and so the configuration becomes super-critical for smaller values of $ \rho $. As just mentioned, this is not the situation we are interested in. Therefore, in Fig.\ \ref{fig:contour_plots_analytic}, comparing the contours in both regularizations, $ q $ was set equal to $ 1 $.

Finally, let us again stress two important observations, already discussed in Sec.~\ref{sec:phenomen}: (i) the contour plots in both regularizations agree in the limit $ R\to 0 $; (ii) the main result---that all degravitating solutions are ruled out by observations---is independent of which regularization scheme is used.
Indeed, the crucial bound \eqref{eq:maxCrossover} can also be derived from equations \eqref{eq:criticalityBound_dynReg} and \eqref{eq:stabilityBound_dynReg}.
\section{Numerical implementation}
\label{ap:numImpl}

\subsection{Algorithm}

Here we present the details of the numerical implementation that is used to solve the bulk brane system.
We focus on the dynamical regularization because it is technically slightly more complicated. The algorithm for the static regularization is simply obtained by discarding the whole evolution of the interior space-time, and using the appropriate junction conditions \eqref{eq:rhoJunctCond} and \eqref{eq:pJunctCondRConst} instead of~\eqref{eq:modFried_dynReg}. Moreover, we only discuss the case $ H_R = 0 $ explicitly. The only difference for the case $ P_\phi = 0 $ is that there is one more dynamical on-brane variable, viz.\ $ R $, which is treated in complete analogy to $ \alpha_0 $, by using the junction conditions \eqref{eq:modFried_pPhi_0}.

The goal is to solve the bulk equations \eqref{eq:einstein_vacuum} in the interior and exterior region, together with the junction conditions \eqref{eq:modFried_dynReg}. The initial data is chosen as described in Sec.~\ref{ap:num_impl} and the time evolution is calculated using the two dynamical equations \eqref{eq:2D_wave} and \eqref{eq:modFried2_dynReg}, whereas the constraint \eqref{eq:modFried1_dynReg} is only enforced at initial time and later used as a consistency check, as discussed in the next section.

For the bulk PDE we use a discretization with fixed equidistant spacing 
\begin{equation}
\label{eq:gridSpacing}
	\Delta \tilde t = \Delta \tilde r \equiv \tilde \epsilon \, , \qquad \Delta t = \Delta r \equiv \epsilon \, .
\end{equation}
Denoting $ \alpha_j^n := \alpha(t^n, r_j) $, the derivatives of $ \alpha $ are approximated by the following finite difference representations
\begin{subequations}
\begin{align}
	\partial_r^2\alpha(t, r) &\to \frac{\alpha_{j+1}^n - 2\alpha_j^n + \alpha_{j-1}^n}{\epsilon^2} \, ,\\
	\partial_r\alpha(t, r) &\to \frac{\alpha_{j+1}^n - \alpha_{j-1}^n}{2\epsilon} \, ,\\
	\partial_t^2\alpha(t, r) &\to \frac{\alpha_j^{n+1} - 2\alpha_j^n + \alpha_{j}^{n-1}}{\epsilon^2}\, .
\end{align}
\end{subequations}
The wave equation \eqref{eq:2D_wave} then allows us to explicitly calculate the next time-step $ \alpha_j^{n+1} $ from the past values $ \alpha_j^n $, $ \alpha_j^{n-1} $:
\begin{equation}
\label{eq:alphaPDENum}
	\alpha^{n+1}_j = - \alpha^{n-1}_j + \alpha^n_{j+1} + \alpha^n_{j- 1} + \left (\alpha^n_{j-1} - \alpha^n_{j+1} \right)  \frac{\epsilon}{2 r_j} \, ,
\end{equation}
and similarly for $ \tilde \alpha $.
The \textit{Courant condition} $ \Delta t / \Delta r \leq 1 $ is satisfied for our choice \eqref{eq:gridSpacing}, and so the scheme is numerically stable \cite{Press:1992}.

Equation \eqref{eq:alphaPDENum} can only be used inside of the spatial domain of integration, \textit{i.e.}, away from the boundaries $ \tilde r \in \{0, \tilde r_0\}, r \in \{ r_0, r_\mathrm{max} \} $. At the axis, $ \tilde\alpha $ is determined by the regularity condition \eqref{eq:cond_axis_alpha} which translates to:
\begin{equation}
	\tilde\alpha^n_0 = \tilde\alpha^n_1 \, .
\end{equation}
At the outer boundary $ r = r_\mathrm{max} $, we simply implement the fixed Dirichlet boundary condition
\begin{equation}
	\alpha^n_{J} = 0 \, .
\end{equation}
In fact, we would like to impose non-reflecting boundary conditions, so that all gravitational waves emitted by the brane leave the domain of integration without any reflections. However, as is well known \cite{Givoli:1991, Hofmann:2013zea}, in two spatial dimensions this condition is nonlocal (in time). Therefore, it is computationally quite expensive and so we simply choose the most primitive alternative of making the domain of integration large enough so that any wave that is reflected at $ r = r_\mathrm{max} $ cannot reach the brane by the end of the numerical simulation.

\begin{figure}
	\includegraphics[width=0.2\textwidth]{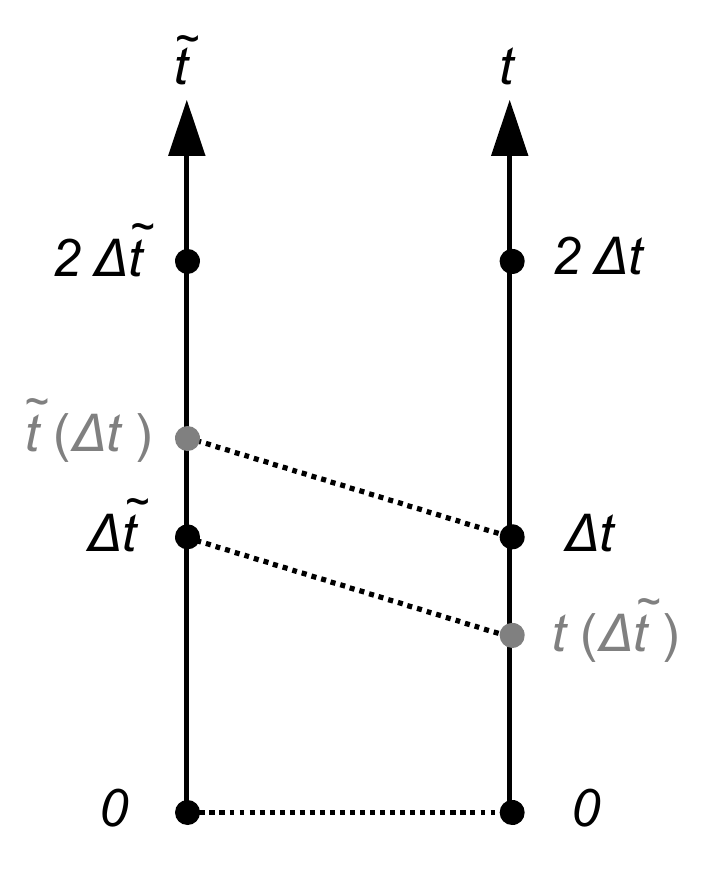}
	\caption{Using two different coordinate patches for the interior and exterior geometry implies that the temporal grid points do not correspond to the same physical time at the position of the brane. The values of $ \tilde\eta_0 $ and $ \eta_0 $ at the gray points, which are needed in \eqref{eq:modFried2_dynReg}, are found by linearly interpolating between the neighboring black points.}
	\label{fig:timeGrid}
\end{figure}

The value at the brane, $ \tilde\alpha_0 = \alpha_0 $ is determined by the dynamical junction condition \eqref{eq:modFried2_dynReg}. However, there is a slight complication because the time steps $ \Delta t $ and $ \Delta \tilde t $ do not correspond to the same physical time steps. (This complication is of course absent in the static regularization.) In fact, the discretized version of equation~\eqref{eq:timeRelation} is
\begin{equation}
	\frac{\Delta t}{\gamma} = \frac{\Delta\tilde t}{\tilde \gamma}\, ,
\end{equation}
and $ \gamma  \neq \tilde\gamma$ whenever there is a modification to the 4D evolution, cf.\ \eqref{eq:modFried1_dynReg}. Now suppose we are given all relevant initial data at the initial time $ \tilde t_i $, $ t_i $ (which we can assume to correspond to the same physical time, and set equal to zero, without loss of generality). Then we use \eqref{eq:modFried2_dynReg} to determine $ \tilde\alpha_0 $ and $ \alpha_0 $ at the next time step, \textit{i.e.}, $ \tilde\alpha_0(\Delta\tilde t) $ and  $ \alpha_0(\Delta t) $. Those we use as the appropriate boundary conditions to solve \eqref{eq:alphaPDENum}, which in turn allows to calculate $ \tilde\eta_0(\Delta\tilde t) $ and  $ \eta_0(\Delta t) $ with the discretized version of
\begin{subequations}
\begin{align}
	\frac{\rd \eta_0}{\rd t} & = \partial_t \eta_0 + \frac{\rd r_0}{\rd t} \partial_r \eta_0 \\
	& = 6r_0 \left[ 2 \left(\partial_t \alpha_0 \right) \left( \partial_r \alpha_0 \right)  +\left (\partial_t \alpha_0 \right)^2 + \left(\partial_r \alpha_0 \right)^2 \right] \, , \nonumber
\end{align}
\end{subequations}
(and similarly for $ \tilde\eta_0 $) where we used \eqref{eq:etaPrime_vac} and \eqref{eq:etaDot_vac} in the limit $ r \to r_0^+ $ (or $ \tilde r \to \tilde r_0^- $).
We now want to iterate this process, but to use \eqref{eq:modFried2_dynReg} again we need $ \tilde\eta_0 $ and $ \eta_0 $ at the same physical time (\textit{i.e.}, both at $ \Delta\tilde t $, or both at $ \Delta t $). Assume that for instance $ \Delta t $ is ``ahead in time'', \textit{i.e.}, $ \tilde t(\Delta t) > \Delta\tilde t $, cf.\ Fig.~\ref{fig:timeGrid}. We then estimate $ \eta_0(t(\Delta \tilde t)) $ by linearly interpolating between $ \eta_0(0) $ and $ \eta_0(\Delta t) $. With this we can repeat the procedure to obtain $ \tilde\eta_0(2\Delta\tilde t) $, from which we get $ \tilde\eta_0(\tilde t(\Delta t)) $---again by linear interpolation. Then we can calculate $ \eta_0(2\Delta t) $ and continue the iteration.

\begin{figure}[h]
	\includegraphics[width=0.25\textwidth]{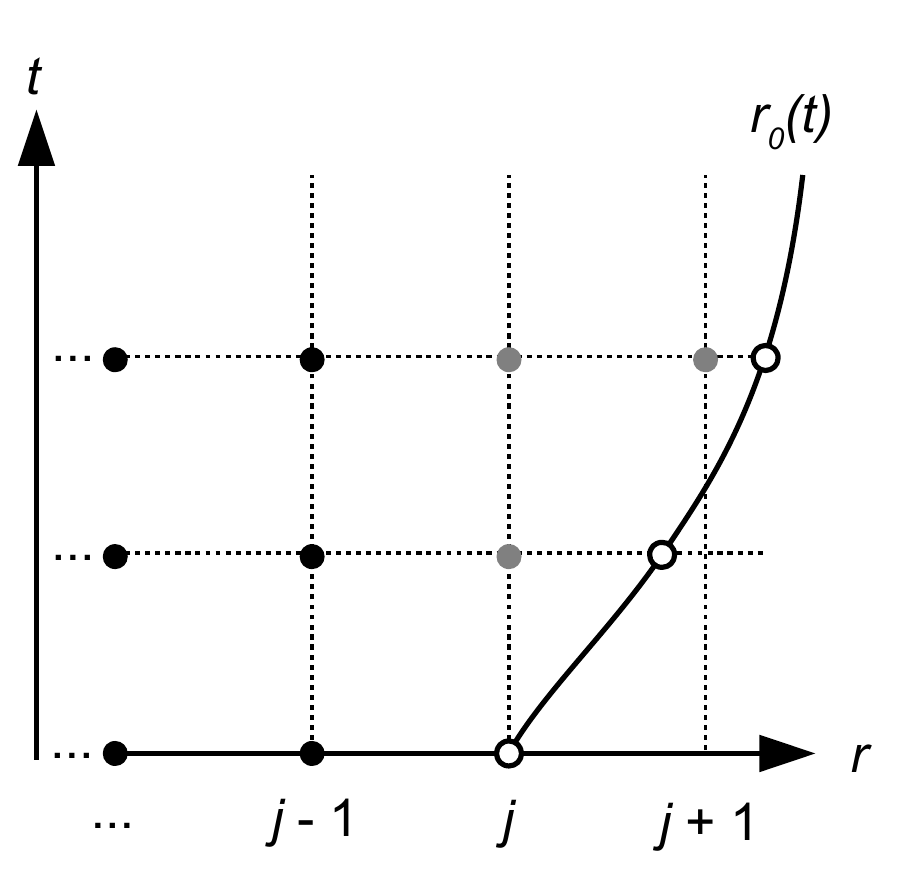}
	\caption{Sketch of the space-time grid. The white points indicate the brane position, which in general does not lie on a grid point, but on which the boundary data for $ \alpha $ is given. Black points are calculated using the wave equation \eqref{eq:alphaPDENum}. For the gray points, this is not possible because of the lack of initial data, so they are obtained by linearly interpolating between the neighboring black and white points.}
	\label{fig:r0Grid}
\end{figure}

A second complication stems from the fact that even though the physical brane circumference $ R $ is kept fixed, the brane's coordinate position $ \tilde r_0 = r_0 $ will be time-dependent for any non-trivial evolution of $ \alpha $, because $ R = r_0 \re^{-3\alpha_0} =\tilde r_0 \re^{-3\tilde\alpha_0} $. But since we use a fixed spatial grid, and the brane moves with a speed less than $ 1 $, this implies that the brane position will lie in between two grid points most of the time. We again solve this problem by linear interpolation: Suppose the brane (say, in the exterior coordinate patch) is initially located at some grid point $ j $, cf.\ Fig.~\ref{fig:r0Grid}. Equation \eqref{eq:modFried2_dynReg} (with the appropriate initial data) gives the new value of $ \alpha_0 $ at the new brane position (which is also determined by $ \alpha_0 $). Assume that the brane moved to the right, as in Fig.~\ref{fig:r0Grid}. Then the new value of $ \alpha $ at $ j $ cannot be obtained using~\eqref{eq:alphaPDENum}, because it would require initial data at the point $ j+1 $, which lies outside the domain of integration. In those cases, we estimate the new value of $ \alpha_j $ by linearly interpolating between the brane value and the new value at the point $ j-1 $ (which \textit{can} be obtained from \eqref{eq:alphaPDENum}). If the brane crosses one spatial grid point, then there are two values of $ \alpha $ which cannot be calculated from \eqref{eq:alphaPDENum}, in which case we determine both of them by linear interpolation.

Finally, we checked that the numerically results are practically unchanged if instead of linear interpolations we use quadratic interpolations everywhere. This shows that the numerical errors are mainly not due to the interpolation, but to the discretization. But those errors are very well under control, as we will discuss next.

\subsection{Error estimates and consistency checks}
\label{ap:numErrors}

\begin{figure*}[hbt]
	\includegraphics[width=0.45\textwidth]{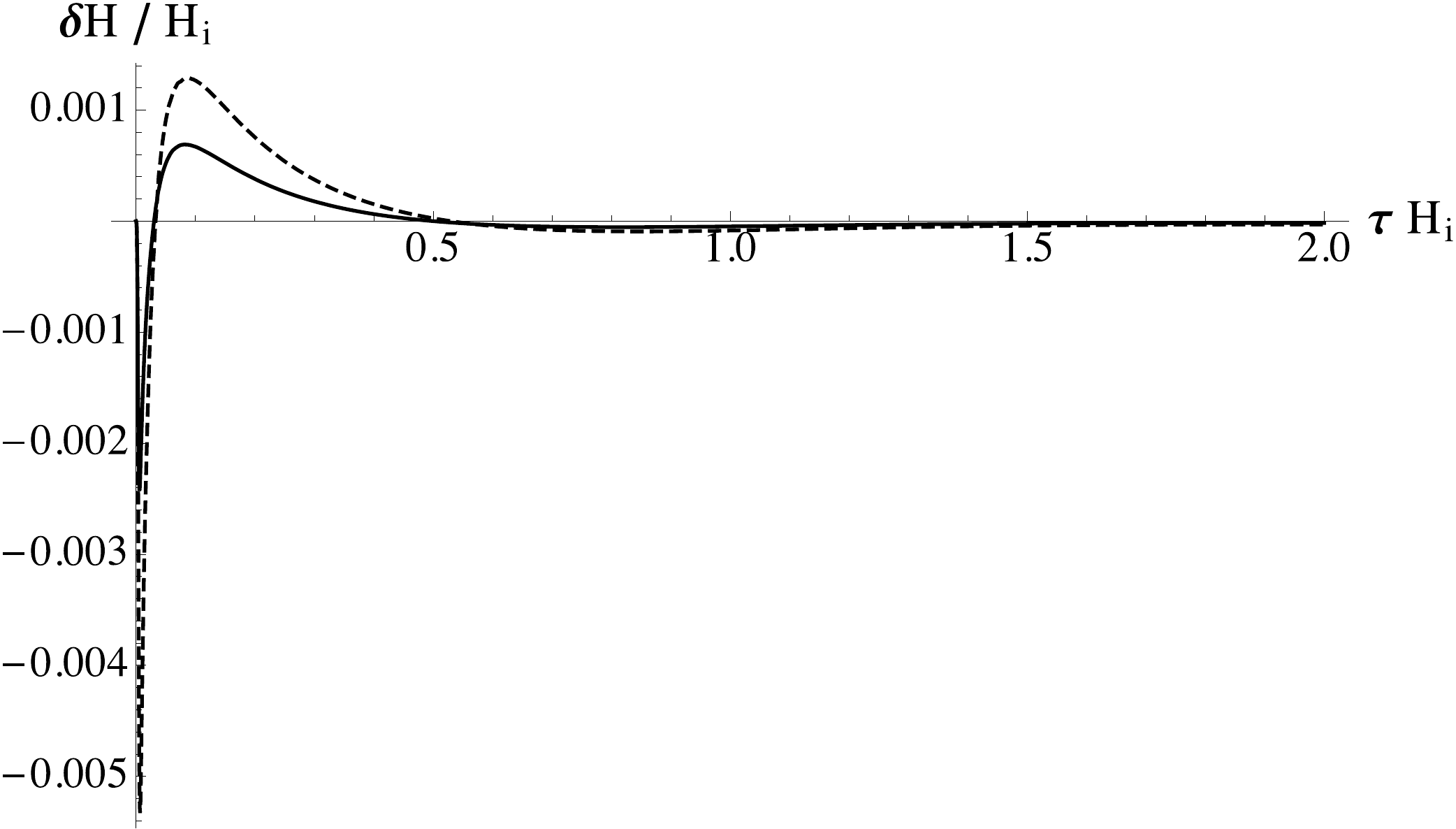}
	\hfill
	\includegraphics[width=0.45\textwidth]{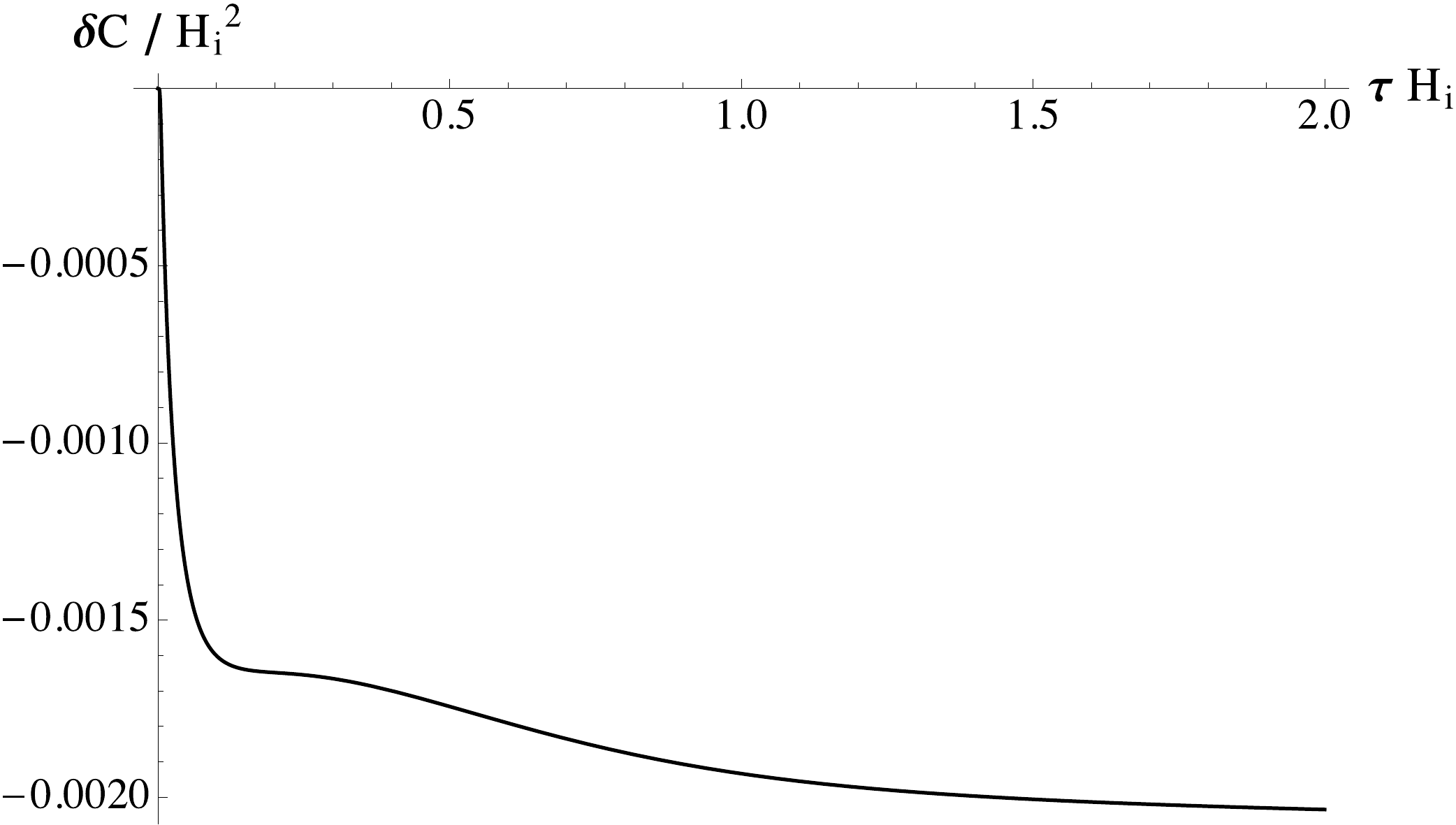}\\
	\vskip 0.5cm
	\includegraphics[width=0.45\textwidth]{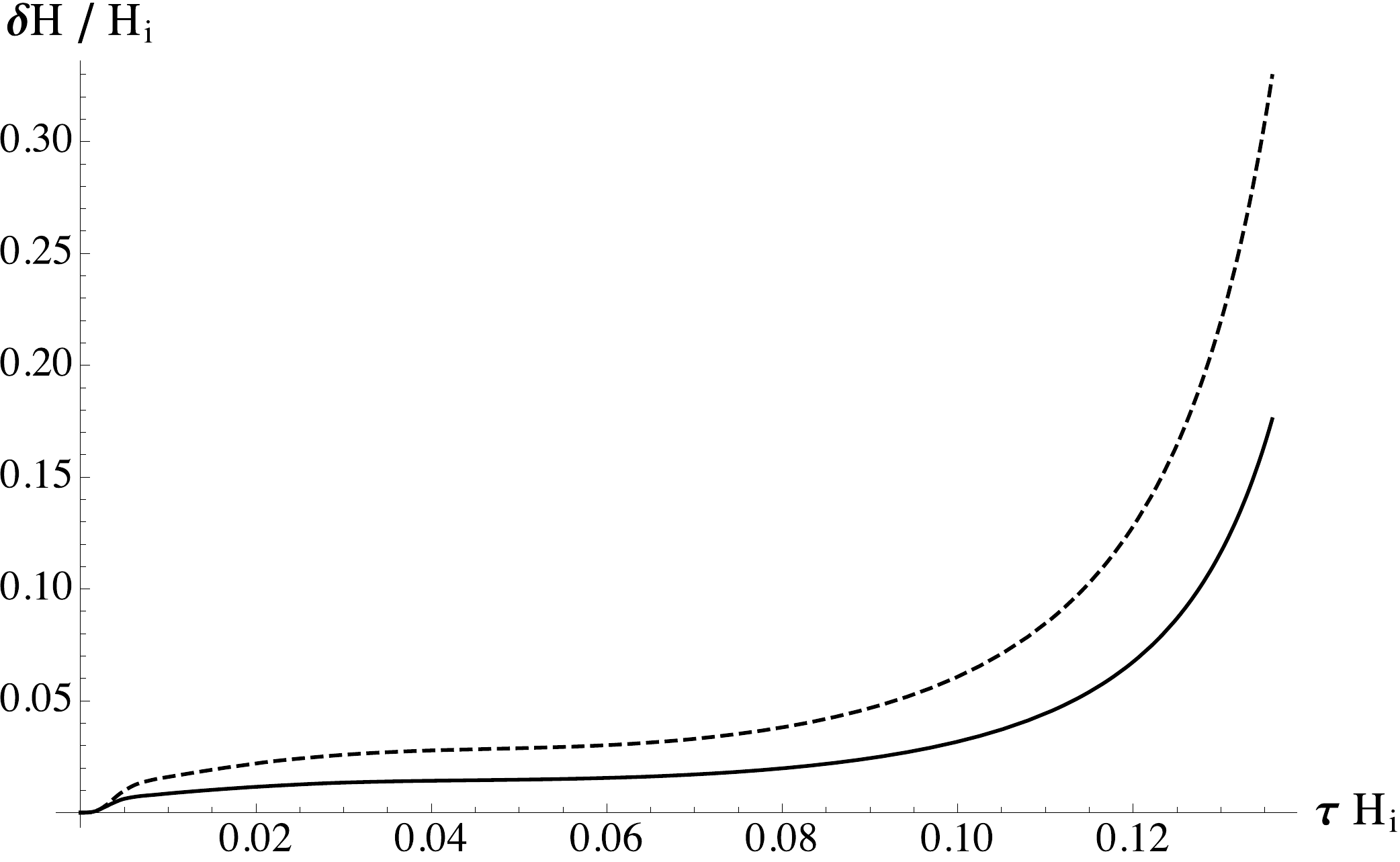}
	\hfill
	\includegraphics[width=0.45\textwidth]{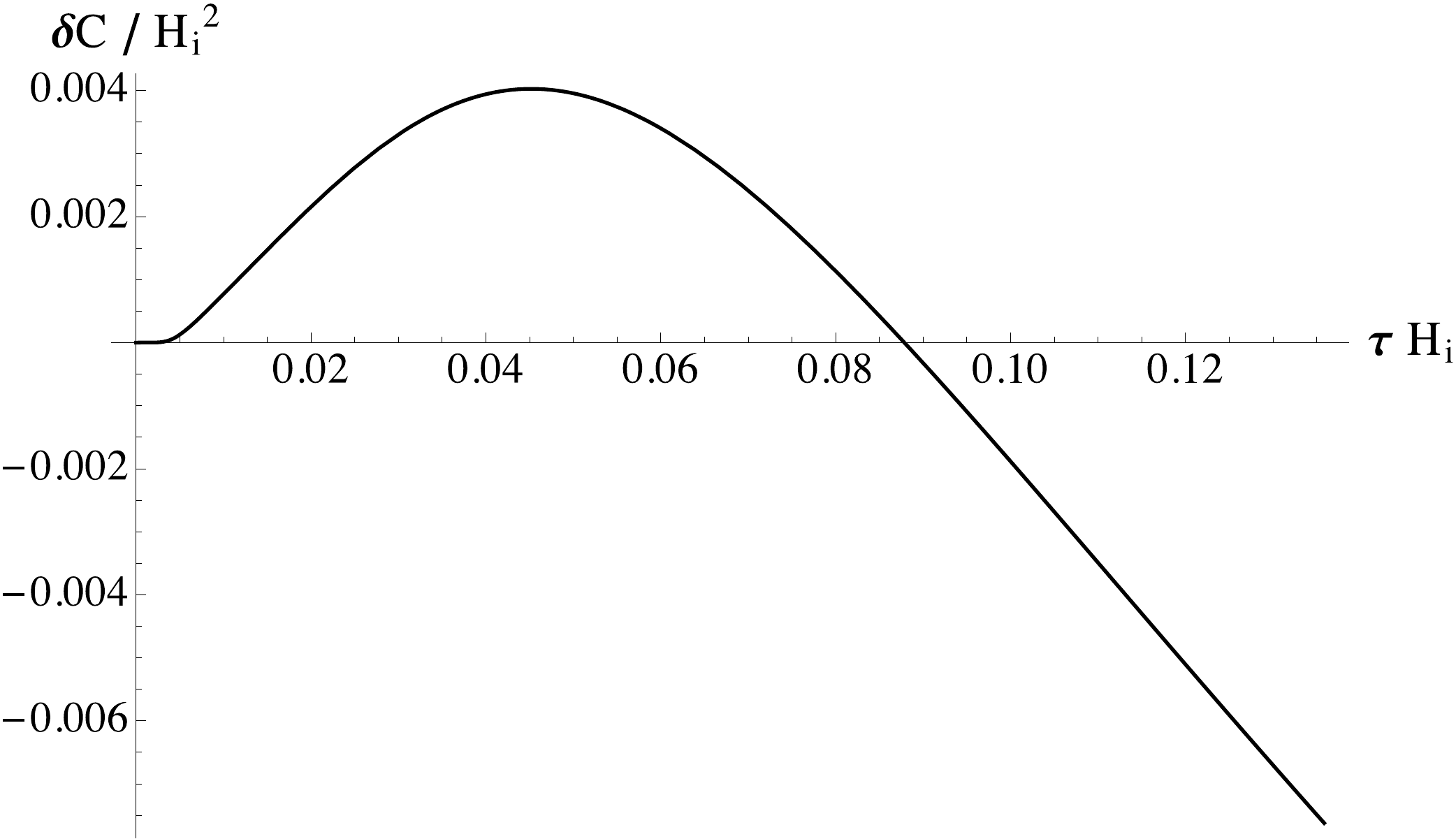}
	\caption{Error estimates of Hubble (left column) and consistency checks (right column) for the degravitating solution presented in Sec.~\ref{sec:degrav_sol} (upper row) and the super-accelerating solution of Sec.~\ref{sec:pathol_sol} (lower row), as explained in the text.}
	\label{fig:numErrors}
\end{figure*}

One way to estimate the numerical uncertainties is to check how much the calculated quantities change when the grid-spacing is made smaller. For instance, one can define an error estimate $ \delta A $ for some quantity $ A $ calculated with grid-spacing $ \epsilon $ as $ \delta A(\epsilon) \equiv A(2\epsilon) - A(\epsilon) $. If $ A(\epsilon) $ converged to its true value linearly in $ \epsilon $ as $ \epsilon \to 0 $, this would give exactly the correct error, for a faster convergence the true error would even be smaller. The plots on the left hand side of Fig.~\ref{fig:numErrors} show the corresponding error of the Hubble parameter for the degravitating solution presented in Sec.~\ref{sec:degrav_sol} (top), and for the super-accelerating solution of Sec.~\ref{sec:pathol_sol} (bottom). In the Hubble plots (Figs.~\ref{fig:hubble_cc_degrav} and \ref{fig:hubble_cc_pathol}) the corresponding error-bars would not exceed the line thickness. The dashed curves depict the error estimates when the grid spacing is doubled; the scaling of the errors is compatible with an (approximately linear) convergence as $ \epsilon \to 0 $.

There are several non-trivial consistency checks that one can perform. The most important one is the constraint equation \eqref{eq:rhoJunctCond}, which is only imposed at the initial time, and should be automatically fulfilled at all later times. Its violation $ \delta C \equiv H^2 - \rho / (3 M_\mathrm{Pl}^2) - (\gamma - 1) / r_c^2 $, measured in units of $ H_i^2 $, is plotted on the right hand side of Fig.~\ref{fig:numErrors}; it is indeed compatible with being zero within the numerical uncertainties. 

We do not explicitly show the corresponding error plots for the solutions in the dynamical regularization presented in Appendix \ref{ap:dynReg}, but we checked that they are all equally well under control. In that case, there is also another non-trivial consistency check, coming from the fact that some quantities (like Hubble) which should be continuous across the brane can be calculated independently from the interior and exterior in our numerical scheme. The difference between them was again found to be compatible with being zero.
\section{Einstein-Rosen coordinates}
\label{ERcoords}

With the assumed symmetries, the bulk metric can be written as 
\begin{equation}
\label{eq:met_cyl_symm}
	\rd s^2_6 = \re^{2(\eta - 3\alpha)} \left( -\rd t^2 \!+ \rd r^2 \right) + \re^{2\alpha} \rd \vec{x}^2 + \re^{-6\alpha} W^2 \rd\phi^2 \; ,
\end{equation}
where $ \eta, \alpha $ and $ W $ are functions of $ (t, r) $.  This is the 6D generalization of a metric describing ``whole-cylinder symmetry'' as discussed in~\cite{Thorne:1965} or \citep[chap. 22]{Stephani} in the case of 4D. This form does not completely fix the $(t,r)$-coordinates since it is still invariant under a transformation  $(t,r) \mapsto (t^*,r^*)$, subject to the condition 
\begin{align} \label{eq:cond_trafo}
	\begin{pmatrix} \partial_t r^* \\ \partial_r r^* \end{pmatrix}
	=\pm
	\begin{pmatrix} \partial_r t^* \\\partial_t t^*\end{pmatrix}\;.
\end{align}
This implies an integrability condition for $r^*$:
\begin{equation}
\label{eq:cond_trafo_2}
\partial_r^2 r^* =  \partial_t^2 r^*\;.
\end{equation}
Away from the brane, the bulk Einstein's equations imply
\begin{equation}
\label{eq:1D_wave}
\partial_r^2 W  = \partial_t^2 W\,,
\end{equation}
where we have used $T^{(6)}_{AB}=0$. This naturally suggests fixing the remaining gauge freedom through
\begin{equation}
\label{eq:ER_gauge}
r^*=W(t,r)\;.
\end{equation}
Dropping asterisks, the metric ansatz becomes~\eqref{eq:met_cyl_symm_2}.

Note that the interpretation of $ r^* $ as a spatial coordinate implicitly assumes that the gradient of $ W $ in the original coordinates \eqref{eq:met_cyl_symm} is space-like. Furthermore, in deriving the junction conditions in Sec.~\ref{sec:junction_cond} we assumed that $ r^* $ gets larger as one moves away form the brane, which in turn requires the gradient of $ W $ to be outward pointing.
One of these assumptions, however, is not true if the energy density localized on the brane is super-critical, as we will now show.
The following discussion partly follows the one in the appendix of \cite{PhysRevD.46.2435}.

We begin with the general cylindrically symmetric ansatz \eqref{eq:met_cyl_symm} for the exterior line element $\rd s_6^2$ which depends on the function $W(t,r)$. For the interior we require the Einstein-Rosen form, \textit{i.e.}, we make the ansatz \eqref{eq:met_int} for $\rd \tilde s_6^2$. (For the case of the static regularization you simply have to set $\tilde \gamma$=1 in the following discussion.)

By introducing the metric function $W(t,r)$, the junction condition \eqref{eq:rhoJunctCond_dynReg} gets generalized to 
\begin{equation}
	\label{eq:match1_gen}
	 \tilde \gamma - \re^{-3\alpha_0}  n^A \partial_A W|_0 = \frac{\hat \rho}{\rho_{\rm crit}} \;,
 \end{equation}
where the normal vector $n^A$ and $\hat \rho$ are defined in \eqref{eq:normal_vec_int} and \eqref{eq:rhoHat_gen}, respectively.   Moreover, a similar equation can be derived from the continuity condition $W_0=\tilde r_0$. By differentiating it with respect to $\tau$ and using  \eqref{eq:timeRelation} we find:
\begin{equation}
	\label{eq:match2_gen}
	\tilde \gamma\,\frac{\rd{\tilde r}_0}{\rd \tilde t} - \re^{-3\alpha_0} t^A \partial_A W|_0 =0 \;.
\end{equation}
Here, $t^A=\gamma \re^{3\alpha_0}(1,\rd{r}_0/ \rd t,0,0,0,0)$ denotes the unit tangent vector on the brane for which $n^At_A=0$. 

According to \eqref{eq:1D_wave}, $W$ obeys a 1D wave equation in the bulk. The general solution can be written as
\begin{equation}
	W(t,r)=W_{+}(t+r)+W_{-}(t-r)\;.
\end{equation}
Plugging this back into \eqref{eq:match1_gen} and \eqref{eq:match2_gen}, we find for $W^{\prime}_{+}$ and $W^{\prime}_{-}$ evaluated at the brane
\begin{subequations}
\label{eq: crit_bound}
\begin{align}
	W^{\prime}_{+}\big|_{0}&=\frac{1}{2\gamma (1+\frac{\rd{ r}_0}{\rd t})}\left[\tilde \gamma\, \left(1+\frac{\rd{\tilde r}_0}{\rd \tilde t}\right)-\frac{\hat \rho}{\rho_{\rm crit}} \right]\,,\\
	W^{\prime}_{-}\big|_{0}&=\frac{1}{2\gamma (1-\frac{\rd{ r}_0}{\rd t})}\left[\frac{\hat \rho}{\rho_{\rm crit}}-\tilde \gamma\, \left(1-\frac{\rd{\tilde r}_0}{\rd \tilde t}\right) \right]\,.
\end{align}
\end{subequations}
These two equations allow to characterize the gradient
\begin{equation}
	\partial_A W|_0=(w_{+}+w_{-},w_{+}-w_{-},0,0,0,0)\;,
\end{equation}
where $w_{+}\equiv W^{\prime}_{+}|_{0}$ and $w_{-}\equiv W^{\prime}_{-}|_{0}$ have been introduced. We distinguish three different cases:

(i)  $w_{+}>0$ and $w_{-}<0\,$: In this regime $\partial_A W|_0$ is space-like and outward pointing. Therefore, it is consistent to introduce a new radial coordinate $r^*=W(t,r)$ in order to implement the Einstein-Rosen form. The condition gets translated via \eqref{eq: crit_bound} into
\begin{equation}
	\frac{\hat \rho}{\rho_{\rm crit}}< \tilde \gamma -  |3H + H_R| R\;.
\end{equation}
As expected, this is precisely the criticality bound \eqref{eq:criticalityBound_dynReg} once we set $H_R=0$.

(ii)  $w_{+}w_{-}>0$: In this regime $\partial_A W|_0$ is time-like. Consequently $W(t,r)$ could play---at least locally at the position of the brane---the role of a new time coordinate but not of a spatial coordinate as assumed for our analysis. This happens for 
\begin{equation}
	\tilde \gamma -  |3H + H_R| R<\frac{\hat \rho}{\rho_{\rm crit}}< \tilde \gamma +  |3H + H_R| R\;.
\end{equation}
This interval vanishes in the static case. In our analysis, it already corresponds to the gray (super-critical) area in the contour plot in Fig.~\ref{fig:contour_plots_analytic}.

(iii)  $w_{+}<0$ and $w_{-}>0$: In this regime $\partial_A W|_0$ is again space-like but inward pointing. Consequently, $W(t,r)$ can play the role of a ``reversed'' radial coordinate. This happens in the static supercritical case discussed in Sec.~\ref{sec:static_sol}. In the general dynamical case, the condition on the energy density becomes
\begin{equation}
	\frac{\hat \rho}{\rho_{\rm crit}}>\tilde \gamma +  |3H + H_R| R\;.
\end{equation}
This regime is also part of the gray area in the contour plot.
\section{Effective field theory bounds}
\label{ap:EFT}
In this section the validity of the EFT description is investigated. We will find that, depending on the value of $R$, there are further bounds on the possible parameters of the model stemming from the requirement of having a valid EFT.

\begin{figure}[htb]
	\subfloat[$ R M_6=0.05 $]{
		\includegraphics[width=0.3\textwidth]{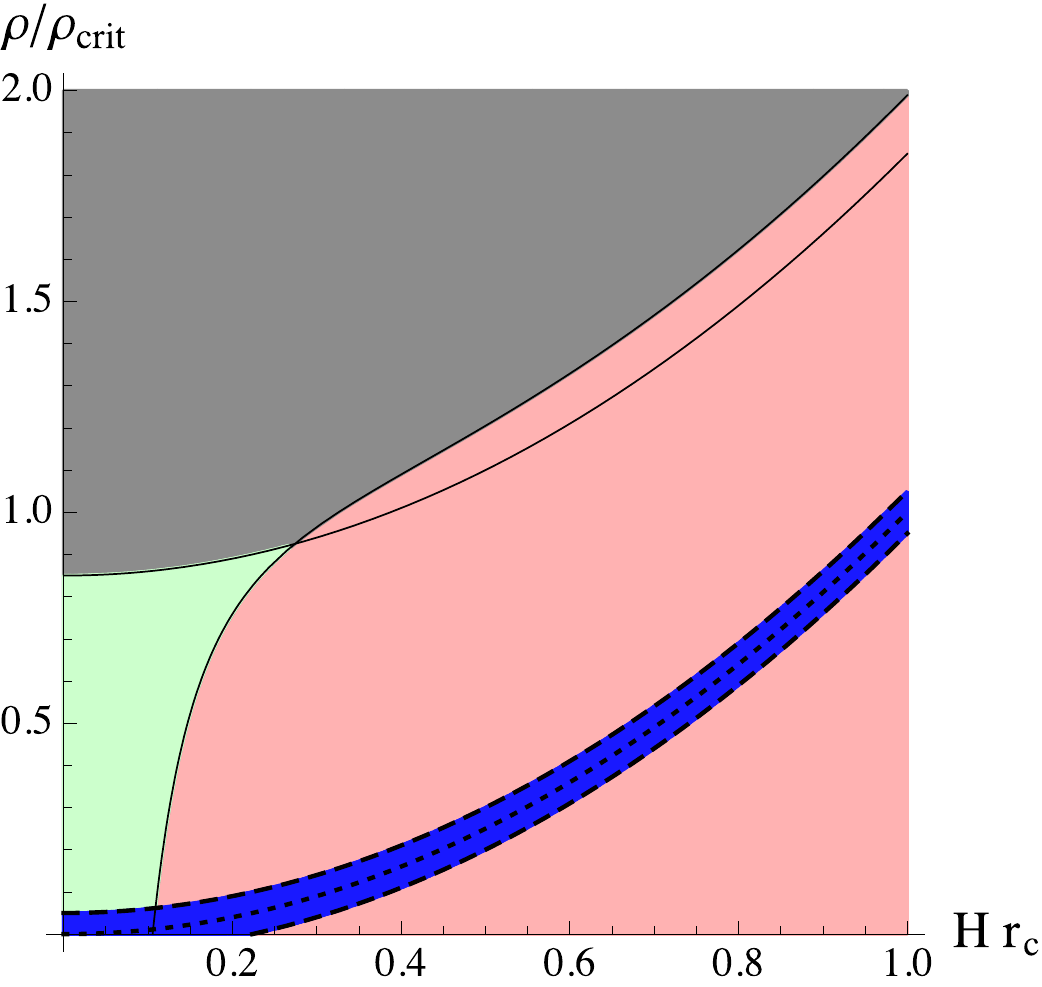}
	}%
	\\
	\subfloat[$ R M_6=0.5 $]{
		\includegraphics[width=0.3\textwidth]{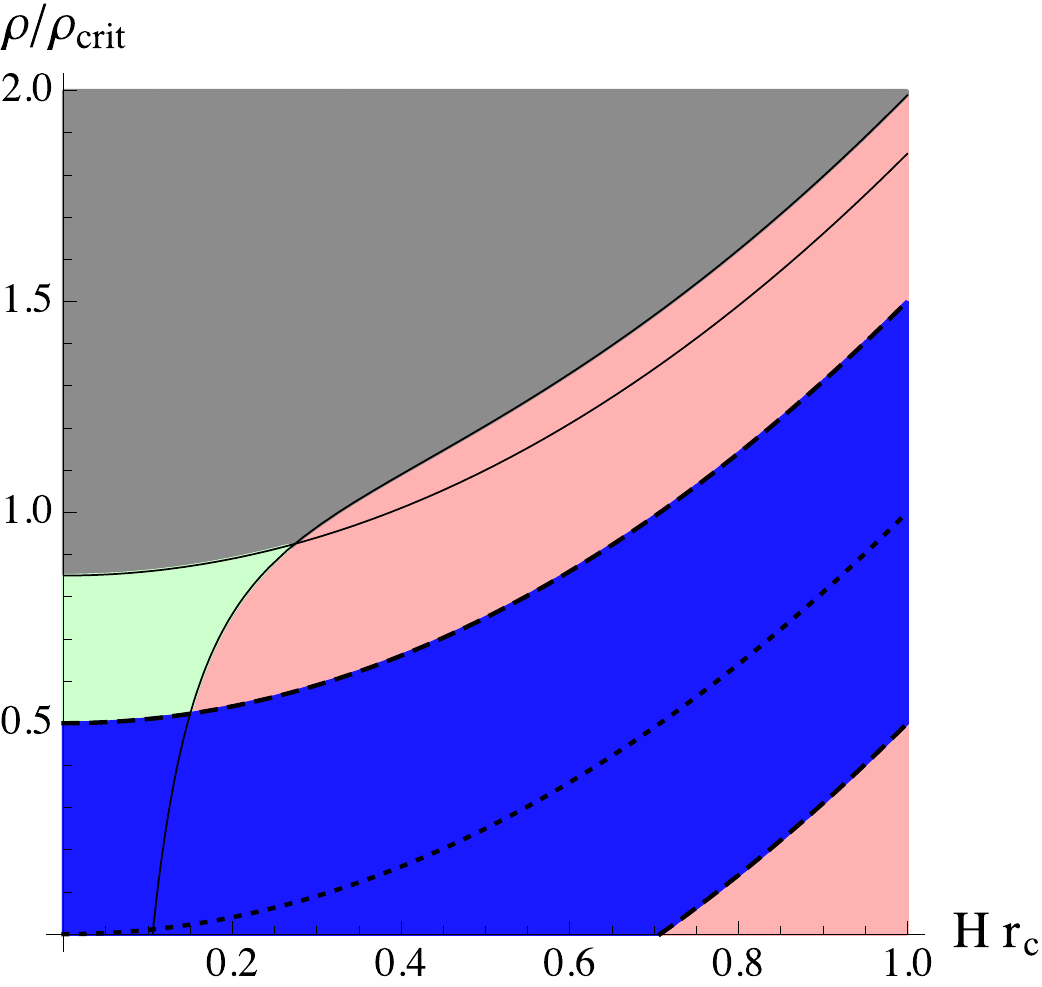}
	}
	\\
	\subfloat[$ R M_6=1.5 $]{
		\includegraphics[width=0.3\textwidth]{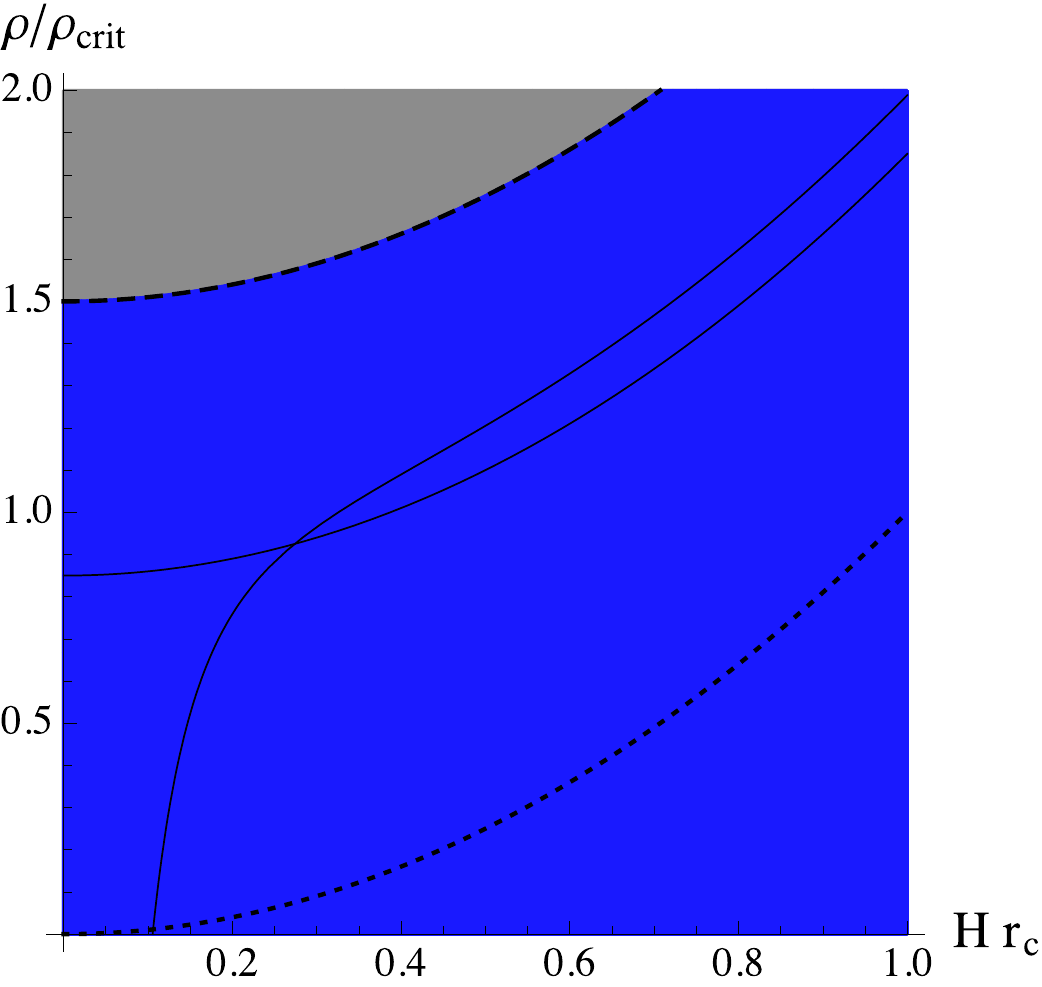}
	}
	\caption{Contour plots in the static regularization for different values of $R M_6$ with $ H R=0.05 $ fixed. The blue area (framed by the dashed lines) corresponds to an estimate of the parameter regime with a valid EFT. The dotted line in the center corresponds to $\mathcal{K}=0$, the dashed lines to $\mathcal{K}=\pm M_6$.
	}
	\label{fig:contour_plots_EFT}
\end{figure}

 Since the fundamental cutoff scale in the bulk is given by $M_6$, the breakdown of the EFT occurs once the bulk curvature terms are of the same order. We can use the extrinsic curvature as a diagnostic tool by comparing it to the $M_6$ scale. To be precise, we focus on the combination $\mathcal{K} \equiv \big([K^c_{\hphantom{c}c}] - [K^0_{\hphantom{0}0}]\big)$ which occurs in the $(0, 0)$ component of the junction conditions \eqref{israel}. Therefore, the dimensionless combination of $\mathcal{K}$ and $M_6$ for a stabilized azimuthal direction $(H_R=0)$ can be evaluated to
\begin{equation}\label{eq:EFT}
\frac{\mathcal{K}}{M_6}=\frac{1}{R M_6} \left[\left(r_c H\right)^2 -\frac{\rho}{\rho_{\rm crit}} \right] \,.
\end{equation}
Once this expression becomes of order unity, we expect the EFT to break down. (At this point higher order operators, which are normally suppressed by $M_6$, would modify the right hand side of \eqref{israel}, thereby invalidating our previous analysis.)  Obviously, this strongly depends on the scale $R$. Fig.~\ref{fig:contour_plots_EFT} visualizes the regime of validity for different values of $R M_6$. Outside the blue area (framed by the dashed lines) the EFT breaks down since  $\mathcal{K}>M_6$. The dotted line corresponds to a vanishing extrinsic curvature and hence to standard 4D evolution as becomes clear from \eqref{eq:EFT}. 

The dependence on $R$ has several consequences: On the one hand, if we are interested in studying the super-critical regime, we have to choose a large radius $R>M_6^{-1}$. On the other hand, for very small radius $R \ll M_6^{-1}$ the super-critical regime cannot be probed within a valid EFT. Moreover, the blue region then only allows for rather small deviations from standard GR.  

\newpage
\bibliography{ER_BIC_v11}

\end{document}